\documentclass[a4paper, 11pt]{article}

\usepackage{graphicx}%
\usepackage{multirow}%
\usepackage{amsmath,amssymb,amsfonts}%
\usepackage{amsthm}%
\usepackage{mathrsfs}%
\usepackage[title]{appendix}%
\usepackage{xcolor}%
\usepackage{textcomp}%
\usepackage{manyfoot}%
\usepackage{booktabs}%
\usepackage{algorithm}%
\usepackage{algorithmicx}%
\usepackage{algpseudocode}%
\usepackage{listings}%
\usepackage{natbib}
\usepackage{rotating}

\usepackage{caption}
\usepackage[a4paper,top=3cm,bottom=4cm,left=3cm,right=3cm,marginparwidth=1.75cm]{geometry}

\usepackage[colorlinks=true, allcolors=blue]{hyperref}

\title{Prior Smoothing for Multivariate Disease Mapping Models}
\author{Garazi Retegui$^{1,2*}$, Mar{\'i}a Dolores Ugarte$^{1,2}$, Jaione Etxeberria$^{1,2}$ and Alan E. Gelfand$^{3}$,\\
	\\
	\small $^1$ Department of Statistics, Computer Science and Mathematics, Public University of \\
	\small Navarre (UPNA), Arrosadia Campus, Pamplona, 31006, Navarra, Spain.\\
	\small $^2$ Institute for Advanced Materials and Mathematics (INAMAT2), Public University \\ 
	\small of Navarre (UPNA), Arrosadia Campus, Pamplona, 31006, Navarra, Spain.\\
	\small $^3$ Department of Statistical Science, Duke University, Old Chemistry Building,\\
	\small Durham, NC 27707, USA.\\
	\small $^{*}$ Corresponding author.\\
	\\
	\small garazi.retegui@unavarra.es\\ \\
	\date{}
}

\begin{document}
	\maketitle
	
	\begin{abstract}
		\noindent To date, we have seen the emergence of a large literature on multivariate disease mapping.  That is, incidence of (or mortality from) multiple diseases is recorded at the scale of areal units where incidence (mortality) across the diseases is expected to manifest dependence.   
		The modeling involves a hierarchical structure: a Poisson model for disease counts (conditioning on the rates) at the first stage, and a specification of a function of the rates using spatial random effects at the second stage.  These random effects are specified as a prior and introduce spatial smoothing to the rate (or risk) estimates.  What we see in the literature is the amount of smoothing induced under a given prior across areal units compared with the observed/empirical risks. Our contribution here extends previous research on smoothing in univariate areal data models.  Specifically, for three different choices of multivariate prior, we investigate both within prior smoothing according to hyperparameters and across prior smoothing.  
		Its benefit to the user is to illuminate the expected nature of departure from perfect fit associated with these priors, providing measures of their smoothing behavior not captured by conventional measures of predictive performance.
		We propose both theoretical and empirical metrics for our investigation and illustrate with both simulated and real data.
	\end{abstract}

	\section{Introduction}\label{S1:Intro}
	
	Disease mapping has emerged as an important area in spatial epidemiology, providing statistical techniques to estimate the geographical distribution of disease risk or rates over sets of areal units within a particular study region. Its applications are diverse. It can be used to identify geographic patterns of a disease and even generate new hypotheses about the causes (aetiology) of the disease. These analyses play an important role in the guide of resource allocation within healthcare systems and in highlighting potential health disparities across populations. 
	This is particularly important in the context of rare diseases or small-area populations, where rates can be extremely unstable and highly sensitive to small changes in disease counts or population size. 
	Historically, disease mapping has relied on neighbor-based modeling, particularly using  conditionally autoregressive (CAR) models \citep{besag1974, besag1991}, which introduce spatial smoothing to stabilize incidence or mortality estimates.
	The modeling framework is hierarchical, assuming a Poisson model for disease counts (conditional on the rates) at the first stage, and incorporating a specification of a function of the rates using spatial random effects in the second stage. These random effects are specified as a prior and introduce the spatial smoothing that results to the rate (or risk) estimates.  For detailed reviews of this modeling see, for example, \cite{banerjee2025}.
	
	Model behavior has historically been investigated with regard to goodness of fit, which measures the discrepancy between the observed data counts and the values estimated by the model, while penalizing for model complexity. 
	Predictive criteria, such as cross-validation and information criteria based on predictive performance, provide a complementary perspective by evaluating how well a model generalizes beyond the data used for fitting. Such criteria can penalize insufficient smoothing when an unsmoothed model overfits noisy area-level observations. However, the amount of smoothing that optimizes predictive performance does not necessarily correspond to the amount of smoothing that best recovers the underlying spatial risk surface. In particular, MSE-optimal shrinkage is known to systematically attenuate the variance of the underlying risk surface, potentially obscuring high-risk areas or spatial clusters even when it provides optimal point prediction. Moreover, the interpretation of predictive performance in spatial settings depends on the validation scheme. In conventional spatial cross-validation, held-out areas typically retain neighboring areas in the training data. Consequently, the resulting predictions are closer to spatial interpolation than to prediction in a completely unobserved region. This is particularly relevant for spatial priors, which explicitly borrow information from neighboring areas and may therefore perform well at interpolation. Thus, the design of spatial cross-validation determines the predictive task being evaluated. Furthermore, a single global predictive score may favor a model that performs well across the majority of areas while oversmoothing a smaller but potentially important high-risk cluster, since the choice of scoring rule implicitly determines which types of prediction errors receive greater weight. Therefore, although predictive performance provides valuable information about the consequences of smoothing, it does not directly characterize the extent of smoothing induced by a spatial prior, nor does predictive optimality necessarily correspond to the degree of smoothing that best preserves the underlying spatial risk surface. In disease mapping, smoothing involves a trade-off between reducing random variation and preserving meaningful spatial heterogeneity.
	Insufficient smoothing results in noisy maps dominated by random variation, while excessive smoothing can obscure the detection of both high- and low-risk areas. Since one of the main goals of disease mapping is to identify spatial patterns that inform public health decisions, it is crucial to understand and assess the extent of smoothing induced by the chosen model to ensure suitable and actionable conclusions.
	In fact, as \citet{stern2000} point out, when the aim is smoothing, evaluating model performance becomes challenging. In the simplest univariate spatial case, the most smoothing would occur if we have no risk factors and every area had the same intercept. The least smoothing (in fact, none)  would occur if each area was fitted independently, again with no risk factors but with its local maximum likelihood estimate (MLE). The various spatial specifications provide something in between -- neighbor-based smoothing. Therefore, one might seek to quantify the smoothing achieved by a given model specification. However, until the recent work of \citet{retegui2025} there had not been any study of the amount of smoothing arising from a choice of prior or a comparison of the amount of smoothing across commonly adopted priors. That study  focuses on the univariate setting and highlights that the choice of spatial prior plays an important role in the amount of smoothing achieved within a dataset, as each spatial prior has its own parameters that shape the smoothing.
	
	In the current paper we intend to extend this study to the multivariate context because, 
	at present, we have seen the emergence of a large literature on multivariate disease mapping. 
	That is,  incidence of (mortality from) multiple diseases is observed at an areal unit where incidence (mortality) across the diseases is expected to exhibit dependence.
	Consequently, a wide range of priors for multivariate disease mapping has been developed over the past two decades. A first class extends the univariate Conditional Autoregressive (CAR) models of \cite{besag1974} to the multivariate setting. These include Multivariate Conditional Autoregressive (MCAR) models, introduced by \cite{mardia1988} and further developed by \cite{carlin2003}, \cite{gelfand2003}, and \cite{jin2005}, as well as models based on conditional univariate distributions, with the general formulation of \cite{sain2011}. A second class comprises coregionalization models, which represent multiple outcomes as linear combinations of latent spatial processes \citep{gelfand2004, macnab2016}. However, some coregionalization approaches rely on an ordering of the responses \citep{jin2007}. To overcome this limitation, \cite{martinez2013} proposed a general coregionalization framework for multivariate areal models that encompasses many previous formulations while avoiding order dependence. Nevertheless, similar to most multivariate areal models, this approach may seriously increase computational burden which makes simultaneous modeling of a moderate to large number of responses unapproachable. To address this limitation, \cite{botella2015} proposed the M-models, which provide a computationally efficient framework that balances model flexibility, identifiability, and computational tractability. These models have been employed to explore the spatial correlation among health outcomes when these correlations are assumed to be initially unknown. Another widely used approach is spatial factor models, which represent multiple diseases through a reduced number of latent spatial factors. Scale parameters (or weights) quantify the contribution of each factor to each disease. Shared component models \citep[SCMs;][]{knorr2001, held2005,Retegui2024} are a particular case of spatial factor models in which the scale parameters are constrained to be positive. Consequently, SCMs are most appropriate when diseases are assumed to share one or more common latent risk factors, but they are less suitable when negative or no associations between spatial patterns are expected. When the observed risk variability within responses is high, the use of splines for smoothing risks is also an attractive approach \citep{vicenteetal2023}. 
	More recently, scalable approaches for high-dimensional multivariate disease mapping have been proposed to address the computational challenges associated with jointly modeling an increasing number of diseases and spatial units \citep{vicente2023}. Despite these methodological advances, little attention has been paid to understanding the smoothing properties induced by multivariate spatial priors. This is the focus of the present work.
	
	Our contribution in this paper is to extend earlier investigation of smoothing for univariate neighbor-based spatial priors to that associated with multivariate neighbor-based spatial priors. We focus on M-models because they provide a computationally efficient and flexible framework for multivariate disease mapping while avoiding some of the computational and identifiability issues associated with more general multivariate models. Moreover, M-models allow the spatial smoothing induced by the neighborhood-based prior to be studied separately from the dependence structure among diseases. Using M-models with three different spatial priors, the iCAR, the LCAR and the L$_j$CAR, we introduce a theoretical measure of smoothness based on the conditional generalized variance. In addition, we propose empirical smoothing measures for the multivariate case, examining whether different weights should be assigned to the smoothing associated with each disease. Similar to the univariate case, we limit the analysis of smoothing behavior to the parameters of the spatial priors and we do not consider the effects of population size, as the population at risk is fixed for a given application and is not a modeling choice. However, to investigate the behavior of the models as the number of areas increases within a spatial region, we vary the level of disaggregation. We employ both simulated and real data sets to quantitatively characterize the extent of smoothing within and across the spatial priors as well as to link the theoretical metrics to the empirical metrics. More precisely, for the simulation studies we consider the joint modeling of two diseases and for the real-data we increase the number of diseases to three. 
	
	The structure of the paper is as follows. In \autoref{Section:Modeling}, we describe the employed Bayesian model and spatial priors used to quantify and compare the extent of smoothing. We also derive the theoretical metric proposed for analyzing the factors influencing the extent of smoothing for each spatial prior, and we present the empirical smoothing metrics used throughout the study. Additionally, to assess whether different weights should be assigned to the smoothing associated with each disease, we discuss the theoretical expected smoothing for the “baseline” Poisson–Gamma model. In \autoref{Section:Simulation}, we present two simulation studies: the first examines the extent of smoothing and the factors influencing the level of expected smoothing for each choice of spatial prior under consideration, while the second compares the extent of smoothing across the proposed priors under different simulated scenarios. Next, in \autoref{Section:RealData}, we analyze a real dataset on colon, stomach, and pancreas cancer mortality counts, along with the corresponding population at risk for females aged 50 and older in continental Spain. Finally, \autoref{Section:Conclusions} summarizes our main contributions and outlines directions for future research.

	\section{Modeling and Methods}\label{Section:Modeling}
	This section presents the multivariate disease mapping model together with the spatial prior structures considered throughout the paper (\autoref{Subsection:Priors}). We next derive the theoretical smoothing metric, based on the conditional generalized variance (\autoref{Subsection:MTCV}). Before introducing the empirical smoothing metrics in \autoref{Section:Empirical}, we investigate the multivariate Poisson–Gamma model in \autoref{Subsection:Poisson-Gamma} to assess whether different weights should be assigned to the smoothing associated with each disease.
	
	\subsection{Model Specification and Spatial Priors} \label{Subsection:Priors}
	We assume the study region $S$ is partitioned into $G$ non-overlapping geographical area units.  Let $O_{ji}$ denote the observed count for disease $j=1,\dots, J$ in area $i=1,2,\dots,G$, and let $n_{i}$ denote the population at risk for geographical area $i$. In this work, rather than explicitly accounting for population size (which, in practice, cannot be controlled), we vary the number of geographical area units $G$ to study its effect on smoothing; as $G$ changes, the corresponding $n_i$ also varies.
	
	We consider the extent of smoothing achieved by a given model specification, for inferring about rates. Let $r_{ji}$ denote the rate for disease $j$ in area $i$. Conditional on the rates, we assume that counts follow an independent Poisson distribution, $O_{ji}\mid r_{ji} \sim Poisson\left(n_{i}r_{ji}\right)$ for each $j = 1, \dots, J$ and $i = 1, \dots, G$. The rates are modeled using a logit–Normal specification, $\textrm{logit} (r_{ji}) = \alpha_j + \theta_{ji}$,
	where $\alpha_j$ is the disease-specific intercept and $\theta_{ji}$ are the spatial main effects. Since we are focused on smoothing, we ignore covariates for our development. 
	
	Defining $\mathbf{r} = \left\{r_{ji}:  j=1, \dots, J; i=1,\dots,G\right\}$ a $J \times G$ matrix, then $\textrm{logit}(\mathbf{r}) = \mathbf{\alpha} \otimes \mathbf{1}'_{G} +\mathbf{\Theta}$ with  $\mathbf{\alpha} = (\alpha_1, \alpha_2, \dots, \alpha_J)'$, $\mathbf{1}_{G}$ a column of ones of size $G$ and $\mathbf{\Theta}=\left\{\theta_{ji}: j=1, \dots, J; i=1,\dots,G\right\}$ the corresponding matrix of spatial effects.
	Recall that, as a spatial model, we interpret this model hierarchically, i.e., the $\theta_{ji}$'s are latent and the model specifies $O_{ji}$ conditionally independent given $\theta_{ji}$ (and $\alpha_j$). We assume the vectorized spatial effects $\textrm{vec}(\mathbf{\Theta}) = \left(\mathbf{\theta}_{\cdot 1}', \dots, \mathbf{\theta}_{\cdot G}'\right)'$  follow a multivariate Normal prior, $\textrm{vec}(\mathbf{\Theta})\sim N_{JG}\left(\mathbf{0},\mathbf{\Sigma}\right)$,   where $\mathbf{\Sigma}$ is the $JG \times JG$ covariance matrix. Then, 
	$\textrm{vec}(\mathbf{\Theta})$ has a joint density proportional to $
	\textrm{exp}\left(- \textrm{vec}\left(\mathbf{\Theta}\right)^{'} \mathbf{\Sigma}^{-1}\textrm{vec}\left(\mathbf{\Theta}\right)\right)$ where $\mathbf{\Sigma}^{-1}$ corresponds to the precision matrix. When $\mathbf{\Sigma} $ is not full rank, we replace $\mathbf{\Sigma}^{-1}$ with $\mathbf{\Sigma}^{-}$ where $^{-}$ denotes the Moore-Penrose generalized inverse.
	
	In what follows we consider three families of multivariate spatial priors for  $\mathbf{\Sigma}$ to investigate the smoothing behavior induced by each through both simulation studies and real data analysis.
	While various multivariate spatial models exist, we follow the approach proposed by \cite{martinez2013} to generalize spatial distributions to the multivariate case. In this framework, the covariance matrix $\mathbf{\Sigma}$ is a combination of the within- and the between-disease covariance matrices, denoted by $\mathbf{\Sigma}_w$ and $\mathbf{\Sigma}_b$, respectively. To enhance computational efficiency and tractability, we adopt the M-model formulation from \citet{botella2015}, defining $\mathbf{\Sigma}_b = \mathbf{MM'}$.  Consequently, $\mathbf{\Theta}$ is expressed as $ \mathbf{\Theta} = \mathbf{M}\epsilon\tilde{\mathbf{\Sigma}}_w'= \mathbf{M}(\tilde{\mathbf{\Sigma}}_w\epsilon')' = \mathbf{M}\phi,$  
	where $\tilde{\mathbf{\Sigma}}_w$ is the lower Cholesky triangle of the matrix $\mathbf{\Sigma}_w$ and $\epsilon$ is a matrix containing independent Gaussian variables.
	
	According to the specification of $\phi$, different multivariate models can be defined. The simplest case assumes a common within-disease covariance structure across all diseases, known as the separable covariance structure. In this setting, $\phi = (\tilde{\mathbf{\Sigma}}_w\epsilon')'=\left[(\tilde{\mathbf{\Sigma}}_w)\epsilon_{1\cdot}: (\tilde{\mathbf{\Sigma}}_w)\epsilon_{2\cdot}:\cdots:(\tilde{\mathbf{\Sigma}}_w)\epsilon_{J\cdot}\right]'  $, and each row of $\phi$ has covariance matrix $\mathbf{\Sigma}_w$. Defining $\mathbf{I}_G$ as the $G \times G$ identity matrix,  we can express $\textrm{vec}\left(\mathbf{\Theta}\right) = \textrm{vec}\left(\mathbf{M}\phi\right) = \left(\mathbf{I}_G \otimes \mathbf{M}\right)\textrm{vec}\left(\phi\right)$
	and the covariance matrix of $\mathbf{\Theta}$ is:
	\begin{equation}\label{Eq2.1}
		\mathbf{\Sigma} = \left(\mathbf{I}_{G} \otimes \mathbf{M}\right)\textrm{cov}\left(\textrm{vec}(\phi)\right)\left( \mathbf{I}_{G} \otimes \mathbf{M}'\right) = \left(\mathbf{I}_{G} \otimes \mathbf{M}\right)\left(\mathbf{\Sigma}_w \otimes \mathbf{I}_J\right)\left( \mathbf{I}_{G} \otimes \mathbf{M}'\right) = \left(\mathbf{\Sigma}_w \otimes \mathbf{\Sigma}_b\right). 
	\end{equation}
	
	Different within-disease covariance structures can be specified to investigate the extent of induced spatial smoothing. In this work, we consider two options for the covariance matrix $\mathbf{\Sigma}_w$ in the separable case presented above—namely, the intrinsic conditional autoregressive prior \cite[iCAR,][]{besag1974} and the Leroux CAR prior \cite[LCAR,][]{leroux2000}, both fixing the precision parameter equal to one to avoid identifiability issues \citep[][Appendix B]{Urdangarin2024}.
	Namely,
	\begin{itemize}
		\item iCAR: $\mathbf{\Sigma}_w = \left(\mathbf{D} - \mathbf{W}\right)^-$, where $\mathbf{W}$ is the spatial proximity matrix defined as $w_{ii} = 0$ and $w_{ij} = 1$ if the geographical units $i$ and $j$ are neighbors and 0 otherwise, and $\mathbf{D} $ is a diagonal matrix whose elements are the number of neighbors of the $i$th area, $\mathbf{D} \equiv diag(w_1^+;\dots;w_G^+)$. Since $\left(\mathbf{D} - \mathbf{W}\right)$ is rank-deficient (of rank $G-1$ for a connected graph), $^-$ denotes the Moore-Penrose generalized inverse.  As noted above, a comparison which we do not explore here is say, between a specification with first order neighbors vs. one with first and second order neighbors. 
		
		\item LCAR: $\mathbf{\Sigma}_w = \left(\lambda \mathbf{R} + (1-\lambda)\mathbf{I_G}\right)^{-1}$, where $\mathbf{R} = \mathbf{D} - \mathbf{W}$, $\mathbf{I_G}$ is the identity matrix of size $G \times G$ and $\lambda\in[0,1]$ is a spatial dependence parameter. $\lambda$ weights the structured and unstructured spatial components. This specification yields the independence case if $\lambda = 0$, and iCAR if $\lambda = 1$.
		
	\end{itemize}
	
	On the other hand, in the most general case, where spatial distributions within diseases are considered different, $\phi$ is defined as $\phi = \left[(\tilde{\mathbf{\Sigma}}_w)_1\epsilon_{1\cdot}:(\tilde{\mathbf{\Sigma}}_w)_2\epsilon_{2\cdot}:\cdots:(\tilde{\mathbf{\Sigma}}_w)_J\epsilon_{J\cdot}\right]'$, with each row $j$ of $\phi$ having covariance matrix $(\mathbf{\Sigma}_w)_j$.
	Again, the covariance matrix of $\mathbf{\Theta}$ is
	\begin{equation}\label{Eq2.2}
		\mathbf{\Sigma} = \left(\mathbf{I}_{G} \otimes \mathbf{M}\right)\textrm{cov}\left(\textrm{vec}(\phi)\right)\left( I_{G} \otimes \mathbf{M}'\right).
	\end{equation}
	In this case, denoting the elements of the matrix $(\mathbf{\Sigma}_w)_j$ by $\left(\Sigma_w^{j}\right)_{lk}$, the covariance
	$\textrm{cov}\left(\textrm{vec}(\phi)\right)$ takes the form 
	\begin{equation}\label{Eq2.3}
		\textrm{cov}\left(\textrm{vec}(\phi)\right) = \left(\begin{array}{ccc}
			\textrm{diag}\left((\Sigma^1_w)_{11}, \dots, (\Sigma^J_w)_{11} \right)  & \cdots & \textrm{diag}\left((\Sigma^1_w)_{1G}, \dots, (\Sigma^J_w)_{1G} \right)\\
			\textrm{diag}\left((\Sigma^1_w)_{21}, \dots, (\Sigma^J_w)_{21} \right) & \cdots & \textrm{diag}\left((\Sigma^1_w)_{2G}, \dots, (\Sigma^J_w)_{2G} \right)\\
			\vdots& \cdots & \vdots\\
			\textrm{diag}\left((\Sigma^1_w)_{G1}, \dots, (\Sigma^J_w)_{G1} \right) & \cdots & \textrm{diag}\left((\Sigma^1_w)_{GG}, \dots, (\Sigma^J_w)_{GG} \right)\\
		\end{array}\right)= \left[D_{lk}\right]^G_{l,k=1}
	\end{equation}
	where $D_{lk}= \textrm{diag}\left((\Sigma^1_w)_{lk}, \dots, (\Sigma^J_w)_{lk} \right)$.
	
	For the non-separable covariance structure, we adopt $L_j$CAR model. In this approach, each disease is modeled using a LCAR prior with its own disease-specific spatial dependence parameter $\lambda_j$. Accordingly, the within-disease covariance matrix for disease $j$ is given by $(\mathbf{\Sigma}_w)_j = \left(\lambda_j \mathbf{R} + (1-\lambda_j)\mathbf{I_G}\right)^{-1}$. Under this specification, the overall covariance matrix is defined as in \autoref{Eq2.2}, and the resulting expression for  $\textrm{cov}(\textrm{vec}(\phi))$ follows from \autoref{Eq2.3}.
	
	Model fitting and inference are performed using the NIMBLE software \citep{devalpine2017}, which provides a flexible framework for implementing Markov chain Monte Carlo (MCMC) algorithms in R. Details on parameterization of the M-model and implementation details are provided in \autoref{Appendix:Model Fitting}.
	
	\subsection{Theoretical Smoothing}\label{Subsection:MTCV}
	
	In \cite{retegui2025}, we analyzed the smoothing effects induced by various spatial priors in the one-dimensional case. There, we argued that the extent of smoothing implied by different priors motivates a closer examination of the conditional variances for the various choices of the neighbor weight matrix which provides the precision matrix associated with each spatial prior. Specifically, the conditional variance is given by the inverse of the $i$th diagonal element of the precision matrix scaled by the variance value of the prior. Summing these conditional variances across all regions $i$ yields a measure of overall theoretical smoothing, denoted as total conditional variance (TCV). Extending this reasoning to the multivariate setting, we need to look at the multivariate conditional distribution of $\mathbf{\theta}_{\cdot i} \mid \mathbf{\theta}_{\cdot -i}, \forall i=1,\dots,G$ to analyze the smoothing behavior. Unlike the scalar case, it is insufficient to consider individual components; instead, we must understand how the values of neighborhood $\mathbf{\Theta}$'s affect smoothing for $\mathbf{\theta}_{\cdot i}$. 
	In particular, we need the $i$th diagonal $J \times J$ block of the inverse covariance matrix or the precision matrix $\mathbf{\Sigma}^{-1}$, denoted by  $(\mathbf{\Sigma}^{-1})_{ii}$. Inverting it yields the conditional covariance matrix
	$\left[\left(\mathbf{\Sigma}^{-1}\right)_{ii}\right]^{-1}.$
	To create a scalar from this matrix, we would typically calculate the generalized variance, i.e., the determinant of the matrix, viewing this as a measure of the concentration of the multivariate conditional distribution. Summing these determinants across all areal units provides a multivariate analogue of the total conditional variance (TCV) given by \citet{retegui2025}. Thus, our final expression for multivariate TCV is
	$\sum_{i} \det\left(\left[\left(\mathbf{\Sigma}^{-1}\right)_{ii}\right]^{-1}\right)$. A smaller TCV value indicates stronger prior-induced smoothing. It is important to note that multivariate TCV reflects only the theoretical properties of the prior. Specifically, it characterises the conditional concentration (or theoretical smoothing) implied by the prior and how this changes as a function of the prior hyperparameters. Consequently, its purpose is to analyse the effect of changes in the hyperparameters within a given modeling setting, rather than to compare smoothing across different datasets or model specifications. For this reason, we complement multivariate TCV with the empirical smoothing metrics presented in \autoref{Section:Empirical}, which quantify the realised posterior smoothing resulting from the joint effect of the prior, the likelihood, and the observed data.  Furthermore, as discussed in \autoref{Appendix:Model Fitting}, M-models are subject to identifiability issues, whereby $\phi$ and $\mathbf{M}$ are not identifiable. However, $\mathbf{\Theta}$, and the covariance matrices $\mathbf{\Sigma_b}$ and $\mathbf{\Sigma}$ are all identifiable \citep[][Appendix B]{Urdangarin2024}.
	
	\begin{table}[b!]
		\centering
		\caption{Theoretical multivariate total conditional variance (MultiTCV) for the spatial priors considered in this paper.}
		\label{tab0}
			\resizebox{\textwidth}{!}{
		\begin{tabular}{c|c|c}
			Spatial Prior  & $\Sigma$  & MultiTCV \\
			\hline
			iCAR & $\left(\mathbf{\Sigma}_w \otimes \mathbf{\Sigma}_b\right)$ & $\sum_{i} \det\left(\left[\left(\left(\mathbf{D} - \mathbf{W}\right) \otimes \mathbf{\Sigma}_b^{-1}\right)_{ii}\right]^{-1}\right)$\\
			LCAR & $\left(\mathbf{\Sigma}_w \otimes \mathbf{\Sigma}_b\right)$ & $\sum_{i} \det\left(\left[\left(  \left(\lambda \mathbf{R} + (1-\lambda)\mathbf{I_G}\right) \otimes \mathbf{\Sigma}_b^{-1}\right)_{ii}\right]^{-1}\right)$\\
			$L_j$CAR & $\left(\mathbf{I}_{G} \otimes \mathbf{M}\right)\textrm{cov}\left(\textrm{vec}(\phi)\right)\left( I_{G} \otimes \mathbf{M}'\right)$ &$\sum_{i} \det\left(\left[\left(\left(\left(\mathbf{I}_{G} \otimes \mathbf{M}\right)\left[D_{lk}\right]^G_{l,k=1}\left( \mathbf{I}_{G} \otimes \mathbf{M}'\right)\right)^{-1}\right)_{ii}\right]^{-1}\right)$\\
			\hline
		\end{tabular}
	}
	\end{table}
	
	The resulting multivariate TCV expressions for the spatial priors presented in \autoref{Subsection:Priors} are given in \autoref{tab0}. Note that, for the iCAR prior, $\Sigma_w$ is rank deficient, with rank $G-1$. Consequently,  $\Sigma = \left(\mathbf{\Sigma}_w \otimes \mathbf{\Sigma}_b\right)$ is also rank deficient, with rank $JG-J$. However, since $\Sigma_b$ is positive definite, the diagonal $J\times J$  blocks of the precision matrix $\Sigma^-$ are themselves positive definite and therefore invertible. As a result, the theoretical metric can be defined as $\sum_{i} \det\left(\left[\left(\left(\mathbf{D} - \mathbf{W}\right) \otimes \mathbf{\Sigma}_b^{-1}\right)_{ii}\right]^{-1}\right)$.

 \subsection{Multivariate Poisson-gamma model}\label{Subsection:Poisson-Gamma}

We briefly consider the Poisson-gamma model with regard to two issues.  Does global disease rate affect posterior empirical smoothing, and therefore different weights should be assigned to the smoothing associated with each disease? Does dependence between diseases affect posterior empirical smoothing?
Although the Poisson–Gamma model does not include spatial dependence, it yields explicit expressions for posterior smoothing. This makes it a useful analytical tool for investigating whether global disease rates and between-disease dependence influence the amount of smoothing, before introducing the spatial models. For illustration, we consider two diseases.

Recall that the Poisson-gamma model assumes $O^{(j)}|\eta^{(j)}_{i} \sim Poisson(E^{(j)}_{i}\eta^{(j)}_{i})$, for $j=1,2$, where $E^{(j)}_{i}$ is the expected number of cases, usually computed under internal standardization, and $\eta^{(j)}_{i}$ is the relative risk for areal unit $i$ and disease $j$.
We introduce dependence between the diseases through dependence between $\eta^{(1)}_{i}$ and $\eta^{(2)}_{i}$.  We do this by writing $\eta^{(1)}_{i} = \phi^{(1)}_{i} + \xi_{i}$ and $\eta^{(2)}_{i} = \phi^{(2)}_{i} + \xi_{i}$.  Here, $\phi^{(1)}_{i} \sim Gamma(a^{(1)}, b)$, $\phi^{(2)}_{i} \sim Gamma(a^{(2)}, b)$, and $\xi_{i} \sim Gamma(c,b)$, all independent.  The shared $\xi_{i}$ makes the $\eta$'s dependent so marginally $O^{(1)}|\eta^{(1)}_{i}$ and $O^{(2)}|\eta^{(2)}_{i}$ are dependent.  In fact, we can straightforwardly calculate the correlation between $\eta^{(1)}_{i}$ and $\eta^{(2)}_{i}$.  It is $\frac{c}{\sqrt{(a^{(1)}+c )(a^{(2)}+c)}}$ and increasing $c$ increases dependence. Further, $\eta^{(1)}_{i} \sim Gamma(a^{(1)} + c, b)$ and $\eta^{(2)}_{i} \sim Gamma(a^{(2)} + c, b)$ so we retain the conjugacy in the modeling.  In addition, $\mu^{(1)} = (a^{(1)}+c)/b$ and $\mu^{(2)} = (a^{(2)}+c)/b$ with $\sigma^{2(1)} = (a^{(1)}+c)/b^{2}$ and $\sigma^{2(2)} = (a^{(2)}+c)/b^{2}$.  So, the posterior mean for disease $j$ in areal unit $i$ becomes
\begin{eqnarray*}
	E\left(\eta^{(j)}_i \mid O^{(j)}_i\right) &=& \frac{a^{(j)} + c + O^{(j)}_i}{b  + E^{(j)}_i} = \frac{\frac{\mu^{2(j)}}{\sigma^{2(j)}} + O^{(j)}_i}{\frac{\mu^{(j)}}{\sigma^{2{(j)}}}+ E^{(j)}_i} =\frac{\frac{\mu^{(j)}}{\sigma^{2(j)}}}{\frac{\mu^{(j)}}{\sigma^{2(j)}} + E^{(j)}_i} \mu^{(j)}  + \frac{E^{(j)}_i}{\frac{\mu^{(j)}}{\sigma^{2(j)} }+E_i^{(j)}}\frac{O^{(j)}_{i}}{E^{(j)}_{i}}\\
	& =&  (1-w^{(j)}_{i})\mu^{(j)}  +  w^{(j)}_{i}\frac{O^{(j)}_{i}}{E^{(j)}_{i}}
\end{eqnarray*}
where $w^{(j)}_{i} = \frac{E^{(j)}_i}{\frac{\mu^{(j)}}{\sigma^{2(j)}}  + E^{(j)}_i}$.

However, the main model considered in this paper is formulated in terms of disease rates rather than relative risks. Specifically, $O^{(j)}_i|r^{(j)}_i \sim Poisson(n_ir^{(j)}_i)$. To relate this formulation to the standard relative risk model, note that $E^{(j)}_{i}\eta^{(j)}_{i} = n_ir^{(j)}_i$, which implies $r^{(j)}_i=\frac{E^{(j)}_{i}}{n_i}\eta^{(j)}_{i}$. Using internal standardization, $E^{(j)}_{i} = n_i\frac{\sum_i O^{(j)}_{i}}{\sum_in_i}$, so that $\frac{E^{(j)}_{i}}{n_i}=\frac{\sum_i O^{(j)}_{i}}{\sum_in_i} = \bar{r}^{(j)}$ where $\bar{r}^{(j)}$ is the average rate for each disease $j$, which is constant across areas. Therefore,
$r_i^{(j)}=\bar{r}^{(j)}\eta_i^{(j)}$, showing that the disease rate is simply the relative risk scaled by the overall average rate. Consequently, the smoothing results derived below for relative risks translate directly to disease rates.
With $\eta^{(j)}_{i}$ having the Gamma distribution above, $E\left(r^{(j)}_i\right) = E\left(\bar{r}^{(j)}\eta^{(j)}_{i}\right)= \bar{r}^{(j)}\mu^{(j)}$ and $var\left(r_i^{(j)}\right) = var\left(\bar{r}^{(j)}\eta^{(j)}_{i}\right) = \left(\bar{r}^{(j)}\right)^{2} \sigma^{2(j)}$. We obtain that
\begin{eqnarray*}
	E\left(r_i^{(j)} \mid O^{(j)}_i\right) &=& \bar{r}^{(j)}E\left(\eta_i^{(j)} \mid O_i^{(j)}\right) = \bar{r}^{(j)}\left( \left(1-w_{i}^{(j)}\right)\mu^{(j)}  +  w_{i}^{(j)}\frac{O_{i}^{(j)}}{E_{i}^{(j)}}\right)\\
	&=& \left(1-w_{i}^{(j)}\right)\bar{r}^{(j)}\mu^{(j)}  +  w_{i}\frac{O_{i}^{(j)}}{n_{i}}.
\end{eqnarray*}

With no smoothing we have $\hat{r}^{(j)}_{i}= O^{(j)}_{i}/n_{i}$. The difference between the posterior mean smoothing for areal unit $i$ and $\hat{r}_{i}^{(j)}$ is $E\left(r^{(j)}_i \mid O^{(j)}_i\right)-\hat{r}^{(j)}_{i} = \frac{\mu^{(j)}}{\bar{r}^{(j)}\sigma^{2(j)} n_i + \mu^{(j)}}\left(\bar{r}^{(j)}\mu^{(j)} - \hat{r}^{(j)}_{i}\right).$
Plugging in $\mu^{(j)}$ and $\sigma^{2(j)}$ as above, we see that $1-w^{(j)}_{i} = \frac{(a^{(j)} + c)/b}{\bar{r}^{(j)}\sigma^{2(j)}n_{i} + \mu^{(j)}} = \frac{b}{\bar{r}^{(j)} n_i + b}$.
We see that the dependence cancels out of the weighting  indicating that it is not necessary to assign different weights to the smoothing associated with each disease.  Moreover, since $1-w^{(j)}_{i} = \frac{b}{\bar{r}^{(j)} n_i + b}$, a smaller average disease rate $\bar{r}^{(j)}$ implies more smoothing. However, the difference term becomes $\bar{r}^{(j)}(a^{(j)}+c)/b - \hat{r}^{(j)}_{i}$.  So, the dependence  will drive the bias in the smoothing; larger values of $c$, i.e., stronger dependence will tend to encourage greater deviations in the difference term.  That is, our smoothing metrics will tend to be larger with more dependence between the diseases.

\subsection{Empirical Smoothing Metrics}\label{Section:Empirical}

Here we provide the empirical metrics we employ to calculate the resulting smoothing under a given multivariate prior. We consider insight gleaned from the Poisson-gamma model to motivate our choices.
We define different measures because smoothing comparisons may be performed across different spatial priors, diseases, or spatial aggregations, and each comparison requires an appropriate empirical metric.

We first define an aggregated relative square smoothness criterion ($SRSS_j$) for each disease $j$, which allows comparisons within the same disease across different spatial priors or modeling specifications. For example, $SRSS_j$ can be used to compare the smoothing of disease $j$ when it is modeled in a pairwise framework with different diseases, or when it is jointly modeled with different combinations of diseases.
With $\hat{r}_{ji} = O_{ji}/n_i$, the aggregated relative square smoothness criterion for each disease $j$ is defined as $$SRSS_j = \sum_{i} RSS_{ji}= \sum_{i} \frac{(E(r_{ji}|O_{ji}) - \hat{r}_{ji})^{2}}{E(r_{ji}|O_{ji})}.$$
In addition, we record the maximum empirical relative square smoothing for each disease $j$, $$maxRSS_{j} = \max_{i}RSS_{ji}=\max_{i} \frac{(E(r_{ji}|O_{ji}) - \hat{r}_{ji})^{2}}{E(r_{ji}|O_{ji})}.$$ This measure allows comparisons within the same disease across different spatial priors or modeling specifications and quantifies the largest area-specific contribution to the overall smoothing.

To facilitate direct comparisons between priors when the same set of diseases is modeled, we further define an overall smoothing measure by summing the disease-specific criteria,
$SRSS = \sum_{j=1}^{J} SRSS_j$
a formulation consistent with the Poisson–gamma model, which suggests that weighting across diseases is unnecessary. The resulting metric allows comparisons of the overall smoothing induced by different priors under the same modeling specification.

To be able to do comparison of spatial priors among different set of diseases, spatial structures, or disaggregation levels, we define two metrics. To compare the smoothing between diseases, we propose the relative smoothing proportion for disease $j$ (RSP$_j$): $$RSP_j = \frac{\sum_{i}\left[E\left(r_{ji}|O_{ji}\right) - \hat{r}_{ji}\right]^2}{\sum_{i}\left(\bar{r}_j - \hat{r}_{ji}\right)^2},$$
where $\bar{r}_{j}$ is the average rate of disease $j$.
This criterion evaluates the degree of smoothing relative to the maximum possible smoothing for disease $j$. On the other hand, if the goal is to compare multivariate models with different priors, we can use the overall metric, smoothing proportion (SP): $$SP =\frac{\sum_{i}\left[E\left(r_{1i}|O_{1i}\right) - \hat{r}_{1i}\right]^2 + \sum_{i}\left[E\left(r_{2i}|O_{2i}\right) - \hat{r}_{2i}\right]^2}{\sum_{i}\left(\bar{r}_1 - \hat{r}_{1i}\right)^2 + \sum_{i}\left(\bar{r}_2 - \hat{r}_{2i}\right)^2}.$$

For the simulation studies, under a given prior $M$, we draw $B$ replicate sets of observations. Consequently, the empirical smoothing metrics are computed as averages across these replicates. Note that, the crude rate $\hat{r}_{ji}^{b}= O^b_{ji}/n_i$ is specific for each replicate $b=1,2,\dots,B$.

\section{Simulation Studies} \label{Section:Simulation}

This section reports the results of two types of simulation studies conducted to examine the amount of smoothing under the multivariate framework for the priors introduced in \autoref{Section:Modeling}. The first study, presented in \autoref{Section:WithinSimulation}, explores both the theoretical and empirical amounts of smoothing associated with each spatial prior, using the multivariate TCV to identify the parameters affecting the expected amount of smoothing. This analysis focuses on understanding how smoothing behaves within individual priors. Then, \autoref{Section:AcrossSimulation} extends the investigation to a comparative setting, assessing differences in smoothing across priors under a variety of simulated scenarios. This second study is designed to mimic real-world data contexts, where several correlated diseases are modeled jointly. Before presenting the results, we describe the simulation scenarios under consideration.

\subsection{Design of the Simulation Scenarios}\label{Section:Scenarios}

These scenarios are designed to reflect different correlation and rate variability between diseases, allowing us to assess how the amount of smoothing varies within and across priors under diverse conditions. To simplify the analysis and make it easier to interpret, we restrict to the analysis of two diseases ($J=2$). We consider four different scenarios that vary the correlation and rate variability between diseases. Specifically, Scenarios 1 and 2 exhibit high correlation between diseases, whereas Scenarios 3 and 4 present low correlation. Within each pair, the scenarios differ by rate variability, with Scenarios 2 and 4 presenting higher variability than Scenarios 1 and 3, respectively. In addition, we explore the effect of varying spatial disaggregation levels across continental Spain. We consider three levels of spatial disaggregation: 47 provinces, and 100, and 300 areas. To define the spatial structures for the 100 and 300 areas, we preserve the proportion of municipalities per province as observed in the real case of continental Spain.

To simulate the data, we use a model-free approach to ensure a fair comparison among the spatial priors. To generate realistic scenarios, we base the simulations on Gaussian Process (GP) surfaces. Specifically, we first define a continuous rate surface using a GP and then discretize it to obtain a finite set of area-level rates. This approach allows us to control the smoothness of the surface, avoiding unrealistically smooth patterns, and it also enables us to generate multiple spatial disaggregation levels from the same underlying surface for each scenario.
Therefore, we first create a high-resolution grid for continental Spain to simulate the true rates for these different spatial units. We then define the rate surface $\mathbf{r}(\mathbf{s})$ on the logit scale, as it is the link function of the Poisson-logitNormal models. Since we are analyzing a multivariate case, we examine the smoothing induced by different spatial priors when studying two diseases simultaneously. For the simulation study, we require two distinct rate surfaces, denoted as $r_j(\mathbf{s})$ for each disease $j=1,2$. 
Furthermore, to compare the amount of smoothing across priors, we define four scenarios by varying the correlation between diseases, achieved by modifying both the spatial variation and the variation in rates. However, the spatial distribution and rate variation of the first disease remain constant across all scenarios. This approach enables us to isolate the effects of changes in the second disease and assess how smoothness varies.

To define the rate surface $\mathbf{r}(\mathbf{s}) = \left[r_1(\mathbf{s}): r_2(\mathbf{s})\right]'$  on the logit scale, we proceed in two steps. First, we construct a bivariate Gaussian Process (GP) to generate a baseline rate surface that is common across all scenarios under consideration. In the second step, given that our simulation study involves only two diseases, we apply a simple coregionalization approach. This allows us to flexibly model both dependent and independent disease structures. Importantly, it is this second step that enables the definition of distinct scenarios. We begin by outlining the procedure for the first step. We generate bivariate pairs over a fine grid of locations $\mathbf{\omega}(\mathbf{s})$ where $\mathbf{\omega}(\mathbf{s}) =\left[\omega_1(\mathbf{s}): \omega_2(\mathbf{s})\right]'$ are independent and follow $\textrm{vec}\left(\mathbf{\omega}\right) \sim N\left(\mathbf{\mu}(\mathbf{\theta}), \mathbf{C}(\mathbf{\theta}) \right)$. To introduce spatial variation, we assume fixed locations, where the mean function $\mathbf{\mu}(\mathbf{\theta}) = \textrm{vec}\left(\left[\mu_1(\theta): \mu_2(\theta)\right]'\right)$ depends on the distance from these locations, decreasing according to an exponential function. Each disease is assigned distinct exposure sites, allowing for variation in spatial patterns across diseases.

To smooth the surfaces, we adopt the spatial covariance matrix, $\mathbf{C}(\mathbf{\theta}) = \sigma_C^2\mathbf{R}(\mathbf{\varphi})$, with $\mathbf{R}(\mathbf{\varphi})$ a correlation matrix whose size corresponds to $J$ times the number of grid points. To simulate the surface, we consider $\mathbf{\omega}(\mathbf{s}) =\left[\omega_1(\mathbf{s}),\omega_2(\mathbf{s})\right]'$ to be independent. Each disease-specific component is modeled using its own spatial correlation matrix, denoted by $R_j(\mathbf{\varphi})$. 
The  $(i,k)$ elements of $R_j(\mathbf{\varphi})$ are computed using a Matérn correlation function,
$\rho(s_i,s_k;\varphi) = \frac{1}{2^{v-1}\Gamma(v)}\left(\mid \mid s_i-s_k\mid \mid \varphi\right)^v K_v\left(\mid \mid s_i-s_k \mid \mid \varphi\right).$
We fix a common smoothness parameter $\nu$ for both diseases while we vary the decay parameter $\varphi$ between them.

\begin{figure}
	\centering
	\includegraphics[width=\linewidth]{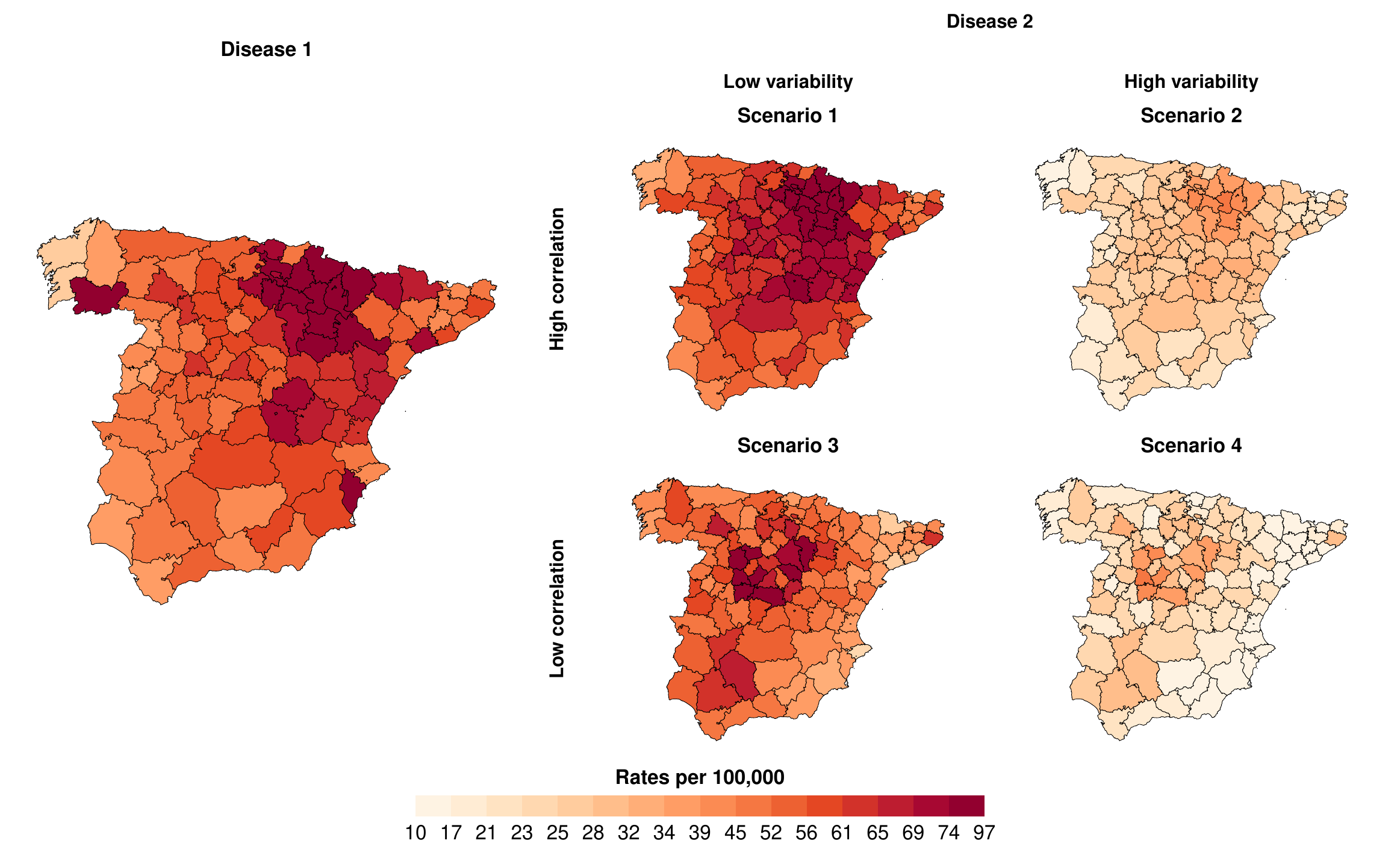}
	\caption{Simulated true rate surfaces for disease 1 and disease 2 across Scenarios 1–4 ($G=100$). Scenarios combine high or low between-disease correlation with high or low rate variability. Disease 1 remains fixed across all four scenarios.}
	\label{fig1}
\end{figure}

In the second step, we implement the coregionalization approach. 
Specifically, we define $\textrm{logit} (\mathbf{r}(\mathbf{s})) = \mathbf{A} \mathbf{\omega}(\mathbf{s})$,
where $\mathbf{A}$ is a $2 \times 2$ matrix that governs the degree of dependence between diseases. This formulation allows us to flexibly simulate both dependent and independent disease scenarios within a unified framework. The definition of the parameters in $\mathbf{A}$ determines the four simulation scenarios outlined previously, which combine high or low between-disease correlation with high or low rate variability. Our first assumption is to fix the value of $a_{11}=1.005$, ensuring that the rate surface for the first disease remains consistent across all scenarios. A value slightly greater than one was selected to calibrate the simulated rates on the original scale. Additionally, we set $a_{12}=0$, meaning that the second disease does not influence the first.  With these constraints, we define different scenarios by varying the parameters $a_{21}$ and $a_{22}$, allowing us to explore different relationships and dependencies between the two diseases while maintaining a fixed reference for the first disease. 
We begin by defining Scenario 1, in which both diseases are spatially correlated. This correlation may arise from shared risk factors or because they represent different health outcomes of the same underlying disease. To model this, we set the parameters $a_{21} = 0.73$ and $a_{22}=0.23$, assigning greater weight to the spatial pattern of disease 1 and thereby inducing a strong dependence between the two diseases. In Scenario 2, we again assume that the diseases are correlated. However, in this case, disease 2 is rarer or less common than disease 1.  To reflect this, we need to reduce the overall intensity of disease 2 while maintaining its spatial correlation with disease 1 on the rate scale, rather than on the logit scale used to define the scenarios. Since the scenarios are defined on the latent logit scale, this requires increasing the corresponding latent-scale parameters. Therefore, we multiply both $a_{21}$ and $a_{22}$ by a factor of 1.12. This adjustment preserves approximately the same level of between-disease dependence on the rate scale while reducing the overall rates of disease 2. 
For Scenario 3, we assume that the diseases are independent, meaning disease 1 does not influence disease 2. To enforce this independence, we set $a_{21}=0$, ensuring that the second disease develops without any contribution from the first. We set $a_{22}=1.015$ so that the rates of disease 2 are of a similar magnitude to those of disease 1.
Finally, in Scenario 4, the diseases remain independent, but disease 2 is rarer or less common than disease 1. To achieve this while preserving the spatial structure of disease 2, we rescale the corresponding latent parameter by multiplying $a_{22}$ by a factor of 1.12. Once we obtain the rate surfaces on the logit scale, we compute the rates for each areal unit of the study region, i.e. $r_{ji}$ for each  $j = 1, 2$ and $i=1,\dots, G$. Specifically, we compute	
$r_{ji} = \frac{1}{|A_i|}\int_{s \in A_i} \frac{1}{(1+\exp(-r_j(s)))} ds \approx \frac{1}{H}\sum_{s \in A_i} \frac{1}{(1+\exp(-r_j(s)))},$ with $|A_i|$ the area of areal unit $A_i$ and $H$ the number of $s$ points in $A_i$. Note that diseases were simulated independently; however, spatial aggregation increases dependence within and across diseases, so in scenarios 3 and 4 low correlations between diseases may still be observed.
\autoref{fig1} displays the simulated true rate surfaces for $G=100$ spatial units across all four scenarios. The corresponding true rate surfaces for $G=47$ and $G=300$ are provided in \autoref{SimStudy}.

Finally, to generate the observed counts $O_{ji}$ from a Poisson distribution, it is necessary to establish the population. Specifically, the population data used in the simulation study have been taken from the real dataset examined in the next section, for each spatial structure. We generate $B = 1000$ datasets for each scenario defined. 

\subsection{Within Simulation Study}\label{Section:WithinSimulation}
In this section, we individually assess the three priors proposed in \autoref{Section:Modeling}. More precisely, we compare theoretical smoothing with empirical smoothing to assess whether the model's smoothing, as evaluated using the multivariate TCV metric, aligns with that arising through the empirical metrics. For the within prior simulation study, we analyze the four different scenarios described above that vary in terms of the correlation and variability between diseases, and we focus in the spatial disaggregation level of $G=100$. 
We provide a detailed analysis of the results for Scenario 1, along with a summary of results for a representative case across all scenarios. 

For each prior, we vary the values of the parameters that influence the degree of smoothing. In all the spatial priors considered, for a given $J$, the covariance matrix $\mathbf{\Sigma_b}$ has $(J(J+1)/2)$ parameters, since it is symmetric. In this simulation study, we focus on the case with two diseases, i.e., $J=2$. Therefore, $\mathbf{\Sigma_b}$ consists of three parameters: $\Sigma_{11}$, $\Sigma_{22}$ and $\Sigma_{12}=\Sigma_{21}$. We expect the first two to affect smoothing like $\sigma^2$ in the univariate case.  However, the effect of $\Sigma_{12}$ will be of most interest since it captures the dependence between the diseases.
We fix the diagonal elements $\Sigma_{11}$ and $\Sigma_{22}$ to 0.0025, 0.04 and 0.25. To define the covariance term $\Sigma_{12}$ between the two diseases, we fix the correlation coefficient $\rho$, using the relationship: $\Sigma_{12} = \rho\sqrt{\Sigma_{11}\Sigma_{22}}$. We consider two values for $\rho$: 0 and 0.7.  We chose to fix the values of $\mathbf{\Sigma_b}$ due to its interpretability, which allows for more intuitive understanding of each parameter. However, to fit the M-models, we must define the matrix $\mathbf{M}$ such that $\mathbf{\Sigma_b}=\mathbf{MM'}$. As discussed above, to avoid overparameterization of the matrix $\mathbf{M}$, we rely on the eigen-decomposition of $\mathbf{\Sigma_b}$ and define $\mathbf{M} = \left(\mathbf{H} diag\left(\sqrt{\kappa_1}, \dots, \sqrt{\kappa_J}\right)\right)'$, where $\kappa_1, \ldots, \kappa_J$ are the eigenvalues and the columns of $H$ are the corresponding eigenvectors.
In addition, the LCAR and L$_j$CAR priors include extra parameters in the within-disease covariance matrix $\mathbf{\Sigma_w}$. More precisely,  we consider $\lambda = 0.2$ and $0.8$ for the LCAR spatial prior and for the $\lambda_j$ of the L$_j$CAR spatial prior, we evaluate all possible combinations of the values $0.2$ and $0.8$.

\begin{table}[h!]
	\caption{Theoretical and empirical smoothing criteria values for the iCAR spatial prior for $G=100$ across different $\Sigma_{11}$, $\Sigma_{22}$ and $\rho$ values.}
	\label{tab1}
	\begin{center}
		\resizebox{\textwidth}{!}{
			\begin{tabular}{lllrrrrrcrrrrr}
				\hline
				&&&\multicolumn{5}{c}{$\rho=0$}&&\multicolumn{5}{c}{$\rho=0.7$}\\
				\cline{4-8}\cline{10-14}
				&&& MultiTCV & SP & RSP & SRSS & MaxRSS && MultiTCV & SP & RSP & SRSS & MaxRSS \\ 
				\hline
				\multicolumn{14}{l}{$\Sigma_{11}=0.0025$}\\[1.2ex]
				&$\Sigma_{22}=0.0025$ & Disease 1 ($j=1$) &  &  & 0.70 & 416.31 & 49.04 &&  &  & 0.67 & 388.66 & 46.22\\ 
				&& Disease 2 ($j=2$) &  &  & 0.72 & 350.93 & 54.77 &&  &  & 0.66 & 316.24 & 50.54  \\ 
				&& Total & 0.0000 & 0.70 & 1.42 & 767.23 & && 0.0000 & 0.65 & 1.33 & 704.89 &   \\ [2.2ex]
				&$\Sigma_{22}=0.04$ & Disease 1 ($j=1$)&  &  & 0.70 & 416.36 & 49.05 & &  &  & 0.71 & 421.82 & 49.77\\ 
				&& Disease 2 ($j=2$)  &  &  & 0.48 & 219.38 & 36.61 &&  &  & 0.47 & 211.14 & 35.64\\ 
				&& Total& 0.0005 & 0.59 & 1.18 & 635.75 & & & 0.0003 & 0.59 & 1.18 & 632.96 &  \\ [2.2ex]
				&$\Sigma_{22}=0.25$ & Disease 1 ($j=1$) &  &  & 0.70 & 416.36 & 49.05 &&  &  & 0.75 & 448.58 & 52.70  \\ 
				&& Disease 2 ($j=2$)&  &  & 0.39 & 176.77 & 24.16 & &  &  & 0.39 & 175.53 & 24.06\\ 
				&& Total & 0.0031 & 0.55 & 1.10 & 593.13 & & & 0.0016 & 0.57 & 1.14 & 624.12 &    \\ [2.7ex]
				\multicolumn{14}{l}{$\Sigma_{11}=0.04$}\\[1.2ex]
				&$\Sigma_{22}=0.0025$ & Disease 1 ($j=1$)&  &  & 0.41 & 234.19 & 33.73 & & &  & 0.41 & 227.71 & 33.40 \\ 
				&& Disease 2 ($j=2$)&  &  & 0.72 & 350.88 & 54.78&&  &  & 0.70 & 344.09 & 54.36 \\ 
				&& Total & 0.0005 & 0.55 & 1.13 & 585.06 & & & 0.0003 & 0.54 & 1.11 & 571.80 &  \\ [2.2ex]
				&$\Sigma_{22}=0.04$ & Disease 1 ($j=1$) &  &  & 0.41 & 234.24 & 33.84 & & &  & 0.41 & 230.83 & 33.88 \\ 
				&& Disease 2 ($j=2$)&  &  & 0.48 & 219.24 & 36.55 &&  &  & 0.47 & 213.43 & 36.75 \\ 
				&& Total & 0.0080 & 0.44 & 0.89 & 453.47 & & & 0.0041 & 0.43 & 0.88 & 444.26 &  \\ [2.2ex]
				&$\Sigma_{22}=0.25$ &Disease 1 ($j=1$)&  &  & 0.41 & 234.19 & 33.77&&  &  & 0.43 & 244.35 & 34.93 \\ 
				&& Disease 2 ($j=2$) &  &  & 0.39 & 176.85 & 24.07&&  &  & 0.39 & 177.08 & 24.82 \\ 
				&& Total  & 0.0497 & 0.40 & 0.81 & 411.04&&& 0.0253 & 0.40 & 0.82 & 421.43 &  \\ [2.7ex]
				\multicolumn{14}{l}{$\Sigma_{11}=0.25$}\\[1.2ex]
				&$\Sigma_{22}=0.0025$ & Disease 1 ($j=1$)&&  &   0.32 & 181.70 & 23.89 &&  &  & 0.32 & 180.41 & 23.94 \\ 
				&& Disease 2 ($j=2$) & & & 0.72 & 350.88 & 54.78 &&  &  & 0.75 & 370.26 & 57.50 \\ 
				&& Total & 0.0031 & 0.49 & 1.04 & 532.58 & && 0.0016 & 0.51 & 1.07 & 550.67 &  \\ [2.2ex]
				&$\Sigma_{22}=0.04$ & Disease 1 ($j=1$) &  &  & 0.32 & 181.61 & 23.82 & &  &  & 0.32 & 181.29 & 24.54\\ 
				&& Disease 2 ($j=2$)&  &  & 0.48 & 219.39 & 36.57 &&  &  & 0.48 & 223.84 & 38.56\\ 
				&& Total & 0.0497 & 0.38 & 0.80 & 401.01 &&& 0.0253 & 0.39 & 0.80 & 405.13 &   \\ [2.2ex]
				&$\Sigma_{22}=0.25$ & Disease 1 ($j=1$) &  &  & 0.32 & 181.63 & 23.85 &&  &  & 0.32 & 186.22 & 26.01 \\ 
				&& Disease 2 ($j=2$) &  &  & 0.39 & 176.84 & 24.16 &&  &  & 0.40 & 181.28 & 26.51 \\ 
				&& Total & 0.3106 & 0.34 & 0.71 & 358.48 &&& 0.1584 & 0.35 & 0.72 & 367.50 &  \\ 
				\hline
			\end{tabular}
		}
	\end{center}
\end{table}

\autoref{tab1} presents the theoretical multivariate TCV metric, denoted as MultiTCV, together with the empirical smoothing criteria presented in \autoref{Section:Empirical} for the iCAR spatial prior. Results are reported for $G=100$ under varying values of $\Sigma_{11}$, $\Sigma_{22}$ and $\rho$. For a given $\rho$ value, the theoretical metric aligns with the empirical criteria, both indicating stronger smoothing as $\Sigma_{11}$ and $\Sigma_{22}$ decrease. Recall that the theoretical and empirical metrics behave reciprocally: while an increase in MultiTCV corresponds to a decrease in smoothing, an increase in the empirical metrics indicates that the amount of smoothing has increased. When $\rho=0$, changes in $\Sigma_{22}$ do not affect the smoothing of disease 1. In contrast, when $\rho$ is different from 0, variations in $\Sigma_{22}$ slightly influence the smoothing of disease 1, with smoothing increasing as $\Sigma_{22}$ grows. This effect becomes more pronounced as $\Sigma_{11}$ decreases. A similar pattern is observed when varying $\Sigma_{11}$ instead of $\Sigma_{22}$. Regarding the role of $\rho$ in the amount of smoothing, an increase in $\rho$ leads to a slight decrease in the empirical smoothing, while the theoretical metric shows the opposite trend, as $\rho$ increases, the MultiTCV decreases, implying more smoothing. Notably, when the MultiTCV reaches the highest values reported in \autoref{tab1}, the behavior of the empirical and theoretical metrics begins to align.

\begin{figure}
	\begin{center}
		\centerline{\includegraphics[width=0.9\linewidth]{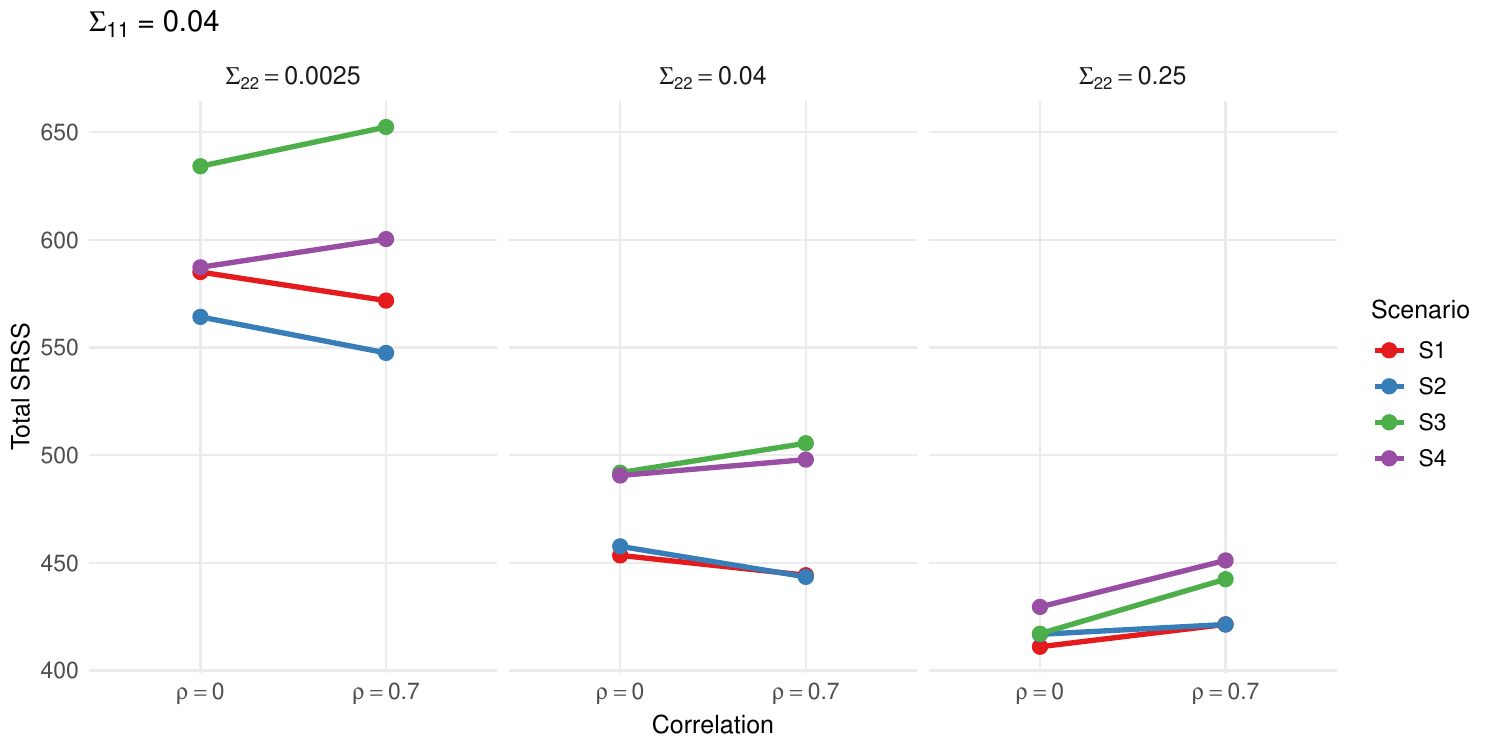}}
	\end{center}
	\caption{Total SRSS values, representing the combined smoothing of both diseases, for $\Sigma_{11}=0.04$ and varying $\Sigma_{22}$ across the four scenarios and different correlation levels.\label{fig2}}
\end{figure}

Since the effect of $\rho$ on the amount of smoothing does not fully align with theoretical expectations, we examine this further in \autoref{fig2}. This figure displays the total SRSS values obtained across the four scenarios for the specific case where $\Sigma_{11}=0.04$, while varying $\Sigma_{22}$. Different colors correspond to different scenarios. Although \autoref{fig2} does not include the MultiTCV, as shown in \autoref{tab1}, increasing $\rho$ leads to a decrease in MultiTCV, which theoretically indicates greater smoothing. This behavior is consistent across all scenarios. However, \autoref{fig2} reveals different empirical smoothing behaviors depending on the scenario and the value of $\Sigma_{22}$. For the smallest $\Sigma_{22}$ values (0.0025 and 0.04), for scenarios 1 and 2, the SRSS decrease as $\rho$ increases from 0 to $0.7$, suggesting that the empirical and theoretical metrics are not aligned. In contrast, for scenarios 3 and 4, the opposite behavior is observed, the empirical and theoretical metrics are consistent, both reflecting increased smoothing as $\rho$ grows.
The disparity in the behavior of the smoothing could be due to the underlying correlation structure between diseases in each scenario. In Scenarios 1 and 2, where the amount of smoothing does not follow the theoretical expectation, the simulated correlation between diseases is high; consequently, setting $\rho=0.7$ yields a value closer to the true correlation. In contrast, in Scenarios 3 and 4, the simulated between disease correlation is low, so increasing $\rho$ results in a more noticeable change in the correlation structure. These findings suggest that the degree of realised posterior smoothing is influenced not only by the prior specification but also by the interaction between the prior, the likelihood, and the observed data, as discussed by \cite{gelman2017}. As discussed in \autoref{Subsection:MTCV}, MultiTCV characterises the theoretical properties of the prior, whereas the empirical metrics quantify realised posterior smoothing. 
As $\Sigma_{22}$ increases, the empirical and theoretical metrics align across all scenarios. Nevertheless, as shown in \autoref{tab1}, the impact of $\rho$ on the amount of smoothing is relatively minor compared with the effects of $\Sigma_{11}$ and $\Sigma_{22}$. 

\begin{table}[h!]
	\caption{Theoretical and empirical smoothing criteria values for the LCAR spatial prior for $G=100$ across different $\Sigma_{11}$, $\Sigma_{22}$, $\lambda$ and $\rho$ values.}
	\label{tab2}
	\begin{center}
		\resizebox{\textwidth}{!}{
			\begin{tabular}{llllrrrrrcrrrrr}
				\hline
				&&&&\multicolumn{5}{c}{$\rho=0$}&&\multicolumn{5}{c}{$\rho=0.7$}\\
				\cline{5-9}\cline{11-15}
				&&&& MultiTCV & SP & RSP & SRSS & MaxRSS && MultiTCV & SP & RSP & SRSS & MaxRSS \\ 
				\hline
				\multicolumn{15}{l}{$\Sigma_{11}=0.0025$}\\[1.2ex]
				&$\Sigma_{22}=0.0025$ &$\lambda=0.2$& Disease 1 ($j=1$) & && 0.78 & 473.16 & 59.30 &  & & & 0.73 & 437.74 & 55.52 \\ 
				&&& Disease 2 ($j=2$)  &  & & 0.80 & 403.14 & 63.11 &&  &  & 0.73 & 364.07 & 59.63\\ 
				& &&Total & 0.0002 & 0.77 & 1.58 & 876.30 & && 0.0001 & 0.72 & 1.46 & 801.81 &  \\ [2.2ex]
				&& $\lambda=0.8$ &  Disease 1 ($j=1$) &  &  & 0.74 & 442.37 & 52.73 &&  &  & 0.70 & 412.24 & 49.53  \\ 
				& && Disease 2 ($j=2$) & &  & 0.75 & 370.69 & 57.88 &&  &  & 0.69 & 332.57 & 53.25\\ 
				& &&Total & 0.0000 & 0.73 & 1.49 & 813.06 & & & 0.0000 & 0.68 & 1.39 & 744.81 & \\ [2.2ex]
				&$\Sigma_{22}=0.25$ &$\lambda=0.2$ & Disease 1 ($j=1$)&  &  & 0.78 & 473.58 & 59.10 &&  &  & 0.82 & 499.67 & 61.02 \\ 
				&&& Disease 2 ($j=2$)   &  &  & 0.35 & 162.31 & 18.71 &&  &  & 0.35 & 160.76 & 18.28 \\ 
				& &&Total & 0.0200 & 0.57 & 1.13 & 635.90 & && 0.0102 & 0.59 & 1.17 & 660.42 &  \\ [2.2ex]
				&& $\lambda=0.8$ &  Disease 1 ($j=1$) &  &  & 0.74 & 443.01 & 52.89 &&  &  & 0.79 & 474.40 & 56.31\\ 
				&&& Disease 2 ($j=2$)  &  &  & 0.36 & 164.04 & 22.92 &&  &  & 0.36 & 162.87 & 22.86\\ 
				& &&Total  & 0.0043 & 0.55 & 1.10 & 607.04 &  && 0.0022 & 0.58 & 1.15 & 637.27 & \\ 
				[2.7ex]
				\multicolumn{15}{l}{$\Sigma_{11}=0.25$}\\[1.2ex]
				&$\Sigma_{22}=0.0025$ &$\lambda=0.2$& Disease 1 ($j=1$) &  &  & 0.27 & 158.98 & 17.09 &&  &  & 0.27 & 157.31 & 17.82  \\ 
				&&& Disease 2 ($j=2$) &  &  & 0.80 & 403.07 & 63.01 &&  &  & 0.82 & 414.12 & 64.01  \\ 
				&&& Total & 0.0200 & 0.51 & 1.07 & 562.05 &  && 0.0102 & 0.52 & 1.09 & 571.44 &\\ [2.2ex]
				&& $\lambda=0.8$ &  Disease 1 ($j=1$)  &  &  & 0.29 & 167.46 & 22.82&&  &  & 0.29 & 166.91 & 22.88  \\ 
				&&& Disease 2 ($j=2$) &  &  & 0.75 & 370.37 & 57.77 &&  &  & 0.78 & 387.48 & 59.89  \\ 
				&&& Total  & 0.0043 & 0.50 & 1.04 & 537.83 & & & 0.0022 & 0.51 & 1.07 & 554.38 &  \\ [2.2ex]
				&$\Sigma_{22}=0.25$ &$\lambda=0.2$& Disease 1 ($j=1$) &  &  & 0.27 & 158.39 & 17.65 &&  &  & 0.29 & 172.37 & 21.97\\ 
				&&& Disease 2 ($j=2$)  &  &  & 0.35 & 160.51 & 18.20 &&  &  & 0.36 & 170.74 & 21.53 \\ 
				&&& Total  & 2.0027 & 0.30 & 0.62 & 318.90 & & & 1.0214 & 0.32 & 0.66 & 343.11 &  \\ [2.2ex]
				&& $\lambda=0.8$ &  Disease 1 ($j=1$) &  &  & 0.29 & 167.74 & 22.55 &&  &  & 0.31 & 176.85 & 25.41 \\ 
				&&& Disease 2 ($j=2$) &  &  & 0.36 & 163.79 & 22.81 &&  &  & 0.37 & 171.38 & 26.07 \\ 
				&&& Total  & 0.4267 & 0.32 & 0.65 & 331.53 & && 0.2176 & 0.33 & 0.68 & 348.23 &  \\ 
				\hline
			\end{tabular}
		}
	\end{center}
\end{table}

Analyzing the smoothing amount of the LCAR spatial prior yields the results presented in \autoref{tab2}. 
As before, results correspond to $G=100$ under varying values of $\Sigma_{11}$, $\Sigma_{22}$ and $\rho$. For the LCAR spatial prior, an additional parameter $\lambda$ is included. We evaluate two values, $0.2$ and $0.8$. We summarize the main findings here and omit the cases with $\Sigma_{11}=0.04$ and $\Sigma_{22}=0.04$, which are included in full in Supplementary Material~B.
As in the iCAR case, for fixed values of $\rho$ and $\lambda$, the theoretical and empirical measures are consistent, both showing stronger smoothing as $\Sigma_{11}$ and $\Sigma_{22}$ decrease. For a given $\lambda$, when $\rho=0$, the smoothing of disease 1 is unaffected by changes in $\Sigma_{22}$, whereas for $\rho=0.7$ variations in $\Sigma_{22}$ slightly influence the smoothing of disease 1, with smoothing increasing as $\Sigma_{22}$ becomes larger. The same pattern holds when varying $\Sigma_{11}$ instead of $\Sigma_{22}$. 
However, the effect of $\rho$ is less stable, and the theoretical and empirical metrics do not always align. For large $\Sigma_{11}$ or $\Sigma_{22}$ values, both measures show consistent behavior: increasing $\rho$ leads to increased smoothing. This is similar to what is observed for the iCAR spatial prior and the effect of $\rho$ varies across the scenarios defined (see  Supplementary Material~B).
In contrast, the impact of $\lambda$ differs across the two measures. Higher $\lambda$ yields slightly less empirical smoothing, whereas the theoretical metric decreases as $\lambda$ increases, implying stronger smoothing. 
This discrepancy disappears when both $\Sigma_{11}$ and $\Sigma_{22}$ are set to 0.25.
To gain further intuition about the role of $\lambda$ parameter in the smoothing amount, we examine a plot analogous to \autoref{fig2}.
This figure, reported in Supplementary Appendix Figure~S4, shows consistent behavior across scenarios, indicating that neither the correlation nor the variability structure of the rates alters how smoothing responds to changes in $\lambda$.

The priors discussed here, iCAR and LCAR, provide separable covariance structures; results for the non-separable covariance structure, L$_j$CAR, are presented in Supplementary Material~B. Nevertheless, the conclusions reached are similar to those previously discussed. $\Sigma_{11}$ and $\Sigma_{22}$ have the largest effect on the degree of smoothing. The effect of $\rho$ varies across scenarios and depends on the values of $\Sigma_{jj}$, for $j=1,2$. However, as $\Sigma_{jj}$ increases, the empirical and theoretical metrics align across all scenarios and less smoothing is observed. $\lambda_1$ and $\lambda_2$ behave reciprocally, the theoretical metric indicates that increasing these parameters corresponds to greater smoothing. Similar to $\rho$, their effects depends on the value of $\Sigma_{jj}$ for $j=1,2$, but as $\Sigma_{jj}$ increases, the empirical and theoretical metrics align across all scenarios.

\subsection{Across Simulation Study} \label{Section:AcrossSimulation}

\begin{table}[b!]
	\caption{Empirical smoothing criteria values obtained by the separable iCAR and LCAR spatial priors, and the non-separable L$_j$CAR spatial prior for $G=100$ and $G=300$.}
	\label{tab3}
	\begin{center}
		\resizebox{\textwidth}{!}{
			\begin{tabular}{ccrrrrrrrrrrrrrr}
				\hline
				&& \multicolumn{4}{c}{iCAR} &&\multicolumn{4}{c}{LCAR}&&\multicolumn{4}{c}{L$j$CAR}\\
				\cline{3-6}\cline{8-11}\cline{13-16}
				&&  SRSS  & maxRSS & RSP & SP &&  SRSS  & maxRSS & RSP & SP&&  SRSS   & maxRSS & RSP & SP\\ 
				\hline
				\multicolumn{11}{l}{\textbf{G=100}}\\
				&\multicolumn{10}{l}{\textbf{Scenario 1}}\\
				&Disease 1 ($j=1$)  & 194.72 & 28.37 & 0.34 & & & 187.14 & 27.65 & 0.33 &&& 185.67 & 27.40 & 0.32 & \\ 
				&Disease 2 ($j=2$)  & 193.20 & 30.58 & 0.42 &  && 182.14 & 28.72 & 0.40 &&& 181.64 &  28.41 & 0.40 &  \\ 
				&Total &  387.92 &  & 0.76   & 0.38 && 369.28 &    & 0.73 & 0.36 && 367.31 &    & 0.72 & 0.36 \\ 
				&&  &  &  &  &  & &&& \\ 
				&\multicolumn{10}{l}{\textbf{Scenario 2}}\\
				&Disease 1 ($j=1$) & 190.60 &  27.32 & 0.33 & & & 180.99 & 25.39 & 0.32 && & 181.10 &  26.26 & 0.31 &   \\ 
				&Disease 2 ($j=2$) &  198.28 & 33.41 & 0.56 & & & 192.79  & 33.32 & 0.55 &&& 189.78 &  31.77 & 0.54 & \\ 
				&Total  & 388.88 &   & 0.89 & 0.38 && 373.78 &  & 0.87 &   0.37 && 370.88 &   &0.85  & 0.36 \\ 
				&&  &  &  &  &  & &&& \\ 
				&\multicolumn{10}{l}{\textbf{Scenario 3}}\\
				&Disease 1 ($j=1$)  &  186.13 &  25.73 & 0.32 &&& 176.71 &  25.34 & 0.31 && & 175.40 & 24.28 & 0.30 &     \\ 
				&Disease 2 ($j=2$) &  207.96 & 33.21 & 0.38 &  && 193.91 &  31.55 & 0.36 &&& 193.68 &  30.85 & 0.35 &   \\ 
				&Total &  394.09 &  &  0.70  & 0.35 && 370.62 &  &  0.67 & 0.33&& 369.08 &    & 0.65 & 0.33  \\ 
				&&  &  &  &  &  &  &&&\\ 
				&\multicolumn{10}{l}{\textbf{Scenario 4}}\\
				&Disease 1 ($j=1$) &  184.43 & 25.79 & 0.32 & & & 175.17 & 25.92 & 0.30 &&&  173.51 & 24.45 & 0.30 & \\ 
				&Disease 2 ($j=2$) & 205.59 &  33.19 & 0.51 & & & 196.62 &  33.12 & 0.47 &&& 194.60 &  31.91 & 0.47 &  \\ 
				&Total & 390.03 &  &  0.83  & 0.36 && 371.79 &  & 0.77 &   0.34 && 368.11 &    & 0.77 & 0.34 \\
				\multicolumn{11}{l}{ }\\
				\multicolumn{11}{l}{\textbf{G=300}}\\
				&\multicolumn{10}{l}{\textbf{Scenario 1}}\\
				&Disease 1 ($j=1$)  & 2613.34 & 258.23 & 0.62 & &&  2542.57 &  262.57 & 0.59 &   && 2537.08 &  266.95 & 0.58  \\ 
				&Disease 2 ($j=2$)& 2539.24 & 254.69 & 0.69 & &&  2438.01 & 253.34 & 0.65 &&& 2452.51 & 267.88  & 0.65  \\ 
				&Total & 5152.57   &  &1.31& 0.65 && 4980.58 &  & 1.24 & 0.62 && 4989.60 &  &  1.23 & 0.62\\ 
				&&  &  &  &  &  &  &&&\\ 
				&\multicolumn{10}{l}{\textbf{Scenario 2}}\\
				&Disease 1 ($j=1$) & 2558.89 & 229.53 & 0.60 &&& 2498.98 & 248.95 & 0.58 & && 2401.00 &  228.87 & 0.56    \\ 
				&Disease 2 ($j=2$) & 2555.48 & 294.93 & 0.80 &&& 2545.97 & 316.01& 0.77 & && 2516.86 & 319.20 & 0.76   \\ 
				&Total & 5114.38   &  &1.40& 0.66 && 5044.94 &  & 1.38 & 0.63 && 4917.85 &  &  1.32 & 0.61\\ 
				&&  &  &  &  &  &  &&&\\ 
				&\multicolumn{10}{l}{\textbf{Scenario 3}}\\
				&Disease 1 ($j=1$) & 2550.90 & 229.54 & 0.60 &&& 2351.21 & 221.44 & 0.55 &&& 2336.43 & 215.64 & 0.54  \\ 
				&Disease 2 ($j=2$) & 2611.31 &  255.47 & 0.62 &&& 2405.39 & 249.22 & 0.56 &&& 2402.52 &  248.63 & 0.56  \\ 
				&Total & 5162.21 &   &1.22& 0.61 && 4756.59 &  & 1.11 & 0.55 && 4738.95 &  & 1.10 & 0.55 \\ 
				&&  &  &  &  &  &  &&&\\ 
				&\multicolumn{10}{l}{\textbf{Scenario 4}}\\
				&Disease 1 ($j=1$) & 2514.01 & 221.09 & 0.59 &&& 2274.20 &  205.36 & 0.53 &&& 2267.62  & 207.54& 0.53 \\
				&Disease 2 ($j=2$) & 2534.87 & 283.75 & 0.73&&& 2455.64 & 302.10 & 0.68 &&& 2472.23 & 299.06 & 0.68\\
				&Total & 5048.88 &  & 1.32 & 0.63 && 4729.83 &   & 1.21 & 0.57 & & 4739.85 &  & 1.21  & 0.57 \\
				\hline
			\end{tabular}
		}
	\end{center}
\end{table}

To compare smoothing across priors, we fit the multivariate Poisson–logitnormal model to each scenario using the iCAR and LCAR separable covariance structures and the non-separable  L$_j$CAR prior (\autoref{Section:Modeling}). Smoothing is assessed using empirical metrics: aggregated and maximum RSS, SRSS, disease-specific relative smoothing proportions (RSP$_j$), and the overall smoothing proportion (SP), as defined in \autoref{Section:Empirical}. 
Results for grids $G=100$ and $G=300$ are presented in \autoref{tab3}, while results for $G=47$ are provided in  Supplementary Material~C. Note that, for comparisons of smoothing within scenarios, all empirical criteria can be used. However, for comparisons across scenarios, only RSP$_j$ and SP are directly comparable. In addition, SRSS$_1$ and maxRSS$_1$ can also be compared across scenarios, since disease 1 remains constant, as explained in \autoref{Section:Scenarios}.

In general, the LCAR and L$_j$CAR spatial priors exhibit less smoothing than iCAR across all scenarios according to the empirical metrics, a trend most clearly reflected in the SRSS values. Scenario~3 shows the lowest smoothing proportions; this scenario is characterized by low spatial correlation and similar rate variability between diseases. The reduction in smoothing as correlation decreases is consistent with the findings of the within prior simulation study. Furthermore, the across prior simulation study shows that rate variability between diseases also influences smoothing. Specifically, Scenario 2 exhibits higher relative smoothing proportions than Scenario 1, as does Scenario 4 compared to Scenario 3. Within each pair, the spatial patterns remain identical, yet the variability in disease rates differs, with the second disease presenting lower rates.
These results suggest that wider disparities in disease rates lead to more pronounced smoothing. Furthermore, disease 1 exhibits consistent smoothing across all scenarios, decreasing only marginally as correlation or rate variability diminishes. 
As the number of areas increases, the overall amount of smoothing increases, and differences in smoothing across priors become more pronounced, though the main conclusions remain consistent. Notably, smoothing does not increase proportionally across diseases; instead, the RSP$_j$ values become more similar among diseases, with the largest relative increase observed for disease 1, which initially exhibited a lesser amount of smoothing. The corresponding 95\% bands for the differences across priors are reported in the Supplementary Material C.

\begin{figure}[h!]
	\centering
	\includegraphics[width=0.95\linewidth]{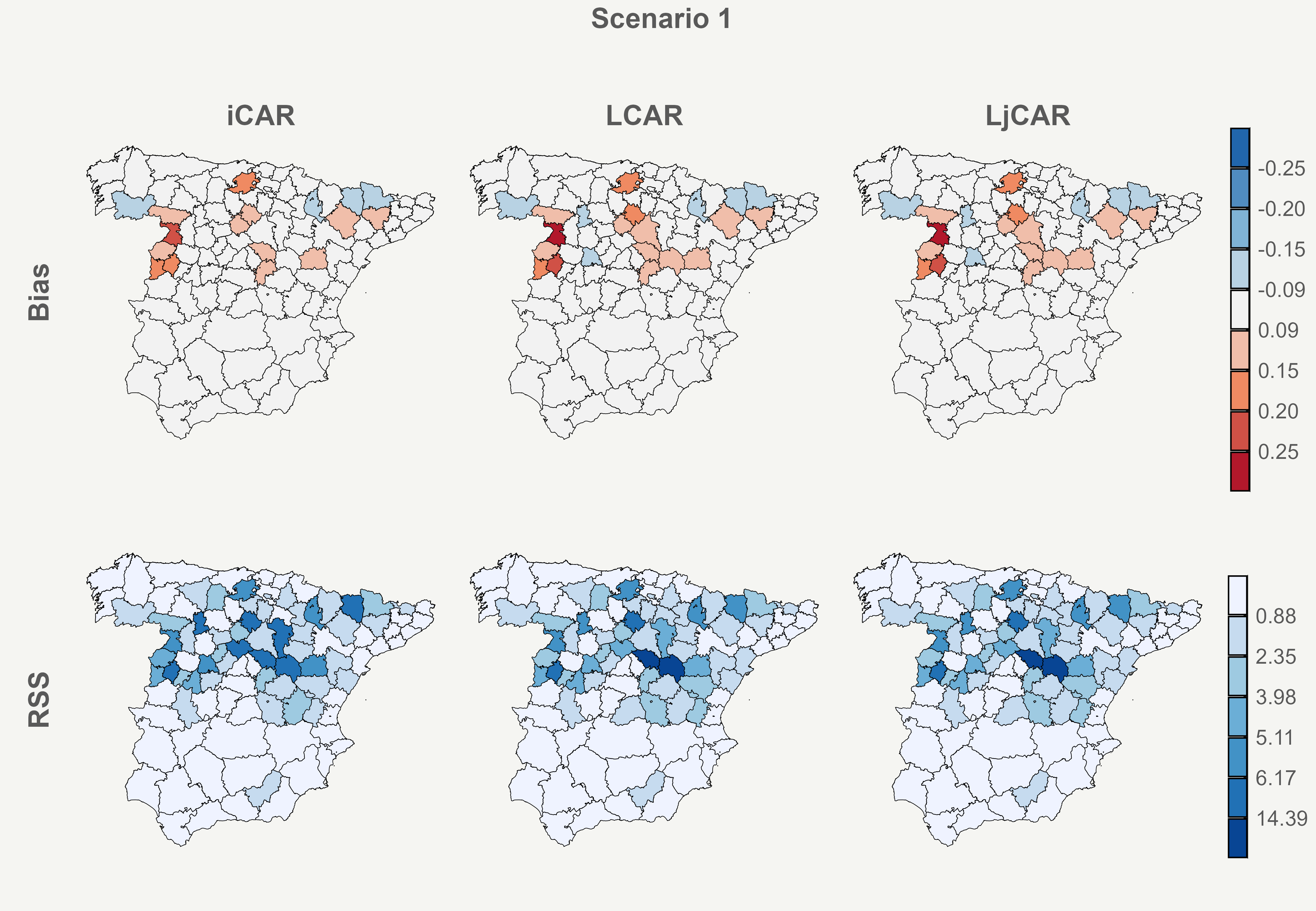}
	\caption{Mean bias across the $B$ replicates of the simulation study and the average across the $B$ replicates of the $RSS_{1i}$  values for the iCAR, LCAR and L$_j$CAR priors in Scenario 1 for disease 1, when analyzing 100 areas.}
	\label{fig3}
\end{figure}

Although the primary goal of this paper is to evaluate, both theoretically and empirically, the degree of smoothing induced by different spatial priors, the simulation study also enables us to assess whether the shift of the posterior estimates away from the crude rates is appropriate relative to the true underlying risk surface.
To this end, we compute the relative bias of the posterior estimated rate, $E(r_{ji}\mid O_{ji})$ with respect to the true simulated rate $r_{ji}$, as $\frac{E(r_{ji}\mid O_{ji}) - r_{ji} }{r_{ji}}$.  
\autoref{fig3} shows the mean bias across the $B$ replicates of the simulation study for the iCAR, LCAR, and L$_j$CAR priors in Scenario 1 for Disease 1, when analyzing 100 areas, together with the corresponding average across the $B$ replicates of the $RSS_{1i}$ values. For the average of $RSS_{1i}$, the legend reports the 50th, 75th, 85th, 90th, 95th, and 97th percentiles. For bias, however, we highlight areas with an absolute bias of at least 9\%, as biases exceeding 10\% are commonly regarded as substantial in the literature.
We observe that areas subject to a low degree of smoothing generally exhibit low bias. However, substantial bias is also observed in some areas subject to a high degree of smoothing. Comparison with the true rate surfaces shown in \autoref{fig1} reveals a clear spatial pattern: areas with high positive bias, where the estimated rate exceeds the true rate, tend to be surrounded by areas with higher rates than the area itself. Conversely, areas with high negative bias tend to be surrounded by areas with lower rates than the area itself. Comparison of the different spatial priors shows that the areas exhibiting larger biases are broadly consistent across priors. These findings are consistent across scenarios and also hold for disease 2; the corresponding figures are provided in the Supplementary Material~C.

\section{Real Data} \label{Section:RealData}

\begin{figure}[t!]
	\centering
	\includegraphics[width=0.95\linewidth]{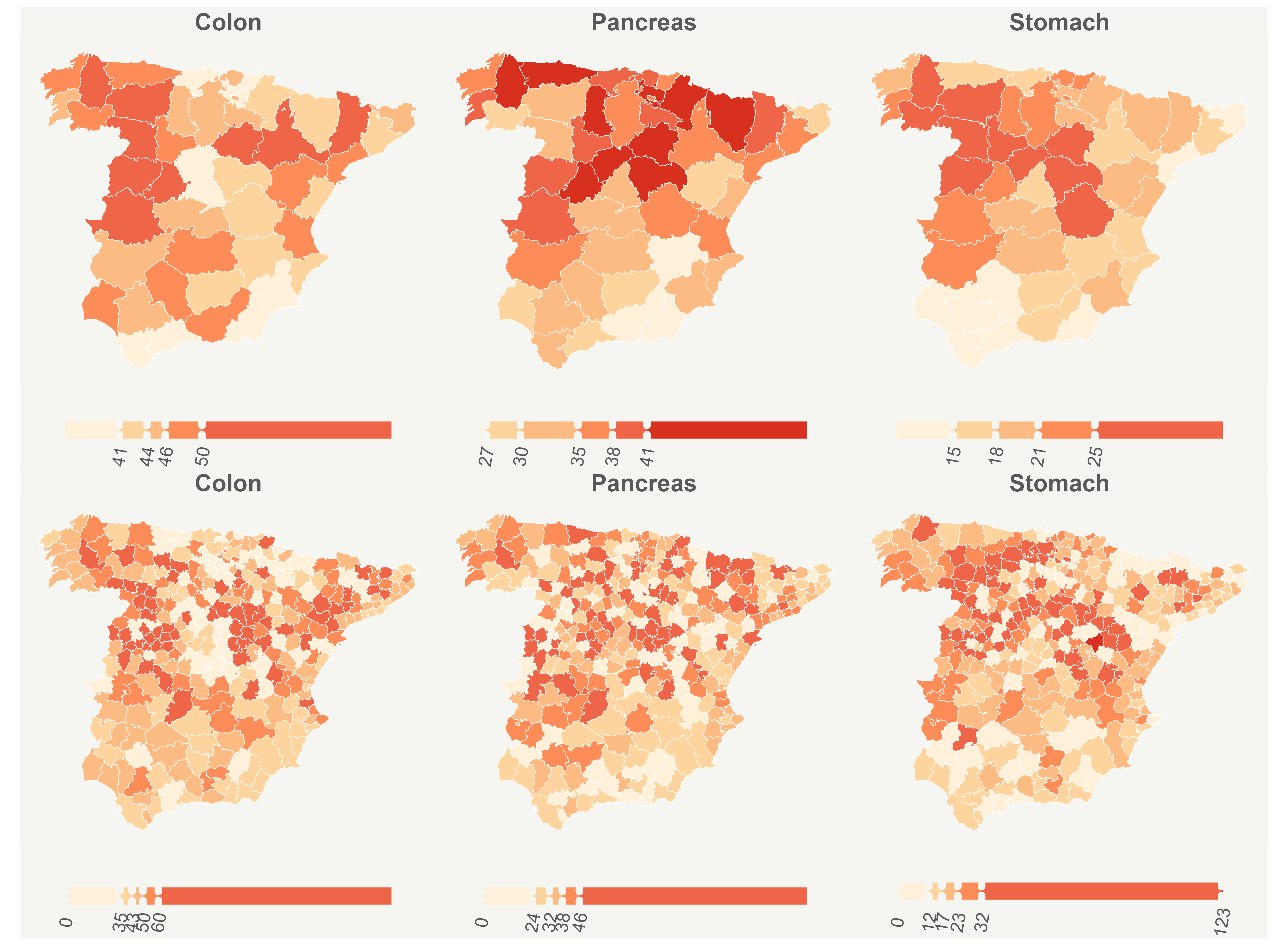}
	\caption{Crude mortality rates per 100,000 inhabitants for colon, pancreatic and stomach cancer in continental Spain, shown for the 47 real provinces ($G = 47$, top panels) and for the disaggregation with $G=300$ (bottom panels).}
	\label{fig4}
\end{figure}

We analyze mortality rates for colon, pancreatic, and stomach cancer in continental Spain at three spatial resolutions: the 47 administrative provinces ($G=47$), and two grids consisting of $G = 100$ and $G = 300$ spatial units. Provincial names and their geographic distributions are shown in Supplementary Material~D. The data consists of mortality counts and population-at-risk figures for Spanish females aged 50+ during 2020–2022, provided by the Spanish Statistical Office (INE). Crude mortality rates per 100,000 inhabitants for the three cancer types are illustrated in \autoref{fig4}, comparing the 47 administrative provinces ($G = 47$) with the disaggregated level of $G = 300$ areas. For $G=47$, rates for colon and pancreatic cancer range approximately between 25 and 70, whereas stomach cancer exhibits lower values, ranging from 10 to 35. When the number of areas increases to $G=300$, the variability in the rates increase: colon and pancreas range from 0 up to 210, and stomach moves between 0 and 125. Regarding spatial patterns, colon and stomach show some similarity for $G=47$, with higher rates concentrated in the central-western to north-western areas. In contrast, pancreas shows higher rates mainly in the central-northern regions of Spain. When the number of areas increases, comparing spatial patterns becomes more challenging.

The real-data analysis confirms the main findings of the simulation studies. Pairwise modeling shows that the LCAR and L$_j$CAR priors consistently induce less smoothing than the iCAR, with similar behavior between them, and that smoothing increases as the number of areas grows. This increase is not proportional across diseases; instead, smoothing levels tend to converge, and each disease exhibits comparable smoothing regardless of the disease pair considered. Extending the analysis to three jointly modeled diseases leads to the same conclusions: the iCAR prior yields the highest smoothing, while L$_j$CAR produces the lowest as $G$ increases. Importantly, modeling additional diseases does not increase the overall proportion of smoothing, as SP values remain stable across pairwise and joint analyses, and disease-specific smoothing levels remain similar, with a slight improvement observed for larger $G$. See Supplementary Material~D for detailed results.

\begin{figure}[h!]
	\begin{center}
		\includegraphics[width=\linewidth]{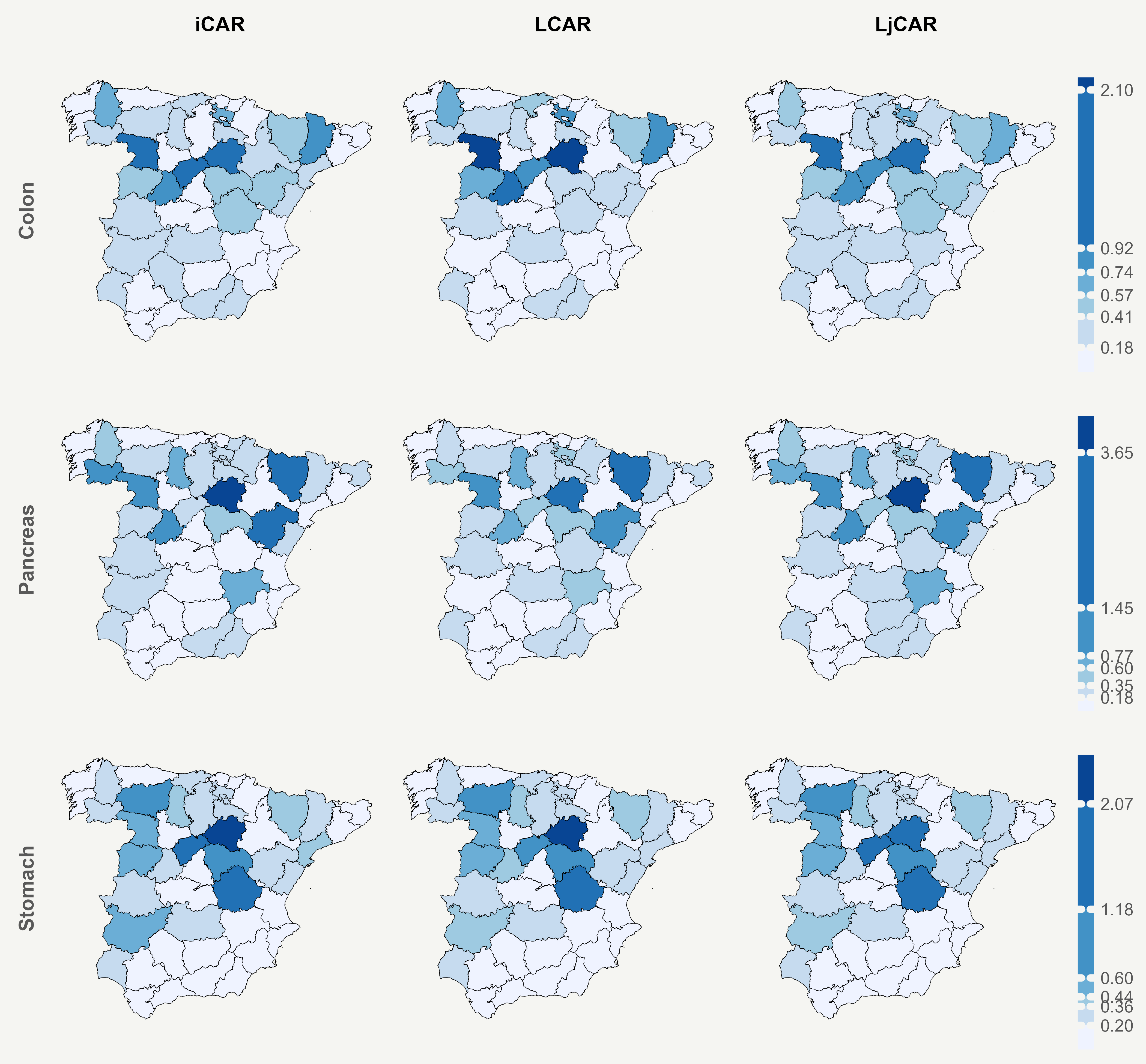}
	\end{center}
	\caption{Area-specific components of the SRSS for each geographical unit $i$ for colon, pancreas and stomach cancer in continental Spain, shown for the 47 real provinces ($G = 47$).The legend reports the 50th, 75th, 85th, 90th, 95th, and 97th percentiles.}
	\label{fig5}
\end{figure}

Beyond comparing overall smoothing levels, \autoref{fig5} displays $RSS_{ji}$ for each disease. Results are shown for models fitting the three diseases jointly; however, the pairwise analyses yield very similar patterns. For clarity of interpretation, we present results for $G=47$. 
The legend of \autoref{fig5} reports the 50th, 75th, 85th, 90th, 95th, and 97th percentiles. All priors identify similar high-smoothing regions, although iCAR yields more areas with larger values. In general the L$_j$CAR produces lower and more homogeneous $RSS_{ji}$ values. Comparing the $RSS_{ji}$ values with the crude disease rates in \autoref{fig4} shows that the
strongest smoothing occurs in areas where rates are unusually high or low relative to their neighbors. For instance, in the case of colon cancer, Soria, Segovia, and Ávila exhibit high levels of smoothing. Soria and Ávila represent high-rate enclaves surrounded by predominantly low-rate areas, whereas Segovia displays the inverse pattern.
As the number of areas increases, drawing general conclusions regarding the spatial distribution of smoothing becomes more challenging. Nevertheless, the pattern observed for $G=47$ is largely preserved, with areas exhibiting strong disparities with their neighbors tend to display higher levels of smoothing. These maps provide a more detailed understanding of smoothing patterns, helping practitioners interpret spatial estimates more critically and identify locations where model-based smoothing may substantially influence the results.

From a practical perspective, the choice of prior depends on the objective of the analysis. If the aim is to obtain a smoother risk surface by reducing the influence of random variation, a prior inducing stronger smoothing may be preferred. In this application, this would point to the iCAR prior. Conversely, if preserving local variation and identifying areas with unusually high or low risks is the priority, a prior inducing less smoothing may be more appropriate. In this case, the choice would lie between the LCAR and L$_j$CAR priors. As shown in \autoref{fig5}, the LCAR prior induces greater smoothing for colon and stomach cancer, whereas the L$_j$CAR prior produces greater smoothing for pancreatic cancer. To guide the final choice of prior, the smoothing diagnostics should be interpreted together with the crude rates, particularly in areas exhibiting the highest smoothing. Examining these areas helps determine whether the observed smoothing is justified, for example because of small populations or unusually high or low observed rates, or whether preserving more local variation may be preferable. The proposed diagnostics therefore provide a practical tool for selecting the most appropriate prior according to both the objectives of the analysis and the characteristics of the data.

\section{Conclusions} \label{Section:Conclusions}

Multivariate disease mapping is substantially more complicated than the univariate version.  Specifically, modeling using disease incidence (mortality) seeks to explain incidence (mortality) rates.  In general form, a prior is selected to capture interaction between diseases through complex modeling of dependence in incidence rates.  However, multivariate disease mapping model priors, as with univariate versions, result in imposing smoothing on the rates, customarily neighbor-based.  In this regard, looking across specification of such priors, two questions of interest arise:  (i) for a given multivariate prior, how does smoothing vary across specification of its parameters? and (ii) under fitting with a given multivariate prior, how much empirical smoothing results?  We have offered potential insight into both questions.  For the first, we have proposed a theoretical metric to capture the implicit smoothing in a multivariate prior given its parameters.  For the latter, we have supplied several empirical measures to enable comparison of extent of smoothing across choice of prior.  

The results of the within prior simulation study show that between-disease variances exert the strongest influence on smoothing; both theoretical and empirical measures indicate that smoothing increases as variances decrease. For the LCAR and L$_j$CAR priors, additional spatial dependence parameters also affect smoothing, with larger values generally associated with greater smoothing. 
The results of the across-prior simulation study further highlight differences between the priors. Generally, all priors exhibit less smoothing in scenarios with low inter-disease correlation, though LCAR and L$_j$CAR consistently yield a lesser amount of smoothing than iCAR. Furthermore, the study shows that variability between disease rates influences smoothing, with greater disparities leading to an increase in the degree of smoothing.
Our results also indicate that smoothing increases with the number of areas, and that differences in smoothing across priors become more pronounced as spatial resolution grows. Notably, smoothing does not scale proportionally across all diseases; rather, disease-specific smoothing proportions tend to become more similar as the number of areas increases.

The real-data analysis supports the findings of the across-prior simulation study.
In addition, modeling two or three diseases jointly yields similar smoothing proportions, both at the disease-specific and overall levels. This suggests that with a moderate number of diseases (e.g., three or four), the degree of smoothing does not deviate substantially from results obtained in pairwise analyses. However, when the number of diseases increases substantially, the behavior of smoothing becomes less predictable. The increased number of hyperparameters could introduce complexities that warrant further investigation. 

To help practitioners understand the level of smoothing in specific areas, we recommend complementing global metrics with maps of the area-specific components of the mean square smoothness criterion for each disease. By examining these maps together with the crude rates, practitioners can identify areas where the estimated risks are being strongly smoothed, investigate whether this is driven by small populations or unstable observed rates, and assess whether the degree of smoothing is appropriate. In particular, our findings indicate that areas whose neighbors have markedly different rates tend to experience higher smoothing. Consequently, these maps provide a more detailed understanding of where smoothing is most pronounced, enabling practitioners to interpret spatial estimates more critically and to identify regions where model-based smoothing may substantially affect inference.
If the primary goal is to obtain a smoother risk surface by reducing random variation, a prior inducing stronger smoothing may be preferred. Conversely, if preserving local variation and identifying areas with unusually high or low risks is the priority, a prior inducing less smoothing may be more appropriate. 
If important discrepancies are observed between  rates of some areas and their neighbors, researchers could consider alternative methods to model discontinuities in the rate surface \citep[see for example,][]{santafe2021}.

In summary, the contribution of this work is to provide a clearer understanding of the smoothing induced by multivariate disease mapping priors. By complementing visual inspection with theoretical and empirical measures of smoothing, practitioners are better able to assess and compare the effects of prior choice and hyperparameter specification. This facilitates more transparent model specification and interpretation in the analysis of areal disease data.  Future work could explore smoothness in the context of dynamic multivariate disease mapping modeling. Autoregressive specification of multivariate rates adds temporal dependence structure to spatial dependence structure.  Further complexity in the model specification results. Appreciation of the effect of prior selection on smoothing over time would be useful for the practitioner.

	\section*{Acknowledgements}
	G.R., J.E., and M.D.U. were supported by Project PID2024-155382OB-I00 funded by MICIU/AEI/10.13039/501100011033 and FEDER, UE. This work was also supported by BIOSTATNET - Proyectos redes de investigaci\'on 2024 - RED2024-153680-T/MICIU/AEI/.
	
	Open access funding provided by Universidad Pública de Navarra.

	\section*{Competing Interests}
	The authors declared no conflicts of interest.

	\section*{Code Availability}
	The code and data supporting this study will be made available upon acceptance at \url{https://github.com/spatialstatisticsupna/Multivariate_Smoothing}.
	
	\appendix
	\section{Model Fitting}\label{Appendix:Model Fitting}
	
	This section provides the implementation details required to fit the multivariate disease mapping models considered in the main manuscript. In particular, we describe the parameterization of the between-disease covariance structure, the constraints required to ensure model identifiability, and the Markov chain Monte Carlo (MCMC) settings used for inference.
	
	To estimate the covariance structure between the spatial patterns of different diseases, we parameterize the between-disease covariance matrix as $\mathbf{\Sigma_b}=\mathbf{MM}'$, where the elements of $\mathbf{M}$ quantify the relationships among spatial effects. Each entry of $\mathbf{M}$ can be interpreted as a regression coefficients of the log-relative risks on
	the rows of $\boldsymbol{\phi}$. Consequently, these elements can be treated as fixed effects and assigned zero-centered normal priors with large variances, which corresponds to assuming a Wishart prior on the covariance matrix, i.e., $\mathbf{\Sigma_b}$, i.e. $\mathbf{\Sigma_b}=\mathbf{MM}'\sim Wishart(J,\sigma^2\mathbf{I}_J)$ \citep[see][for more details]{botella2015}.
	However, this parameterization introduces $J\times J$ parameters in the matrix $\mathbf{M}$, whereas only $J\times(J+1)/2$ parameters are needed to determine the covariance matrix $\mathbf{\Sigma_b}$. To avoid overparameterization, and following the approach of \citet{vicente2023}, we adopt the Bartlett decomposition. Specifically, if $\mathbf{\Sigma_b}$ is a $J \times J$ between-disease covariance matrix such that $\mathbf{\Sigma_b}\sim Wishart(v,\mathbf{V})$, it can be factorized as  $\mathbf{\Sigma_b}=\mathbf{LAA'L'}$, where $\mathbf{L}$ is the Cholesky factor of $\mathbf{V}$, and
	$\mathbf{A}=\left[\begin{array}{cccc}
		c_1 & 0 &  \cdots & 0  \\
		n_{21} & c_2 & \cdots & 0\\
		\vdots & \vdots&\ddots & \vdots\\
		n_{J1} & n_{J2} & \cdots & c_J\\
	\end{array}\right].$
	The elements of $\mathbf{A}$ are independently distributed and the diagonal elements $c_j$ follow $c^2_j \sim \chi^2_{v-j+1}$, whereas the off-diagonal elements follow $n_{jl}\sim N(0,1)$ for $j,l=1, \dots, J$ with $j>l$. This decomposition reduces the number of free parameters to $J\times(J+1)/2$. When $\mathbf{V}=\mathbf{I}_J$, it follows that $\mathbf{L}=\mathbf{I}_J$. Finally, to avoid order dependence among diseases, we compute $\mathbf{\Sigma_b}=\mathbf{AA}'$ and subsequently derive $\mathbf{M}=\left(\mathbf{H}diag\left(\sqrt{\kappa_1},\dots,\sqrt{\kappa_J}\right)\right)'$, where $\kappa_1,\dots, \kappa_J$ are the eigenvalues of the between-diseases covariance matrix $\mathbf{\Sigma_b}$ and the columns of $\mathbf{H}$ are the corresponding eigenvectors. Consistent with \cite{vicente2023}, and following a sensitivity analysis of different values of $v$, we set the degrees of freedom to $v=J+2$, to make the prior slightly more informative. To complete model specification, flat priors were used for the intercepts $\alpha_j$ and for the spatial dependence parameter $\lambda$ we used an uniform distribution ranging from 0 to 1.
	
	In addition, the proposed models include disease-specific intercepts, while the rows of $\phi$ represent the spatial random effects. Since the spatial random effects implicitly contain an intercept, identifiability issues arise unless suitable constraints are imposed. A common approach is to enforce sum-to-zero constraints on the spatial random effects \citep{eberly2000}.
	
	Notably, M-models suffer from identifiability issues. As highlighted by \citet{botella2015}, any orthogonal transformation applied simultaneously to the rows of $\phi$ and the rows of $\mathbf{M}$ results in an equivalent decomposition of $\mathbf{\Theta}$, implying that $\phi$ and $\mathbf{M}$ themselves are not identifiable. Consequently, statistical inference should focus on identifiable quantities such as $\mathbf{\Theta}$ and the covariance matrix $\mathbf{\Sigma_b}$. The Bartlett decomposition described above is used to parameterize and estimate $\mathbf{\Sigma_b}$ efficiently, but it does not alter this identifiability property.
	We perform model fitting and inference using the NIMBLE software \citep{devalpine2017}. NIMBLE is an algorithm library that provides MCMC in R. Three MCMC chains were run for each model, each consisting of 80,000 iterations, with the first 20,000 discarded as burn-in. One out of every 75 iterations was retained, resulting in a total of 2,400 iterations for inference. Convergence was assessed using the Gelman--Rubin diagnostic and the effective sample size (ESS), both of which are implemented in the R package \texttt{coda}. Graphical checks of the chains and their autocorrelations were also performed to further assess convergence.

	\bibliography{BIB2}

\begin{thebibliography}{26}
\expandafter\ifx\csname natexlab\endcsname\relax\def\natexlab#1{#1}\fi
\providecommand{\url}[1]{\texttt{#1}}
\providecommand{\href}[2]{#2}
\providecommand{\path}[1]{#1}
\providecommand{\DOIprefix}{doi:}
\providecommand{\ArXivprefix}{arXiv:}
\providecommand{\URLprefix}{URL: }
\providecommand{\Pubmedprefix}{pmid:}
\providecommand{\doi}[1]{\href{http://dx.doi.org/#1}{\path{#1}}}
\providecommand{\Pubmed}[1]{\href{pmid:#1}{\path{#1}}}
\providecommand{\bibinfo}[2]{#2}
\ifx\xfnm\relax \def\xfnm[#1]{\unskip,\space#1}\fi
\bibitem[{Banerjee et~al.(2025)Banerjee, Gelfand and Carlin}]{banerjee2025}
\bibinfo{author}{Banerjee, S.}, \bibinfo{author}{Gelfand, A.E.},
  \bibinfo{author}{Carlin, B.P.}, \bibinfo{year}{2025}.
\newblock \bibinfo{title}{{Hierarchical Modeling and Analysis for Spatial Data
  (3rd ed.)}}.
\newblock \bibinfo{publisher}{Chapman and Hall/CRC}.
\newblock \DOIprefix\doi{https://doi.org/10.1201/9781003401728}.
\bibitem[{Besag(1974)}]{besag1974}
\bibinfo{author}{Besag, J.}, \bibinfo{year}{1974}.
\newblock \bibinfo{title}{{Spatial interaction and the statistical analysis of
  lattice systems}}.
\newblock \bibinfo{journal}{{Journal of the Royal Statistical Society Series B:
  Statistical Methodology}} \bibinfo{volume}{36}, \bibinfo{pages}{192--225}.
\bibitem[{Besag et~al.(1991)Besag, York and Molli{\'e}}]{besag1991}
\bibinfo{author}{Besag, J.}, \bibinfo{author}{York, J.},
  \bibinfo{author}{Molli{\'e}, A.}, \bibinfo{year}{1991}.
\newblock \bibinfo{title}{{Bayesian image restoration, with two applications in
  spatial statistics}}.
\newblock \bibinfo{journal}{Annals of the Institute of Statistical Mathematics}
  \bibinfo{volume}{43}, \bibinfo{pages}{1--20}.
\bibitem[{Botella-Rocamora et~al.(2015)Botella-Rocamora, Martinez-Beneito and
  Banerjee}]{botella2015}
\bibinfo{author}{Botella-Rocamora, P.}, \bibinfo{author}{Martinez-Beneito,
  M.A.}, \bibinfo{author}{Banerjee, S.}, \bibinfo{year}{2015}.
\newblock \bibinfo{title}{{A unifying modeling framework for highly
  multivariate disease mapping}}.
\newblock \bibinfo{journal}{Statistics in Medicine} \bibinfo{volume}{34},
  \bibinfo{pages}{1548--1559}.
\bibitem[{Carlin et~al.(2003)Carlin, Banerjee et~al.}]{carlin2003}
\bibinfo{author}{Carlin, B.P.}, \bibinfo{author}{Banerjee, S.}, et~al.,
  \bibinfo{year}{2003}.
\newblock \bibinfo{title}{{Hierarchical multivariate CAR models for
  spatio-temporally correlated survival data}}.
\newblock \bibinfo{journal}{Bayesian Statistics} \bibinfo{volume}{7},
  \bibinfo{pages}{45--63}.
\bibitem[{Eberly and Carlin(2000)}]{eberly2000}
\bibinfo{author}{Eberly, L.E.}, \bibinfo{author}{Carlin, B.P.},
  \bibinfo{year}{2000}.
\newblock \bibinfo{title}{Identifiability and convergence issues for markov
  chain monte carlo fitting of spatial models}.
\newblock \bibinfo{journal}{Statistics in medicine} \bibinfo{volume}{19},
  \bibinfo{pages}{2279--2294}.
\bibitem[{Gelfand et~al.(2004)Gelfand, Schmidt, Banerjee and
  Sirmans}]{gelfand2004}
\bibinfo{author}{Gelfand, A.E.}, \bibinfo{author}{Schmidt, A.M.},
  \bibinfo{author}{Banerjee, S.}, \bibinfo{author}{Sirmans, C.},
  \bibinfo{year}{2004}.
\newblock \bibinfo{title}{Nonstationary multivariate process modeling through
  spatially varying coregionalization}.
\newblock \bibinfo{journal}{Test} \bibinfo{volume}{13},
  \bibinfo{pages}{263--312}.
\bibitem[{Gelfand and Vounatsou(2003)}]{gelfand2003}
\bibinfo{author}{Gelfand, A.E.}, \bibinfo{author}{Vounatsou, P.},
  \bibinfo{year}{2003}.
\newblock \bibinfo{title}{{Proper multivariate conditional autoregressive
  models for spatial data analysis}}.
\newblock \bibinfo{journal}{Biostatistics} \bibinfo{volume}{4},
  \bibinfo{pages}{11--15}.
\bibitem[{Gelman et~al.(2017)Gelman, Simpson and Betancourt}]{gelman2017}
\bibinfo{author}{Gelman, A.}, \bibinfo{author}{Simpson, D.},
  \bibinfo{author}{Betancourt, M.}, \bibinfo{year}{2017}.
\newblock \bibinfo{title}{The prior can often only be understood in the context
  of the likelihood}.
\newblock \bibinfo{journal}{Entropy} \bibinfo{volume}{19},
  \bibinfo{pages}{555}.
\bibitem[{Held et~al.(2005)Held, Nat{\'a}rio, Fenton, Rue and
  Becker}]{held2005}
\bibinfo{author}{Held, L.}, \bibinfo{author}{Nat{\'a}rio, I.},
  \bibinfo{author}{Fenton, S.E.}, \bibinfo{author}{Rue, H.},
  \bibinfo{author}{Becker, N.}, \bibinfo{year}{2005}.
\newblock \bibinfo{title}{{Towards joint disease mapping}}.
\newblock \bibinfo{journal}{Statistical Methods in Medical Research}
  \bibinfo{volume}{14}, \bibinfo{pages}{61--82}.
\bibitem[{Jin et~al.(2007)Jin, Banerjee and Carlin}]{jin2007}
\bibinfo{author}{Jin, X.}, \bibinfo{author}{Banerjee, S.},
  \bibinfo{author}{Carlin, B.P.}, \bibinfo{year}{2007}.
\newblock \bibinfo{title}{Order-free co-regionalized areal data models with
  application to multiple-disease mapping}.
\newblock \bibinfo{journal}{Journal of the Royal Statistical Society Series B:
  Statistical Methodology} \bibinfo{volume}{69}, \bibinfo{pages}{817--838}.
\bibitem[{Jin et~al.(2005)Jin, Carlin and Banerjee}]{jin2005}
\bibinfo{author}{Jin, X.}, \bibinfo{author}{Carlin, B.P.},
  \bibinfo{author}{Banerjee, S.}, \bibinfo{year}{2005}.
\newblock \bibinfo{title}{{Generalized hierarchical multivariate CAR models for
  areal data}}.
\newblock \bibinfo{journal}{Biometrics} \bibinfo{volume}{61},
  \bibinfo{pages}{950--961}.
\bibitem[{Knorr-Held and Best(2001)}]{knorr2001}
\bibinfo{author}{Knorr-Held, L.}, \bibinfo{author}{Best, N.G.},
  \bibinfo{year}{2001}.
\newblock \bibinfo{title}{{A shared component model for detecting joint and
  selective clustering of two diseases}}.
\newblock \bibinfo{journal}{Journal of the Royal Statistical Society Series A:
  Statistics in Society} \bibinfo{volume}{164}, \bibinfo{pages}{73--85}.
\bibitem[{Leroux et~al.(2000)Leroux, Lei and Breslow}]{leroux2000}
\bibinfo{author}{Leroux, B.G.}, \bibinfo{author}{Lei, X.},
  \bibinfo{author}{Breslow, N.}, \bibinfo{year}{2000}.
\newblock \bibinfo{title}{{Estimation of disease rates in small areas: a new
  mixed model for spatial dependence}}, in: \bibinfo{editor}{Halloran, M.E.},
  \bibinfo{editor}{Berry, D.} (Eds.), \bibinfo{booktitle}{Statistical Models in
  Epidemiology, the Environment, and Clinical Trials}.
  \bibinfo{publisher}{Springer New York}, pp. \bibinfo{pages}{179--191}.
\bibitem[{MacNab(2016)}]{macnab2016}
\bibinfo{author}{MacNab, Y.C.}, \bibinfo{year}{2016}.
\newblock \bibinfo{title}{Linear models of coregionalization for multivariate
  lattice data: Order-dependent and order-free cmcars}.
\newblock \bibinfo{journal}{Statistical methods in medical research}
  \bibinfo{volume}{25}, \bibinfo{pages}{1118--1144}.
\bibitem[{Mardia(1988)}]{mardia1988}
\bibinfo{author}{Mardia, K.}, \bibinfo{year}{1988}.
\newblock \bibinfo{title}{{Multi-dimensional multivariate Gaussian Markov
  random fields with application to image processing}}.
\newblock \bibinfo{journal}{Journal of Multivariate Analysis}
  \bibinfo{volume}{24}, \bibinfo{pages}{265--284}.
\bibitem[{Martinez-Beneito(2013)}]{martinez2013}
\bibinfo{author}{Martinez-Beneito, M.A.}, \bibinfo{year}{2013}.
\newblock \bibinfo{title}{{A general modelling framework for multivariate
  disease mapping}}.
\newblock \bibinfo{journal}{Biometrika} \bibinfo{volume}{100},
  \bibinfo{pages}{539--553}.
\bibitem[{Retegui et~al.(2024)Retegui, Etxeberria and Ugarte}]{Retegui2024}
\bibinfo{author}{Retegui, G.}, \bibinfo{author}{Etxeberria, J.},
  \bibinfo{author}{Ugarte, M.D.}, \bibinfo{year}{2024}.
\newblock \bibinfo{title}{Multivariate bayesian models with flexible shared
  interactions for analyzing spatio-temporal patterns of rare cancers}.
\newblock \bibinfo{journal}{Environmental and Ecological Statistics} ,
  \bibinfo{pages}{977--1007}\DOIprefix\doi{https://doi.org/10.1007/s10651-024-00630-w}.
\bibitem[{Retegui et~al.(2025)Retegui, Gelfand, Etxeberria and
  Ugarte}]{retegui2025}
\bibinfo{author}{Retegui, G.}, \bibinfo{author}{Gelfand, A.E.},
  \bibinfo{author}{Etxeberria, J.}, \bibinfo{author}{Ugarte, M.D.},
  \bibinfo{year}{2025}.
\newblock \bibinfo{title}{On prior smoothing with discrete spatial data in the
  context of disease mapping}.
\newblock \bibinfo{journal}{Statistical Methods in Medical Research} ,
  \bibinfo{pages}{09622802251362659}\DOIprefix\doi{https://doi.org/10.1177/09622802251362659}.
\bibitem[{Sain et~al.(2011)Sain, Furrer and Cressie}]{sain2011}
\bibinfo{author}{Sain, S.R.}, \bibinfo{author}{Furrer, R.},
  \bibinfo{author}{Cressie, N.}, \bibinfo{year}{2011}.
\newblock \bibinfo{title}{{A spatial analysis of multivariate output from
  regional climate models}}.
\newblock \bibinfo{journal}{The Annals of Applied Statistics} ,
  \bibinfo{pages}{150--175}.
\bibitem[{Santaf{\'e} et~al.(2021)Santaf{\'e}, Adin, Lee and
  Ugarte}]{santafe2021}
\bibinfo{author}{Santaf{\'e}, G.}, \bibinfo{author}{Adin, A.},
  \bibinfo{author}{Lee, D.}, \bibinfo{author}{Ugarte, M.D.},
  \bibinfo{year}{2021}.
\newblock \bibinfo{title}{Dealing with risk discontinuities to estimate cancer
  mortality risks when the number of small areas is large}.
\newblock \bibinfo{journal}{Statistical Methods in Medical Research}
  \bibinfo{volume}{30 (1)}, \bibinfo{pages}{6--21}.
\bibitem[{Stern and Cressie(2000)}]{stern2000}
\bibinfo{author}{Stern, H.S.}, \bibinfo{author}{Cressie, N.},
  \bibinfo{year}{2000}.
\newblock \bibinfo{title}{{Posterior predictive model checks for disease
  mapping models}}.
\newblock \bibinfo{journal}{Statistics in Medicine} \bibinfo{volume}{19},
  \bibinfo{pages}{2377--2397}.
\bibitem[{Urdangarin et~al.(2024)Urdangarin, Goicoa, Kneib and
  Ugarte}]{Urdangarin2024}
\bibinfo{author}{Urdangarin, A.}, \bibinfo{author}{Goicoa, T.},
  \bibinfo{author}{Kneib, T.}, \bibinfo{author}{Ugarte, M.D.},
  \bibinfo{year}{2024}.
\newblock \bibinfo{title}{A simplified spatial+ approach to mitigate spatial
  confounding in multivariate spatial areal models}.
\newblock \bibinfo{journal}{Spatial Statistics} \bibinfo{volume}{59},
  \bibinfo{pages}{100804}.
\newblock \URLprefix
  \url{https://www.sciencedirect.com/science/article/pii/S2211675323000799},
  \DOIprefix\doi{https://doi.org/10.1016/j.spasta.2023.100804}.
\bibitem[{de~Valpine et~al.(2017)de~Valpine, Turek, Paciorek, Anderson-Bergman,
  Lang and Bodik}]{devalpine2017}
\bibinfo{author}{de~Valpine, P.}, \bibinfo{author}{Turek, D.},
  \bibinfo{author}{Paciorek, C.J.}, \bibinfo{author}{Anderson-Bergman, C.},
  \bibinfo{author}{Lang, D.T.}, \bibinfo{author}{Bodik, R.},
  \bibinfo{year}{2017}.
\newblock \bibinfo{title}{{Programming with models: writing statistical
  algorithms for general model structures with NIMBLE}}.
\newblock \bibinfo{journal}{Journal of Computational and Graphical Statistics}
  \bibinfo{volume}{26}, \bibinfo{pages}{403--413}.
\bibitem[{Vicente et~al.(2023a)Vicente, Adin, Goicoa and Ugarte}]{vicente2023}
\bibinfo{author}{Vicente, G.}, \bibinfo{author}{Adin, A.},
  \bibinfo{author}{Goicoa, T.}, \bibinfo{author}{Ugarte, M.D.},
  \bibinfo{year}{2023}a.
\newblock \bibinfo{title}{High-dimensional order-free multivariate spatial
  disease mapping}.
\newblock \bibinfo{journal}{Statistics and Computing} \bibinfo{volume}{33},
  \bibinfo{pages}{104}.
\bibitem[{Vicente et~al.(2023b)Vicente, Goicoa and Ugarte}]{vicenteetal2023}
\bibinfo{author}{Vicente, G.}, \bibinfo{author}{Goicoa, T.},
  \bibinfo{author}{Ugarte, M.D.}, \bibinfo{year}{2023}b.
\newblock \bibinfo{title}{Multivariate bayesian spatio-temporal models to
  analyze crimes against women}.
\newblock \bibinfo{journal}{Biostatistics} \bibinfo{volume}{24},
  \bibinfo{pages}{562--584}.

\end{thebibliography}

\newpage	
	\renewcommand{\thetable}{S\arabic{table}}
	\setcounter{table}{0}
	\renewcommand{\thefigure}{S\arabic{figure}}
	\setcounter{figure}{0}
	
	\begin{appendices}
		
		\section{Design of the Simulation Scenarios}\label{SimStudy}
		
\begin{figure}[h!]
			\centering
			\includegraphics[width=\linewidth]{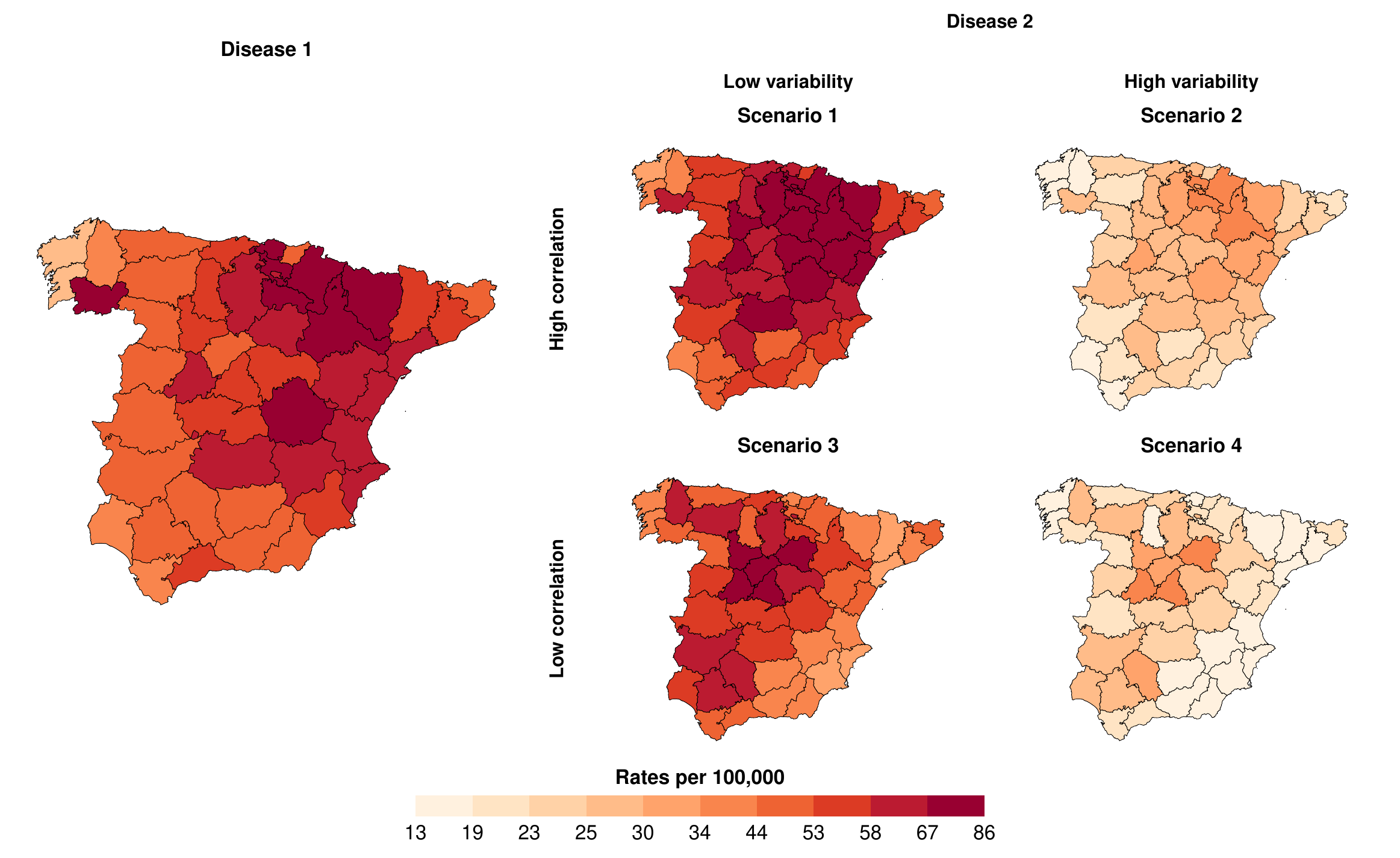}
			\caption{Simulated true rate surfaces for Disease 1 and Disease 2 across Scenarios 1–4 ($G=47$). Scenarios combine high or low between-disease correlation with high or low rate variability. Disease 1 remains fixed across all four scenarios.}
			\label{fig0.1}
		\end{figure}

\begin{figure}[t!]
			\centering
			\includegraphics[width=\linewidth]{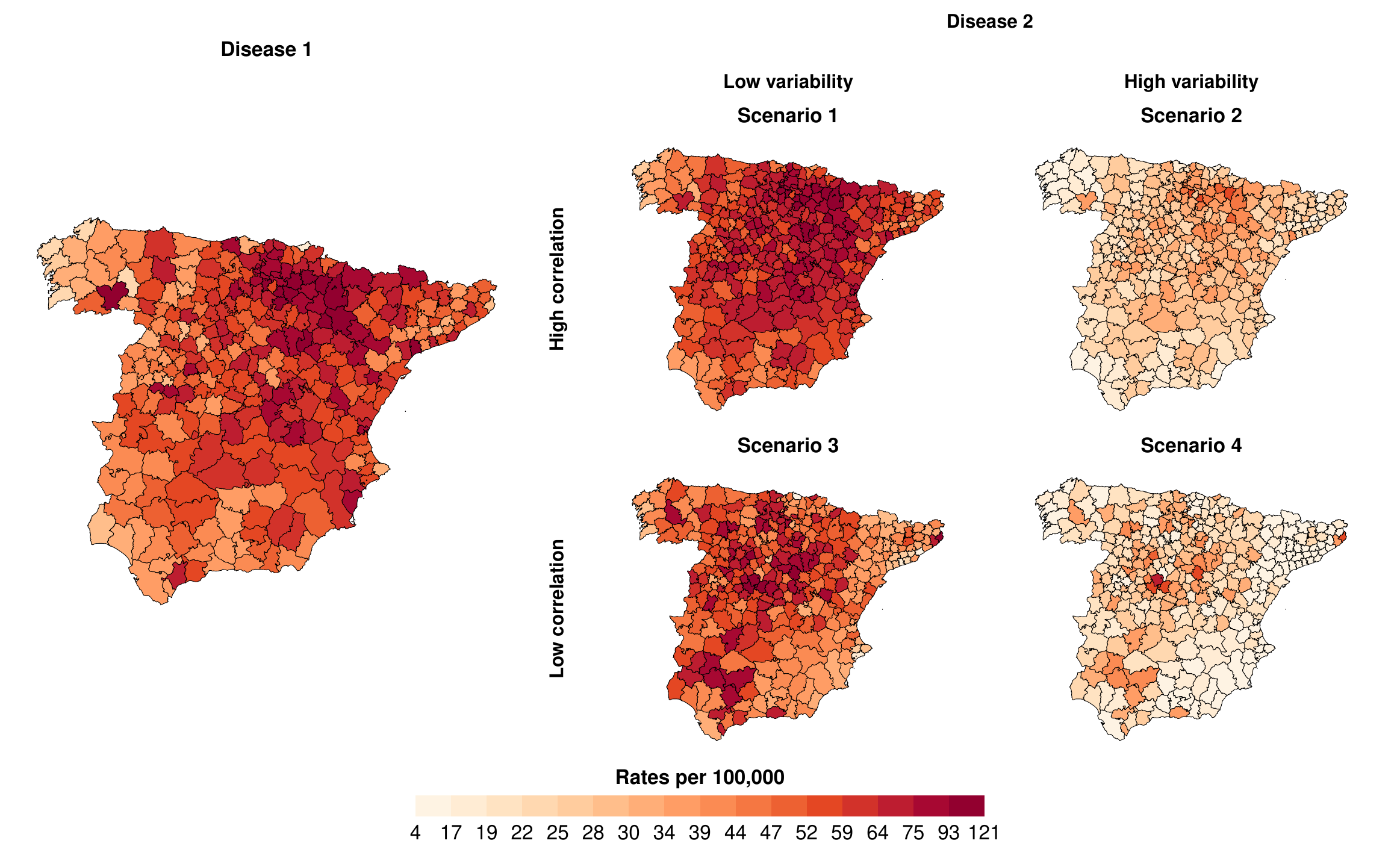}
			\caption{Simulated true rate surfaces for Disease 1 and Disease 2 across Scenarios 1–4 ($G=300$). Scenarios combine high or low between-disease correlation with high or low rate variability. Disease 1 remains fixed across all four scenarios.}
			\label{fig0.2}
		\end{figure}

		This section contains additional figures of the simulated true rate surfaces. Specifically, \autoref{fig0.1} presents the true rate surfaces for the four simulation scenarios described in Section 3.1 of the main manuscript for G=47, while \autoref{fig0.2} presents the corresponding surfaces for G=300. The true rate surface for Disease 1 is the same across all scenarios, whereas the true rate surface for Disease 2 is varied to generate scenarios with different levels of between-disease correlation and variability.
		
	\newpage
\section{Within Simulation Study}\label{SectionSM:Within}

\begin{table}[t!]
	\centering
	\caption{\label{tabA1}Theoretical and empirical smoothing criteria values for the iCAR spatial prior for $G=100$ across different $\Sigma_{11}$, $\Sigma_{22}$ and $\rho$ values under Scenario 2.}
	\resizebox{\textwidth}{!}{
		\begin{tabular}{lll|rrrrrcrrrrr}
			\hline
			&&&\multicolumn{5}{c}{$\rho=0$}&&\multicolumn{5}{c}{$\rho=0.7$}\\
			\cline{4-8}\cline{10-14}
			&&& MultiTCV & SP & RSP & SRSS & maxRSS && MultiTCV & SP & RSP & SRSS & maxRSS \\ 
			\hline
			\multicolumn{14}{l}{$\Sigma_{11}=0.0025$}\\[1.2ex]
			&$\Sigma_{22}=0.0025$ & Disease 1 ($j=1$) &  &  & 0.70 & 423.20 & 53.78 & & &  & 0.67 & 403.22 & 51.64 \\ 
			&& Disease 2 ($j=2$) &  &  & 0.86 & 327.71 & 53.41  &&  &  & 0.79 & 292.68 & 48.72 \\ 
			&& Total & 0.0000 & 0.36 & 1.56 & 750.91 && & 0.0000 & 0.34 & 1.46 & 695.90 &  \\ 
			&& &  &  &  &  &&&&&&&  \\ 
			&$\Sigma_{22}=0.04$ & Disease 1 ($j=1$) &  &  & 0.70 & 423.12 & 53.78& &  &  & 0.70 & 425.85 & 54.08 \\ 
			&&Disease 2 ($j=2$)  &  &  & 0.63 & 221.09 & 37.24 &&  &  & 0.60 & 207.79 & 35.14 \\ 
			&&Total & 0.0005 & 0.33 & 1.33 & 644.21 &&  & 0.0003 & 0.33 & 1.30 & 633.64 &  \\ 
			&& &  &  &  &  &&&&&&&  \\ 
			&$\Sigma_{22}=0.25$ & Disease 1 ($j=1$) &  &  & 0.70 & 423.28 & 53.80& &  &  & 0.74 & 452.92 & 56.88 \\ 
			&& Disease 2 ($j=2$) &  &  & 0.52 & 180.44 & 27.33 &&  &  & 0.52 & 177.38 & 26.77 \\ 
			&& Total& 0.0031 & 0.32 & 1.22 & 603.73 &&  & 0.0016 & 0.34 & 1.26 & 630.30 &  \\ 
			& &&  &  &  &  &&&&&&&  \\ 
			\multicolumn{14}{l}{$\Sigma_{11}=0.04$}\\[1.2ex]
			&$\Sigma_{22}=0.0025$ & Disease 1 ($j=1$)&  &  & 0.41 & 236.60 & 36.96  && & & 0.40 & 231.96 & 36.46\\
			&& Disease 2 ($j=2$) &&   & 0.86 & 327.65 & 53.41  &&  && 0.83 & 315.54 & 52.08\\
			&& Total & 0.0005 & 0.25 & 1.27 & 564.24 &&& 0.0003 & 0.24 & 1.23 & 547.50 &\\
			& &&  &  &  &  &&&&&&&  \\ 
			&$\Sigma_{22}=0.04$ & Disease 1 ($j=1$)&  &  & 0.41 & 236.65 & 36.94& &  &  & 0.40 & 232.08 & 36.64 \\ 
			&& Disease 2 ($j=2$) &  &  & 0.63 & 221.01 & 37.25 &&  &  & 0.60 & 211.40 & 36.50 \\ 
			&& Total & 0.0080 & 0.22 & 1.04 & 457.66 & & & 0.0041 & 0.22 & 1.00 & 443.48 &  \\ 
			& &&  &  &  &  &&&&&&&  \\ 
			&$\Sigma_{22}=0.25$ & Disease 1 ($j=1$) & & & 0.41 & 236.54 & 36.99 &  &&  & 0.42 & 242.30 & 37.29 \\ 
			&& Disease 2 ($j=2$)&  & & 0.52 & 180.27 & 27.30 &  &&  & 0.52 & 179.08 & 27.27 \\ 
			&& Total& 0.0497 & 0.21 & 0.93 & 416.81 & && 0.0253 & 0.21 & 0.94 & 421.37 &  \\ 
			& &&  &  &  &  &&&&&&&  \\ 
			\multicolumn{14}{l}{$\Sigma_{11}=0.25$}\\[1.2ex]
			&$\Sigma_{22}=0.0025$ & Disease 1 ($j=1$)&  &  & 0.31 & 184.14 & 26.11 &&  &  & 0.31 & 183.03 & 25.99  \\ 
			&& Disease 2 ($j=2$) &  &  & 0.86 & 327.76 & 53.42 &&  &  & 0.87 & 334.90 & 54.52 \\ 
			&& Total & 0.0031 & 0.21 & 1.17 & 511.90 &&& 0.0016 & 0.21 & 1.18 & 517.93 &  \\ 
			& &&  &  &  &  &&&&&&&  \\ 
			&$\Sigma_{22}=0.04$ & Disease 1 ($j=1$)  &  &  & 0.31 & 183.89 & 26.07 &&  &  & 0.31 & 182.75 & 26.02 \\ 
			&& Disease 2 ($j=2$) &  &  & 0.63 & 221.09 & 37.19 &&  &  & 0.63 & 223.42 & 38.08 \\ 
			&& Total & 0.0497 & 0.19 & 0.94 & 404.98 & && 0.0253 & 0.19 & 0.94 & 406.18 & \\ 
			& &&  &  &  &  &&&&&&&  \\ 
			&$\Sigma_{22}=0.25$ & Disease 1 ($j=1$)  &  &  & 0.31 & 184.02 & 26.08 &&  &  & 0.32 & 185.01 & 26.53 \\ 
			&& Disease 2 ($j=2$) &  &  & 0.52 & 180.37 & 27.44 &&  &  & 0.52 & 183.57 & 28.40  \\ 
			&& Total & 0.3106 & 0.17 & 0.83 & 364.39 & && 0.1584 & 0.18 & 0.84 & 368.57 &  \\ 
			\hline
		\end{tabular}
	}
\end{table}

In this section, we present additional results from the within-prior simulation study described in Section 3.3 of the main paper. Specifically, we include detailed tables for Scenarios 2, 3, and 4 under the iCAR prior, which were discussed but not shown in the main text. We also report results for the LCAR prior with $\Sigma_{11}=0.04$ or $\Sigma_{22}=0.04$, which were omitted from the main paper to streamline the presentation, along with a figure illustrating the effect of $\rho$ and $\lambda$, also discussed but not displayed previously. Finally, we present results for the L$_j$CAR prior, which were only briefly summarized in the main text. For all tables presented in this section, we additionally report 95\% intervals for the distribution of the empirical smoothing criteria.

\begin{table}[t!]
	\centering
	\caption{\label{tabA2} Theoretical and empirical smoothing criteria values for the iCAR spatial prior for $G=100$ across different $\Sigma_{11}$, $\Sigma_{22}$ and $\rho$ values under Scenario 3.}
	\resizebox{\textwidth}{!}{
		\begin{tabular}{lll|rrrrrcrrrrr}
			\hline
			&&&\multicolumn{5}{c}{$\rho=0$}&&\multicolumn{5}{c}{$\rho=0.7$}\\
			\cline{4-8}\cline{10-14}
			&&& MultiTCV & SP & RSP & SRSS & maxRSS && MultiTCV & SP & RSP & SRSS & maxRSS \\ 
			\hline
			\multicolumn{14}{l}{$\Sigma_{11}=0.0025$}\\[1.2ex]
			&$\Sigma_{22}=0.0025$ & Disease 1 ($j=1$) &&& 0.69 & 400.99 & 48.61 &  & & & 0.72 & 413.87 & 49.29\\ 
			&& Disease 2 ($j=2$) &  &  & 0.70 & 412.84 & 64.72 & &  &  & 0.70 & 411.48 & 62.39 \\ 
			&& Total & 0.0000 & 0.68 & 1.39 & 813.83 & && 0.0000 & 0.69 & 1.42 & 825.35 &  \\ 
			& &&  &  &  &  &&&&&&&  \\ 
			&$\Sigma_{22}=0.04$ & Disease 1 ($j=1$)&   &  & 0.69 & 401.10 & 48.60 &&  &  & 0.74 & 432.37 & 51.94 \\ 
			&& Disease 2 ($j=2$) &  &  & 0.49 & 270.32 & 44.96 &&  &  & 0.48 & 265.37 & 42.97 \\ 
			&& Total & 0.0005 & 0.58 & 1.18 & 671.42 & && 0.0003 & 0.60 & 1.23 & 697.74 &   \\ 
			& &&  &  &  &  &&&&&&&  \\ 
			&$\Sigma_{22}=0.25$ & Disease 1 ($j=1$)&  &  & 0.69 & 401.09 & 48.64 &&  &  & 0.76 & 446.20 & 54.19 \\ 
			&& Disease 2 ($j=2$) &  &  & 0.36 & 195.93 & 27.52 &&  &  & 0.36 & 195.03 & 27.18 \\ 
			&& Total & 0.0031 & 0.52 & 1.05 & 597.01 & && 0.0016 & 0.55 & 1.12 & 641.23 &   \\ 
			& &&  &  &  &  &&&&&&&  \\ 
			\multicolumn{14}{l}{$\Sigma_{11}=0.04$}\\[1.2ex]
			&$\Sigma_{22}=0.0025$ & Disease 1 ($j=1$)&  &  & 0.40 & 221.40 & 30.29 &&  &  & 0.40 & 222.00 & 30.07 \\ 
			&& Disease 2 ($j=2$)  &  &  & 0.70 & 412.80 & 64.69 &&  &  & 0.72 & 430.43 & 66.20  \\ 
			&& Total & 0.0005 & 0.53 & 1.09 & 634.20 & && 0.0003 & 0.54 & 1.13 & 652.43 &  \\ 
			& &&  &  &  &  &&&&&&&  \\ 
			&$\Sigma_{22}=0.04$ & Disease 1 ($j=1$)&&   & 0.40 & 221.52 & 30.36 &&  &  & 0.42 & 230.49 & 30.49 \\ 
			&& Disease 2 ($j=2$)  &  &  & 0.49 & 270.38 & 44.94 &&  &  & 0.50 & 275.06 & 44.51\\ 
			&& Total & 0.0080 & 0.43 & 0.89 & 491.90 & && 0.0041 & 0.44 & 0.91 & 505.55 &    \\ 
			& &&  &  &  &  &&&&&&&  \\ 
			&$\Sigma_{22}=0.25$ & Disease 1 ($j=1$)&  &  & 0.40 & 221.45 & 30.28 &&  &  & 0.44 & 242.09 & 31.56 \\ 
			&& Disease 2 ($j=2$)  &  &  & 0.36 & 195.64 & 27.39 &&  &  & 0.36 & 200.33 & 27.81 \\ 
			&& Total & 0.0497 & 0.37 & 0.76 & 417.09 &&& 0.0253 & 0.39 & 0.80 & 442.42 &   \\
			& &&  &  &  &  &&&&&&&  \\ 
			\multicolumn{14}{l}{$\Sigma_{11}=0.25$}\\[1.2ex]
			&$\Sigma_{22}=0.0025$ & Disease 1 ($j=1$)&  &  & 0.31 & 173.52 & 20.81 &&  &  & 0.31 & 173.73 & 20.79  \\ 
			&& Disease 2 ($j=2$) &  &  & 0.70 & 412.74 & 64.69 &&  &  & 0.74 & 443.41 & 68.36 \\ 
			&& Total & 0.0031 & 0.48 & 1.01 & 586.26 & && 0.0016 & 0.50 & 1.05 & 617.13 &  \\ 
			& &&  &  &  &  &&&&&&&  \\ 
			&$\Sigma_{22}=0.04$ & Disease 1 ($j=1$) &  &  & 0.31 & 173.43 & 20.86 &&  &  & 0.32 & 177.68 & 21.18 \\ 
			&& Disease 2 ($j=2$) &  &  & 0.49 & 270.37 & 44.96 &&  &  & 0.52 & 289.22 & 47.33 \\ 
			&& Total & 0.0497 & 0.38 & 0.80 & 443.80 & && 0.0253 & 0.40 & 0.83 & 466.90 &  \\ 
			&&  &  &  &  &&&&&&&  \\ 
			&$\Sigma_{22}=0.25$ & Disease 1 ($j=1$) &  &  & 0.31 & 173.50 & 20.74 &&  &  & 0.33 & 184.24 & 22.38 \\ 
			&& Disease 2 ($j=2$) &  &  & 0.36 & 195.85 & 27.49 &&  &  & 0.38 & 209.47 & 30.31 \\ 
			&& Total & 0.3106 & 0.32 & 0.67 & 369.35 & && 0.1584 & 0.34 & 0.71 & 393.71 &  \\ 
			\hline
		\end{tabular}
	}
\end{table}

\autoref{tabA1}, \ref{tabA2} and \ref{tabA3} show the theoretical multivariate TCV (MultiTCV) and the empirical smoothing metrics from Section~2.3 of the main text for the iCAR prior in Scenarios~2, 3, and~4, respectively, with $G=100$ and varying $\Sigma_{11}$, $\Sigma_{22}$, and $\rho$.
Across all scenarios and for a fixed $\rho$, the theoretical and empirical metrics are consistent, both indicating stronger smoothing as $\Sigma_{11}$ and $\Sigma_{22}$ decrease. Recall that the metrics behave reciprocally: larger MultiTCV values correspond to less smoothing. When $\rho=0$, changes in $\Sigma_{22}$ do not affect the smoothing of disease~1; when $\rho\neq0$, increasing $\Sigma_{22}$ slightly increases smoothing for disease~1, particularly for smaller $\Sigma_{11}$. An analogous pattern holds when varying $\Sigma_{11}$ instead of $\Sigma_{22}$. This is consistent with the results reported for Scenario~1 in the main paper.
The effect of $\rho$ varies by scenario. In Scenario~2 (\autoref{tabA1}), increasing $\rho$ slightly reduces empirical smoothing, whereas the theoretical metric shows the opposite trend; however, alignment emerges at larger MultiTCV values. In Scenarios~3 and~4 (\autoref{tabA2} and \ref{tabA3}), empirical and theoretical metrics align across all values of $\Sigma_{11}$ and $\Sigma_{22}$. This behavior is also observed in Figure~1 of the main text.

\begin{table}[t!]
	\centering
	
	\caption{\label{tabA3} Theoretical and empirical smoothing criteria values for the iCAR spatial prior for $G=100$ across different values of $\Sigma_{11}$, $\Sigma_{22}$ and $\rho$ under Scenario 4.}
	\resizebox{\textwidth}{!}{
		\begin{tabular}{lll|rrrrrcrrrrr}
			\hline
			&&&\multicolumn{5}{c}{$\rho=0$}&&\multicolumn{5}{c}{$\rho=0.7$}\\
			\cline{4-8}\cline{10-14}
			&&& MultiTCV & SP & RSP & SRSS & maxRSS && MultiTCV & SP & RSP & SRSS & maxRSS \\ 
			\hline
			\multicolumn{14}{l}{$\Sigma_{11}=0.0025$}\\[1.2ex]
			&$\Sigma_{22}=0.0025$ & Disease 1 ($j=1$) &  &  & 0.69 & 411.19& 52.17 &&  &  & 0.71 & 419.81 & 52.41\\ 
			&& Disease 2 ($j=2$)  &  &  & 0.82 & 362.41 & 58.93 &&  &  & 0.84 & 365.01 & 57.13 \\ 
			&& Total  & 0.0000 & 0.30 & 1.52 & 773.60 &&& 0.0000 & 0.31 & 1.55 & 784.82 &  \\ 
			& &&  &  &  &  &&&&&&&  \\ 
			&$\Sigma_{22}=0.04$ & Disease 1 ($j=1$)&  &  & 0.69 & 411.11 & 52.14 &&  &  & 0.74 & 441.39 & 54.83\\ 
			&& Disease 2 ($j=2$) &  &  & 0.64 & 265.75 & 43.90 &&  &  & 0.64 & 259.18 & 40.78 \\ 
			&& Total & 0.0005 & 0.29 & 1.34 & 676.86 &&& 0.0003 & 0.30 & 1.38 & 700.57 &   \\ 
			& &&  &  &  &  &&&&&&&  \\ 
			&$\Sigma_{22}=0.25$ & Disease 1 ($j=1$) &  &  & 0.69 & 411.25 & 52.19 &&  &  & 0.76 & 456.53 & 57.11 \\ 
			&& Disease 2 ($j=2$) &  &  & 0.50 & 204.69 & 30.12 &&  &  & 0.50 & 202.89 & 28.77 \\ 
			&& Total& 0.0031 & 0.27 & 1.20 & 615.95 &&& 0.0016 & 0.29 & 1.26 & 659.42 &  \\ 
			& &&  &  &  &  &&&&&&&  \\ 
			\multicolumn{14}{l}{$\Sigma_{11}=0.04$}\\[1.2ex]
			&$\Sigma_{22}=0.0025$ & Disease 1 ($j=1$)&  &  & 0.40 & 224.87 & 31.81 &&  &  & 0.40 & 224.74 & 31.59 \\ 
			&& Disease 2 ($j=2$)  &  &  & 0.82 & 362.39 & 58.96 &&  &  & 0.85 & 375.65 & 60.00 \\ 
			&& Total & 0.0005 & 0.21 & 1.22 & 587.26 &&& 0.0003 & 0.21 & 1.25 & 600.39 &   \\ 
			& &&  &  &  &  &&&&&&&  \\ 
			&$\Sigma_{22}=0.04$ & Disease 1 ($j=1$) &  &  & 0.40 & 224.81 & 31.79 &  &  &  & 0.41 & 230.24 & 31.84\\ 
			&& Disease 2 ($j=2$)&  &  & 0.64 & 265.76 & 43.86 &&  &  & 0.65 & 267.72 & 42.29\\ 
			&& Total & 0.0080 & 0.19 & 1.04 & 490.57 &&& 0.0041 & 0.19 & 1.06 & 497.96 &  \\ 
			& &&  &  &  &  &&&&&&&  \\ 
			&$\Sigma_{22}=0.25$ & Disease 1 ($j=1$) &  &  & 0.40 & 224.74 & 31.75 &&  &  & 0.43 & 241.84 & 33.05\\ 
			&& Disease 2 ($j=2$) &  &  & 0.50 & 204.80 & 30.17 &&  &  & 0.52 & 209.30 & 29.22\\ 
			&& Total & 0.0497 & 0.18 & 0.90 & 429.54 &&& 0.0253 & 0.19 & 0.94 & 451.14 &  \\ 
			& &&  &  &  &  &&&&&&&  \\ 
			\multicolumn{14}{l}{$\Sigma_{11}=0.25$}\\[1.2ex]
			&$\Sigma_{22}=0.0025$ & Disease 1 ($j=1$) &  &  & 0.31 & 175.46 & 22.29 &&  &  & 0.31 & 175.60 & 22.33\\ 
			&& Disease 2 ($j=2$) &  &  & 0.82 & 362.43 & 58.95 &&  &  & 0.86 & 382.78 & 61.58\\ 
			&& Total  & 0.0031 & 0.18 & 1.13 & 537.89 & && 0.0016 & 0.18 & 1.17 & 558.38 &  \\ 
			& &&  &  &  &  &&&&&&&  \\ 
			&$\Sigma_{22}=0.04$ & Disease 1 ($j=1$) &  &  & 0.31 & 175.49 & 22.29 &&  &  & 0.31 & 177.55 & 22.41 \\ 
			&& Disease 2 ($j=2$) &  &  & 0.64 & 265.63 & 43.83 &&  &  & 0.67 & 278.49 & 44.85 \\ 
			&& Total  & 0.0497 & 0.16 & 0.95 & 441.12 & && 0.0253 & 0.16 & 0.98 & 456.03 & \\ 
			& &&  &  &  &  &&&&&&&  \\ 
			&$\Sigma_{22}=0.25$ & Disease 1 ($j=1$) &  &  & 0.31 & 175.50 & 22.33 &&&& 0.32 & 182.65 & 23.16\\
			&& Disease 2 ($j=2$) &  &  & 0.50 & 204.82 & 30.07 &&&& 0.53 & 218.18 & 31.59\\
			&& Total & 0.3106 & 0.15 & 0.81 & 380.32 & &&0.1584 & 0.15 & 0.85 & 400.83 &  \\ 
			\hline
		\end{tabular}
	}
\end{table}

\begin{table}[t!]
	\centering
	\caption{95\% intervals of the distribution of the empirical smoothing criteria across the simulation replicates for $G=100$ areas under the iCAR spatial prior for different values of $\Sigma_{11}$, $\Sigma_{22}$, and $\rho$ in Scenarios 1 and 2. Results are reported for the overall model rather than separately for each disease.}\label{tabA3.1}
	\resizebox{0.99\textwidth}{!}{
		\begin{tabular}{lll|cccrccc}
			\hline
			&&&\multicolumn{3}{c}{$\rho=0$}&&\multicolumn{3}{c}{$\rho=0.7$}\\
			\cline{4-6}\cline{8-10}
			&&& SRSS & RSP & SP  &&  SRSS & RSP & SP \\ 
			\hline
			\multicolumn{10}{l}{\bf Scenario 1}\\[1.2ex]
			\multicolumn{10}{l}{$\Sigma_{11}=0.0025$}\\[1.2ex]
			&$\Sigma_{22}=0.0025$ & Total  & (604.24, 974.05) & (1.34, 1.51) & (0.67, 0.76) &  & (547.3, 900.44) & (1.22 , 1.42) & (0.61 , 0.72) \\ 
			& &  &  &  &  &  &  &  \\ 
			&$\Sigma_{22}=0.04$ & Total& (512.25, 791.07) & (1.06, 1.31) & (0.53, 0.67) &  & (510.66,  90.22) & (1.04, 1.32) & (0.52, 0.67) \\ 
			&  &  &  &  &  &  &  \\ 
			&$\Sigma_{22}=0.25$ & Total & (488.71, 725.07) & (1.00,  1.21) & (0.49,  0.63) &  & (518.08,  761.82) & (1.04, 1.25) & (0.51,  0.66) \\ 
			&  &  &  &  &  &  &  \\ 
			\multicolumn{10}{l}{$\Sigma_{11}=0.04$}\\[1.2ex]
			&$\Sigma_{22}=0.0025$ & Total & (465.15,  779.43) & (1.03,  1.25) & (0.49,  0.63) &  & (451.72, 768.91) & (1.00,  1.23) & (0.47,  0.62) \\ 
			&  &  &  &  &  &  &  \\ 
			&$\Sigma_{22}=0.04$ & Total  & (350.87,  589.87) & (0.75,  1.03) & (0.37,  0.51) &  & (333.82,  576.94) & (0.73,  1.01) & (0.36,  0.50) \\ 
			&  &  &  &  &  &  &  \\ 
			&$\Sigma_{22}=0.25$ & Total & (326.17,  510.41) & (0.67,  0.94) & (0.34,  0.47) &  & (332.38,  528.66) & (0.69,  0.95) & (0.34,  0.48) \\ 
			&  &  &  &  &  &  &  \\ 
			\multicolumn{10}{l}{$\Sigma_{11}=0.25$}\\[1.2ex]
			&$\Sigma_{22}=0.0025$ & Total & (424.33,  721.30) & (0.96,  1.13) & (0.44,  0.58) &  & (439.13,  747.34) & (0.99,  1.16) & (0.45,  0.59) \\ 
			&  &  &  &  &  &  &  \\ 
			&$\Sigma_{22}=0.04$ & Total & (319.25,  537.06) & (0.68,  0.92) & (0.3,  0.46) &  & (321.68,  543.94) & (0.69, 0.92) & (0.34,  0.48) \\ 
			&  &  &  &  &  &  &  \\ 
			&$\Sigma_{22}=0.25$ & Total & (296.07,  438.12) & (0.60,  0.83) & (0.30,  0.41) &  & (294.82,  475.18) & (0.61,  0.85) & (0.30,  0.42) \\ 
			&  &  &  &  &  &  &  \\ 
			\multicolumn{10}{l}{\bf Scenario 2}\\[1.2ex]
			\multicolumn{10}{l}{$\Sigma_{11}=0.0025$}\\[1.2ex]
			&$\Sigma_{22}=0.0025$ & Total & (586.10,  952.01) & (1.49,  1.63) & (0.70,  0.79) &  & (541.50,  888.58) & (1.37,  1.54) & (0.66,  0.76) \\ 
			&  &  &  &  &  &  &  \\ 
			&$\Sigma_{22}=0.04$ & Total & (501.55,  821.92) & (1.21,  1.44) & (0.64,  0.74) &  & (494.04,  812.10) & (1.17,  1.43) & (0.63,  0.74) \\ 
			&  &  &  &  &  &  &  \\ 
			&$\Sigma_{22}=0.25$ & Total & (470.33,  777.16) & (1.11,  1.33) & (0.61,  0.72) &  & (493.18,  801.76) & (1.14,  1.37) & (0.64,  0.75) \\ 
			&  &  &  &  &  &  &  \\ 
			\multicolumn{10}{l}{$\Sigma_{11}=0.04$}\\[1.2ex]
			&$\Sigma_{22}=0.0025$ & Total & (436.85,  747.33) & (1.18,  1.40) & (0.44,  0.61) &  & (423.50,  728.93) & (1.14,  1.38) & (0.42,  0.60) \\ 
			&  &  &  &  &  &  &  \\ 
			&$\Sigma_{22}=0.04$ & Total & (353.08,  614.60) & (0.91,  1.17) & (0.39,  0.56) &  & (338.23,  595.51) & (0.86,  1.14) & (0.37,  0.55) \\ 
			&  &  &  &  &  &  &  \\ 
			&$\Sigma_{22}=0.25$ & Total& (329.43,  553.65) & (0.80,  1.08) & (0.36,  0.53) &  & (334.78,  553.14) & (0.80,  1.08) & (0.37,  0.53) \\ 
			&  &  &  &  &  &  &  \\ 
			\multicolumn{10}{l}{$\Sigma_{11}=0.25$}\\[1.2ex]
			&$\Sigma_{22}=0.0025$ & Total & (396.75,  699.83) & (1.10,  1.26) & (0.37,  0.52) &  & (402.15,  713.93) & (1.12,  1.28) & (0.37,  0.52) \\ 
			&  &  &  &  &  &  &  \\ 
			&$\Sigma_{22}=0.04$ & Total& (315.43,  542.05) & (0.81,  1.05) & (0.33,  0.46) &  & (316.92,  555.61) & (0.81,  1.07) & (0.32,  0.45) \\ 
			&  &  &  &  &  &  &  \\ 
			&$\Sigma_{22}=0.25$ & Total & (295.89,  480.40) & (0.72,  0.96) & (0.30,  0.43) &  & (297.02,  490.76) & (0.72,  0.98) & (0.31,  0.43) \\ 
			\hline
		\end{tabular}
	}
\end{table}

\begin{table}[t!]
	\centering
	\caption{95\% intervals of the distribution of the empirical smoothing criteria across simulation replicates for $G=100$ areas under the iCAR spatial prior, for different values of $\Sigma_{11}$, $\Sigma_{22}$, and $\rho$ in Scenarios 3 and 4. Results are reported for the overall model rather than separately for each disease.}\label{tabA3.2}
	\resizebox{\textwidth}{!}{
		\begin{tabular}{lll|cccrccc}
			\hline
			&&&\multicolumn{3}{c}{$\rho=0$}&&\multicolumn{3}{c}{$\rho=0.7$}\\
			\cline{4-6}\cline{8-10}
			&&& SRSS & RSP & SP  &&  SRSS & RSP & SP \\ 
			\hline
			\multicolumn{10}{l}{\bf Scenario 3}\\[1.2ex]
			\multicolumn{10}{l}{$\Sigma_{11}=0.0025$}\\[1.2ex]
			&$\Sigma_{22}=0.0025$ & Total & (662.37,  1020.55) & (1.30,  1.51) & (0.65,  0.75) &  & (679.78,  1011.75) & (1.32,  1.54) & (0.66,  0.77) \\ 
			&  &  &  &  &  &  &  \\ 
			&$\Sigma_{22}=0.04$ & Total & (548.10,  839.56) & (1.07,  1.31) & (0.54,  0.66) &  & (572.36,  867.87) & (1.12,  1.35) & (0.57,  0.68) \\ 
			&  &  &  &  &  &  &  \\ 
			&$\Sigma_{22}=0.25$ & Total & (493.32,  741.37) & (0.98,  1.14) & (0.48,  0.60) &  & (531.00,  789.44) & (1.04,  1.21) & (0.51,  0.64) \\ 
			&  &  &  &  &  &  &  \\ 
			\multicolumn{10}{l}{$\Sigma_{11}=0.04$}\\[1.2ex]
			&$\Sigma_{22}=0.0025$ & Total & (504.18,  820.60) & (0.99,  1.23) & (0.48,  0.62) &  & (527.33,  841.95) & (1.03,  1.25) & (0.50,  0.63) \\ 
			&  &  &  &  &  &  &  \\ 
			&$\Sigma_{22}=0.04$ & Total & (391.33,  633.55) & (0.77,  1.04) & (0.38,  0.52) &  & (407.19,  642.93) & (0.78,  1.07) & (0.39,  0.53) \\ 
			&  &  &  &  &  &  &  \\ 
			&$\Sigma_{22}=0.25$ & Total& (343.58,  526.46) & (0.67,  0.88) & (0.33,  0.44) &  & (361.87,  563.21) & (0.68,  0.93) & (0.34,  0.47) \\ 
			&  &  &  &  &  &  &  \\ 
			\multicolumn{10}{l}{$\Sigma_{11}=0.25$}\\[1.2ex]
			&$\Sigma_{22}=0.0025$ & Total& (466.32,  765.88) & (0.90,  1.11) & (0.44,  0.57) &  & (494.09, 813.87) & (0.96,  1.15) & (0.46,  0.60) \\ 
			&  &  &  &  &  &  &  \\ 
			&$\Sigma_{22}=0.04$ & Total& (354.03,  572.77) & (0.68,  0.91) & (0.34,  0.46) &  & (368.30,  612.05) & (0.71,  0.95) & (0.35,  0.48) \\ 
			&  &  &  &  &  &  &  \\ 
			&$\Sigma_{22}=0.25$ & Total & (311.46,  448.87) & (0.59,  0.75) & (0.29,  0.37) &  & (314.63,  498.29) & (0.59,  0.81) & (0.29,  0.41) \\ 
			&  &  &  &  &  &  &  \\ 
			\multicolumn{10}{l}{\bf Scenario 4}\\[1.2ex]
			\multicolumn{10}{l}{$\Sigma_{11}=0.0025$}\\[1.2ex]
			&$\Sigma_{22}=0.0025$ & Total & (622.14,  952.90) & (1.43,  1.60) & (0.68,  0.78) &  & (629.12,  949.14) & (1.47,  1.63) & (0.70,  0.79) \\ 
			&  &  &  &  &  &  &  \\ 
			&$\Sigma_{22}=0.04$ & Total & (539.65,  817.42) & (1.22,  1.44) & (0.63,  0.74) &  & (564.52,  841.30) & (1.28,  1.48) & (0.67,  0.76) \\ 
			&  &  &  &  &  &  &  \\ 
			&$\Sigma_{22}=0.25$ & Total & (490.53,  732.92) & (1.09,  1.30) & (0.59,  0.71) &  & (534.88,  784.42) & (1.16,  1.36) & (0.65,  0.75) \\ 
			&  &  &  &  &  &  &  \\ 
			\multicolumn{10}{l}{$\Sigma_{11}=0.04$}\\[1.2ex]
			&$\Sigma_{22}=0.0025$ & Total & (445.35,  771.52) & (1.11,  1.32) & (0.42,  0.58) &  & (455.24,  779.75) & (1.15,  1.36) & (0.42,  0.59) \\ 
			&  &  &  &  &  &  &  \\ 
			&$\Sigma_{22}=0.04$ & Total& (371.18,  631.26) & (0.91,  1.18) & (0.38,  0.53) &  & (384.31,  641.52) & (0.93,  1.19) & (0.40,  0.54) \\ 
			&  &  &  &  &  &  &  \\ 
			&$\Sigma_{22}=0.025$ & Total & (337.59,  544.92) & (0.77,  1.02) & (0.36,  0.50) &  & (363.09,  575.45) & (0.82,  1.07) & (0.38,  0.52) \\ 
			&  &  &  &  &  &  &  \\ 
			\multicolumn{10}{l}{$\Sigma_{11}=0.25$}\\[1.2ex]
			&$\Sigma_{22}=0.0025$ & Total  & (412.53,  695.34) & (1.04,  1.21) & (0.35,  0.49) &  & (430.19,  720.96) & (1.09,  1.25) & (0.35,  0.51) \\ 
			&  &  &  &  &  &  &  \\ 
			&$\Sigma_{22}=0.04$ & Total & (345.30,  578.18) & (0.85,  1.07) & (0.32,  0.44) &  & (359.50,  591.29) & (0.87,  1.09) & (0.33,  0.46) \\ 
			&  &  &  &  &  &  &  \\ 
			&$\Sigma_{22}=0.025$ & Total & (302.68,  475.80) & (0.71,  0.92) & (0.30,  0.41) &  & (327.07,  513.26) & (0.74,  0.98) & (0.32,  0.44) \\ 
			\hline
		\end{tabular}
	}
\end{table}

\autoref{tabA3.1} and \ref{tabA3.2} present the 95\% intervals of the distribution of the empirical smoothing criteria across simulation replicates for $G=100$ areas under the iCAR spatial prior, for different values of $\Sigma_{11}$, $\Sigma_{22}$, and $\rho$ in Scenarios 1 and 2, and Scenarios 3 and 4, respectively. Given the large number of criteria considered, we report only the intervals corresponding to the total smoothing achieved by the model. As the values of $\Sigma_{11}$ and $\Sigma_{22}$ increase, narrower intervals are generally observed for the SRSS across all scenarios. For the RSP and SP criteria, increasing $\Sigma_{22}$ from 0.0025 to 0.04 generally results in little or no overlap between the corresponding intervals. A similar pattern is observed when increasing $\Sigma_{11}$ from 0.0025 to 0.04. When changing from $\rho=0$ to $\rho\neq0$, the corresponding intervals overlap. However, examining the endpoints of the SRSS intervals reveals a decrease in their values for Scenarios 1 and 2 when either $\Sigma_{11}$ or $\Sigma_{22}$ takes smaller values. This pattern is consistent with the behavior of the mean values reported in the main paper and in the tables above. In contrast, this behavior is not observed in Scenarios 3 and 4, where the endpoints of the SRSS intervals increase as $\rho$ increases.


\begin{table}[t!]
	\centering
	\caption{\label{tabA4}Theoretical and empirical smoothing criteria values for the LCAR spatial prior for $G=100$ across different values of $\Sigma_{11}$, $\Sigma_{22}$, $\rho$ and $\lambda$ under Scenario 1.}
	\resizebox{\textwidth}{!}{
		\begin{tabular}{lll|rrrrrcrrrrr}
			\hline
			&&&\multicolumn{5}{c}{$\rho=0$}&&\multicolumn{5}{c}{$\rho=0.7$}\\
			\cline{4-8}\cline{10-14}
			&&& MultiTCV & SP & RSP & SRSS & maxRSS && MultiTCV & SP & RSP & SRSS & maxRSS \\ 
			\hline
			\multicolumn{14}{l}{$\Sigma_{11}=0.0025$}\\[1.2ex]
			$\Sigma_{22}=0.04$ &$\lambda=0.2$ & Disease 1 ($j=1$) &   &  & 0.78 & 473.36 & 59.20 &&  &  & 0.77 & 464.80 & 57.31\\ 
			&& Disease 2 ($j=2$)  &  &  & 0.43 & 202.80 & 34.32 &&  &  & 0.41 & 193.37 & 33.99\\ 
			& &Total& 0.0032 & 0.61 & 1.21 & 676.16 & && 0.0016 & 0.59 & 1.18 & 658.18 &  \\ 
			& &  &  &  &  & &&&&&&& \\
			& $\lambda=0.8$ &  Disease 1 ($j=1$) &   &  & 0.74 & 443.04 & 52.93 &&  &  & 0.75 & 444.48 & 52.93 \\ 
			&& Disease 2 ($j=2$) &  &  & 0.46 & 213.30 & 36.45 &&  &  & 0.45 & 204.96 & 35.58\\ 
			& &Total  & 0.0007 & 0.60 & 1.21 & 656.34 & && 0.0003 & 0.60 & 1.19 & 649.44 &  \\ 
			& &  &  &  &  & &&&&&&& \\ 
			& &  &  &  &  & &&&&&&& \\
			\multicolumn{14}{l}{$\Sigma_{11}=0.04$}\\[1.2ex]
			$\Sigma_{22}=0.0025$ &$\lambda=0.2$& Disease 1 ($j=1$)  &  &  & 0.37 & 213.25 & 31.47 &&  &  & 0.35 & 202.65 & 30.61 \\ 
			&& Disease 2 ($j=2$) &  &  & 0.80 & 403.27 & 63.04 &&  &  & 0.76 & 381.73 & 60.95  \\ 
			& &Total   & 0.0032 & 0.56 & 1.17 & 616.52 & && 0.0016 & 0.53 & 1.11 & 584.38 & \\ 
			& &  &  &  &  & &&&&&&& \\
			& $\lambda=0.8$ &  Disease 1 ($j=1$) &  &  & 0.40 & 227.20 & 33.41 &&  &  & 0.39 & 219.30 & 32.89 \\ 
			&& Disease 2 ($j=2$) &  &  & 0.75 & 370.13 & 57.68 &&  &  & 0.73 & 357.23 & 56.40 \\ 
			& &Total   & 0.0007 & 0.55 & 1.15 & 597.33 & && 0.0003 & 0.54 & 1.12 & 576.53 &  \\ 
			& &  &  &  &  & &&&&&&& \\
			$\Sigma_{22}=0.04$ &$\lambda=0.2$& Disease 1 ($j=1$) &  &  & 0.37 & 213.25 & 31.46 &&  &  & 0.36 & 210.20 & 32.05 \\ 
			&& Disease 2 ($j=2$) &  &  & 0.43 & 201.65 & 34.20 &&  &  & 0.42 & 198.03 & 35.61  \\ 
			& &Total & 0.0513 & 0.39 & 0.80 & 414.90 & && 0.0261 & 0.38 & 0.79 & 408.23 &  \\ 
			& &  &  &  &  & &&&&&&& \\
			& $\lambda=0.8$ &  Disease 1 ($j=1$) &  &  & 0.40 & 226.86 & 33.26 &&  &  & 0.40 & 224.87 & 33.54 \\ 
			&& Disease 2 ($j=2$) &  &  & 0.47 & 213.09 & 36.36 &&  &  & 0.45 & 207.87 & 36.73 \\ 
			& &Total & 0.0109 & 0.42 & 0.86 & 439.94 &&& 0.0056 & 0.42 & 0.85 & 432.74 & \\ 
			& &  &  &  &  & &&&&&&& \\
			$\Sigma_{22}=0.25$ &$\lambda=0.2$& Disease 1 ($j=1$) &  &  & 0.37 & 213.36 & 31.37 &&  &  & 0.40 & 231.34 & 34.77 \\ 
			&& Disease 2 ($j=2$) &  &  & 0.35 & 161.53 & 17.76 &&  &  & 0.36 & 162.71 & 18.60  \\ 
			&& Total  & 0.3204 & 0.35 & 0.72 & 374.89 & && 0.1634 & 0.37 & 0.75 & 394.05 &  \\ 
			& &  &  &  &  & &&&&&&& \\
			& $\lambda=0.8$ &  Disease 1 ($j=1$) &  &  & 0.40 & 227.29 & 33.61 & &  &  & 0.42 & 239.96 & 34.80\\ 
			&& Disease 2 ($j=2$) &  &  & 0.36 & 164.13 & 23.25 &&  &  & 0.36 & 164.20 & 23.75  \\ 
			&& Total  & 0.0683 & 0.37 & 0.76 & 391.42 & && 0.0348 & 0.39 & 0.79 & 404.16 &   \\ 
			& &  &  &  &  & &&&&&&& \\
			& &  &  &  &  & &&&&&&& \\
			\multicolumn{14}{l}{$\Sigma_{11}=0.25$}\\[1.2ex]
			$\Sigma_{22}=0.04$ &$\lambda=0.2$& Disease 1 ($j=1$) &  &  & 0.27 & 158.31 & 17.17 &&  &  & 0.28 & 161.82 & 18.74\\ 
			&& Disease 2 ($j=2$) &  &  & 0.43 & 202.05 & 34.39 &&  &  & 0.46 & 216.61 & 38.81 \\ 
			&& Total & 0.3204 & 0.34 & 0.70 & 360.36 & && 0.1634 & 0.36 & 0.74 & 378.43 & \\ 
			& &  &  &  &  & &&&&&&& \\
			& $\lambda=0.8$ &  Disease 1 ($j=1$) &  &  & 0.29 & 168.20 & 23.10 &&  &  & 0.29 & 168.37 & 23.57\\ 
			&& Disease 2 ($j=2$) &  &  & 0.47 & 213.30 & 36.35 &&  &  & 0.47 & 220.39 & 38.83  \\ 
			&& Total & 0.0683 & 0.37 & 0.76 & 381.50 & && 0.0348 & 0.37 & 0.77 & 388.76 & \\ 
			\hline
		\end{tabular}
	}
\end{table}

\autoref{tabA4} presents the omitted results for LCAR prior with $\Sigma_{11}=0.04$ or $\Sigma_{22}=0.04$. Results show consistent behavior with the conclusions drawn in the main text. In these settings, increasing $\lambda$ still leads to slightly lower empirical smoothing, while the theoretical metric decreases, indicating stronger smoothing. This discrepancy vanishes when either $\Sigma_{11}$ or $\Sigma_{22}$ is set to 0.25.

\begin{figure}[b!]
	\centering
	\includegraphics[width=0.9\linewidth]{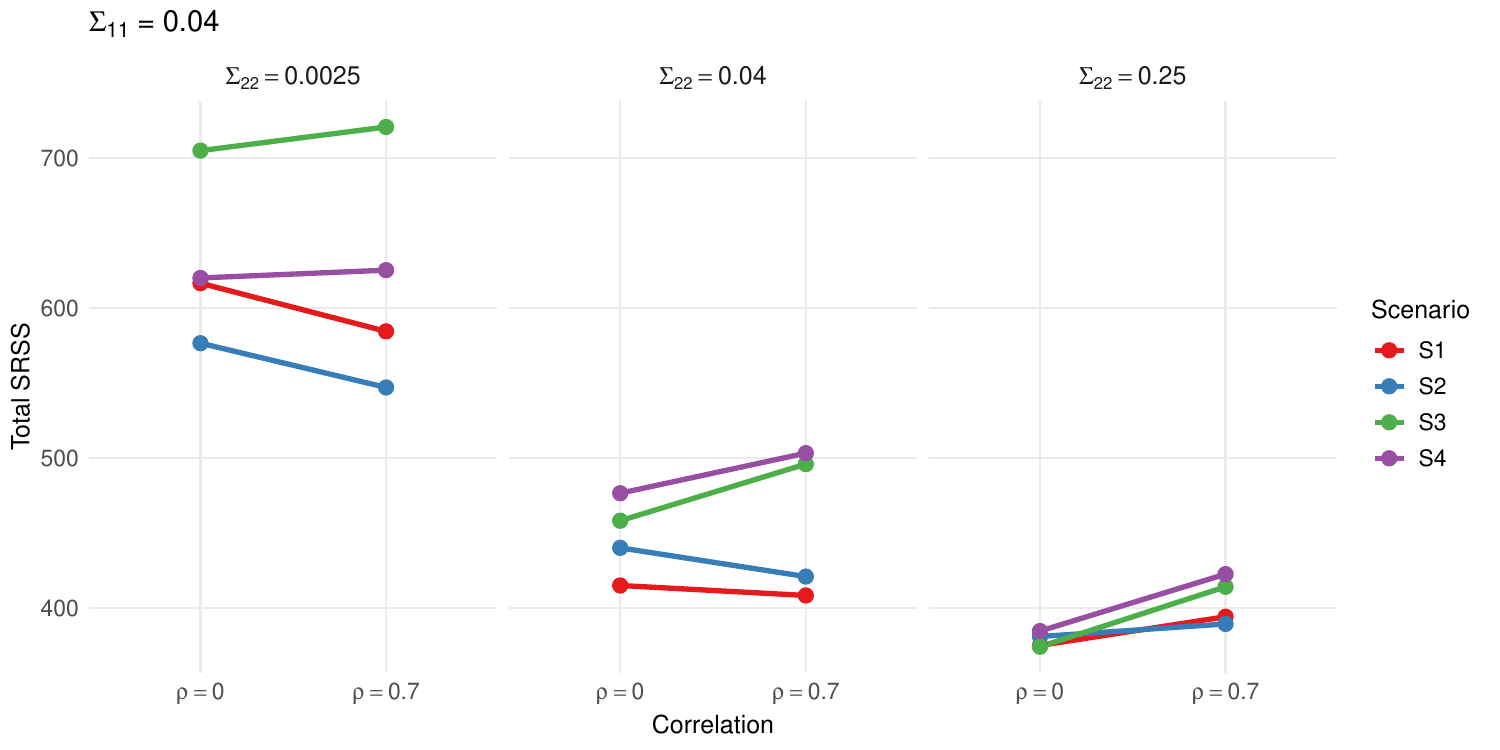}
	\caption{Total SRSS values, representing the combined smoothing of both diseases, for $\lambda=0.2$ and $\Sigma_{11}=0.04$, while varying $\Sigma_{22}$ across the four scenarios and different $\rho$ values.}
	\label{figA1}
\end{figure}

As in the iCAR case, the behavior of the $\rho$ parameter under the LCAR prior does not fully agree with the theoretical metric. To explore this, \autoref{figA1} displays total SRSS values across the four scenarios for $\Sigma_{11}=0.04$ and $\lambda=0.2$, while varying $\Sigma_{22}$. Although MultiTCV is not shown in the figure, \autoref{tabA4} indicates that increasing $\rho$ consistently decreases MultiTCV, implying stronger theoretical smoothing. Empirical behavior, however, varies by scenario and by $\Sigma_{22}$. For small values  of $\Sigma_{22}$ (0.0025 and 0.04), SRSS decreases with increasing $\rho$ in Scenarios~1 and~2, indicating disagreement between empirical and theoretical metrics, while Scenarios~3 and~4 show the opposite pattern. As $\Sigma_{22}$ increases, empirical and theoretical measures align across all scenarios. This behavior mirrors that observed for the iCAR prior and is consistent across different values of $\lambda$.

\begin{figure}[t!]
	\centering
	\includegraphics[width=0.9\linewidth]{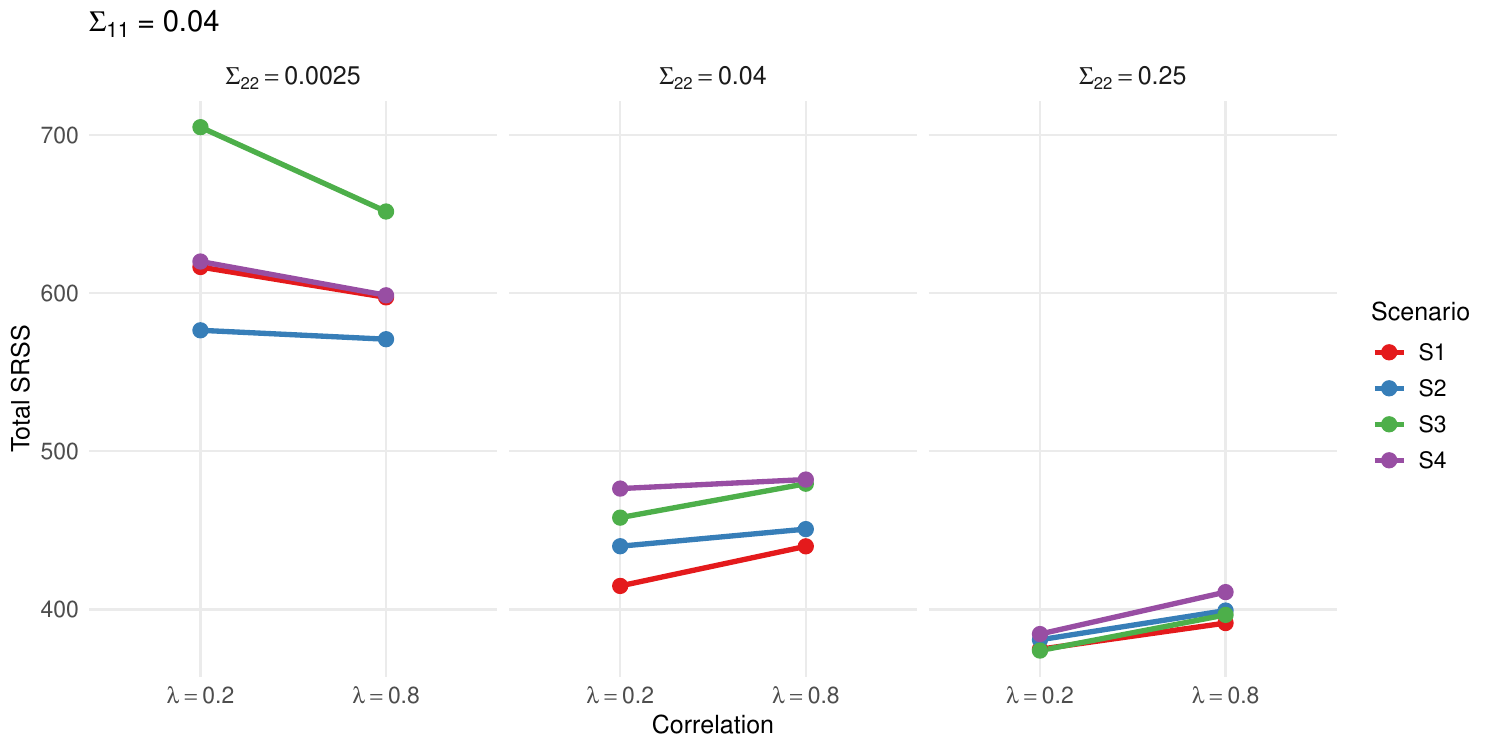}
	\caption{Total SRSS values, representing the combined smoothing of both diseases, for $\rho=0$, $\Sigma_{11}=0.04$ and varying $\Sigma_{22}$ across the four scenarios and different $\lambda$ values.}
	\label{figA2}
\end{figure}

As seen in Table~2 of the main text, the impact of $\lambda$ differs across the two measures when analyzing the LCAR spatial prior. More precisely, higher $\lambda$ yields slightly less empirical smoothing, whereas the theoretical metric decreases as $\lambda$ increases, implying stronger smoothing. To gain further intuition about the role of the $\lambda$ parameter in the smoothing amount, \autoref{figA2} displays the total SRSS values obtained across the four scenarios for the specific case where $\Sigma_{11}=0.04$ and $\rho=0$, while varying $\Sigma_{22}$. Different colors correspond to different scenarios. \autoref{figA2} does not include the MultiTCV but, as \autoref{tabA4} shows, increasing $\lambda$ decreases the MultiTCV, which theoretically indicates greater smoothing. This behavior is consistent across all scenarios. \autoref{figA2} illustrates that all scenarios behave similarly, indicating that neither the correlation nor the variability structure of the rates alters how smoothing responds to changes in $\lambda$. In addition, as $\Sigma_{22}$ increases, the theoretical and empirical metrics align, all scenarios exhibit higher smoothing as $\lambda$ increases. Moreover, the smoothing levels also become more similar across scenarios as $\Sigma_{22}$ increases. A comparable pattern is observed when varying $\Sigma_{11}$ or $\rho$.

\begin{table}[t!]
	\centering
	\caption{95\% intervals of the distribution of the empirical smoothing criteria across simulation replicates for $G=100$ areas under the LCAR spatial prior, for different values of $\Sigma_{11}$, $\Sigma_{22}$, $\rho$ and $\lambda$ in Scenario 1. Results are reported for the overall model rather than separately for each disease.}\label{tabA4.1}
	\resizebox{\textwidth}{!}{
		\begin{tabular}{llll|cccrccc}
			\hline
			&&&&\multicolumn{3}{c}{$\rho=0$}&&\multicolumn{3}{c}{$\rho=0.7$}\\
			\cline{5-7}\cline{9-11}
			&&&& SRSS & RSP & SP  &&  SRSS & RSP & SP \\ 
			\hline
			\multicolumn{11}{l}{$\Sigma_{11}=0.0025$}\\[1.2ex]
			&$\Sigma_{22}=0.0025$ & $\lambda=0.2$ & Total & (695.76,  1116.53) & (1.53,  1.64) & (0.76,  0.82) &  & (626.94,  1027.58) & (1.39,  1.53) & (0.69,  0.77) \\ 
			&  &  &  &  &  &  &  &  \\ 
			& & $\lambda=0.8$ & Total& (642.68,  1029.81) & (1.42,  1.56) & (0.71,  0.78) &  & (577.17,  950.30) & (1.30,  1.47) & (0.65,  0.74) \\ 
			&  &  &  &  &  &  &  \\ 
			&$\Sigma_{22}=0.04$ & $\lambda=0.2$ & Total & (541.43,  861.30) & (1.13,  1.29) & (0.57,  0.68) &  & (532.49,  838.10) & (1.07,  1.27) & (0.54,  0.67) \\ 
			&  &  &  &  &  &  &  \\ 
			& & $\lambda=0.8$ & Total & (531.87,  829.18) & (1.11,  1.32) & (0.55,  0.68) &  & (528.86,  819.62) & (1.06,  1.31) & (0.54,  0.68) \\ 
			&  &  &  &  &  &  &  \\ 
			&$\Sigma_{22}=0.25$ & $\lambda=0.2$ & Total & (523.99,  785.43) & (1.06,  1.24) & (0.51,  0.66) &  & (549.02,  818.67) & (1.08,  1.25) & (0.53,  0.68) \\ 
			&  &  &  &  &  &  &  \\ 
			& & $\lambda=0.8$ & Total  & (498.02,  755.35) & (1.01,  1.20) & (0.50,  0.64) &  & (524.37,  779.29) & (1.06,  1.24) & (0.52,  0.67) \\ 
			&  &  &  &  &  &  &  \\ 
			\multicolumn{11}{l}{$\Sigma_{11}=0.04$}\\[1.2ex]
			&$\Sigma_{22}=0.0025$ & $\lambda=0.2$ & Total  & (486.57,  808.98) & (1.10,  1.23) & (0.51,  0.63) &  & (457.81,  778.01) & (1.02,  1.19) & (0.48,  0.60) \\ 
			&  &  &  &  &  &  &  \\ 
			& & $\lambda=0.8$ & Total  & (474.40,  790.43) & (1.06,  1.25) & (0.50,  0.63) &  & (450.80,  777.31) & (1.01,  1.22) & (0.47,  0.62) \\ 
			&  &  &  &  &  &  &  \\ 
			&$\Sigma_{22}=0.04$ & $\lambda=0.2$ & Total  & (321.36,  529.98) & (0.69,  0.89) & (0.34,  0.44) &  & (314.54,  513.52) & (0.67,  0.91) & (0.33,  0.45) \\ 
			&  &  &  &  &  &  &  \\ 
			& & $\lambda=0.8$ & Total & (340.33,  572.09) & (0.72,  0.99) & (0.36,  0.49) &  & (326.99,  556.59) & (0.70,  0.99) & (0.35,  0.48) \\ 
			&  &  &  &  &  &  &  \\ 
			&$\Sigma_{22}=0.25$ & $\lambda=0.2$ & Total & (298.63,  457.86) & (0.62,  0.85) & (0.31,  0.42) &  & (316.27,  508.59) & (0.65,  0.89) & (0.32,  0.45) \\ 
			&  &  &  &  &  &  &  \\ 
			& & $\lambda=0.8$ & Total & (311.60,  493.70) & (0.65,  0.89) & (0.33,  0.44) &  & (315.58,  503.54) & (0.65,  0.93) & (0.32,  0.46) \\ 
			&  &  &  &  &  &  &  \\ 
			\multicolumn{11}{l}{$\Sigma_{11}=0.25$}\\[1.2ex]
			&$\Sigma_{22}=0.0025$ & $\lambda=0.2$ & Total  & (450.73,  763.27) & (1.02,  1.13) & (0.45,  0.59) &  & (460.62,  765.40) & (1.04,  1.16) & (0.46,  0.60) \\ 
			&  &  &  &  &  &  &  \\ 
			& & $\lambda=0.8$ & Total  & (425.40,  732.45) & (0.98, 1.13) & (0.44,  0.58) &  & (443.37,  750.17) & (1.00,  1.15) & (0.44,  0.59) \\ 
			&  &  &  &  &  &  &  \\ 
			&$\Sigma_{22}=0.04$ & $\lambda=0.2$ & Total  & (287.28,  462.52) & (0.62,  0.79) & (0.30,  0.40) &  & (306.07,  496.25) & (0.65,  0.85) & (0.31,  0.43) \\ 
			&  &  &  &  &  &  &  \\ 
			& & $\lambda=0.8$ & Total  & (299.65,  518.61) & (0.65,  0.87) & (0.32,  0.44) &  & (308.16,  523.28) & (0.66,  0.89) & (0.32,  0.46) \\ 
			&  &  &  &  &  &  &  \\ 
			&$\Sigma_{22}=0.25$ & $\lambda=0.2$ & Total  & (281.50,  365.43) & (0.54,  0.71) & (0.27,  0.35) &  & (282.75,  407.92) & (0.58,  0.76) & (0.29,  0.38) \\ 
			&  &  &  &  &  &  &  \\ 
			& & $\lambda=0.8$ & Total  & (269.23,  406.14) & (0.55,  0.77) & (0.27,  0.38) &  & (275.42,  460.93) & (0.57,  0.80) & (0.29,  0.40) \\ 
			\hline
		\end{tabular}
	}
\end{table}

Similar to the iCAR spatial prior, \autoref{tabA4.1} presents the 95\% intervals of the distribution of the empirical smoothing criteria across simulation replicates for $G=100$ areas under the LCAR spatial prior, for different values of $\Sigma_{11}$, $\Sigma_{22}$, $\rho$, and $\lambda$. These results complement those presented in the main text and in the previous table. When comparing scenarios while keeping $\lambda$ fixed, the results are similar to those obtained under the iCAR prior: increasing $\Sigma_{11}$ or $\Sigma_{22}$ generally leads to narrower SRSS intervals across all scenarios. For the RSP and SP criteria, increasing either $\Sigma_{11}$ or $\Sigma_{22}$ from 0.0025 to 0.04 generally results in little or no overlap between the corresponding intervals. As $\lambda$ increases, smaller values of $\Sigma_{11}$ or $\Sigma_{22}$ generally lead to narrower intervals, whereas larger values tend to produce wider intervals, particularly for the SRSS. Finally, when comparing intervals across increasing values of $\rho$, they generally overlap, as observed for the iCAR prior. However, the endpoints of the intervals decrease when either $\Sigma_{11}$ or $\Sigma_{22}$ takes smaller values.

\begin{sidewaystable}
	\caption{\label{tabA5}Theoretical and empirical smoothing criteria values for the L$_j$CAR spatial prior for $G=100$ across different values of $\Sigma_{11}$, $\Sigma_{22}$, $\lambda_1$, $\lambda_2$ and $\rho$ under Scenario 1. }
	\resizebox{\textheight}{!}{
		\begin{tabular}{lllll|rrrrrcrrrrrcl|rrrrrcrrrrr}
			\hline
			&&&&&\multicolumn{5}{c}{$\rho=0$}&&\multicolumn{5}{c}{$\rho=0.7$}&&\multicolumn{5}{c}{$\rho=0$}&&\multicolumn{5}{c}{$\rho=0.7$}\\
			\cline{6-10}\cline{12-16}\cline{19-23}\cline{25-29}
			&&&&& MultiTCV & SP & RSP & SRSS & maxRSS && MultiTCV & SP & RSP & SRSS & maxRSS &&& MultiTCV & SP & RSP & SRSS & maxRSS && MultiTCV & SP & RSP & SRSS & maxRSS \\ 
			\hline
			\multicolumn{16}{l}{$\Sigma_{11}=0.0025$} & \multicolumn{13}{l}{$\Sigma_{11}=0.25$}\\[1.2ex]
			&$\Sigma_{22}=0.0025$ &$\lambda_1=0.2$&$\lambda_2=0.2$& Disease 1 ($j=1$)&  &  & 0.78 & 473.36 & 59.11 &&  &  & 0.73 & 437.14 & 55.49 & & &  &  & 0.27 & 159.07 & 17.29   &&  &  & 0.27 & 158.45 & 17.93  \\ 
			&&&& Disease 2 ($j=2$) &  &  & 0.80 & 402.89 & 62.88 &&  &  & 0.73 & 364.46 & 59.76 & & &  &  & 0.80 & 403.23 & 63.13 &&  &  & 0.82 & 414.42 & 64.05\\ 
			&&& &Total & 0.0002 & 0.77 & 1.58 & 876.25 && & 0.0001 & 0.72 & 1.46 & 801.59 & & & & 0.0200 & 0.51 & 1.07 & 562.31  & && 0.0102 & 0.52 & 1.09 & 572.88 &  \\ 
			& & & &  &  &  & &&&&&&&& & & & &  &  &  & &&&&\\
			&&& $\lambda_2=0.8$ &  Disease 1 ($j=1$)&  &  & 0.74 & 442.39 & 52.92 &&  &  & 0.73 & 437.58 & 55.27 & &  & & & 0.27 & 160.86 & 17.93 &&  &  & 0.27 & 157.73 & 16.88 \\ 
			&&&& Disease 2 ($j=2$)&  &  & 0.80 & 403.00 & 62.84&&  &  & 0.74 & 365.03 & 59.92 &&&  &  & 0.75 & 370.47 & 57.81 &&  &  & 0.78 & 388.81 & 60.43 \\ 
			&&&&Total  & 0.0001 & 0.76 & 1.54 & 845.40 & && 0.0000 & 0.72 & 1.47 & 802.61 &  &&& 0.0091 & 0.49 & 1.03 & 531.33  &&& 0.0047 & 0.50 & 1.05 & 546.55 &   \\ 
			& & & &  &  &  & &&&&&&&& & & & &  &  &  & &&&&\\
			&&$\lambda_1=0.8$&$\lambda_2=0.2$& Disease 1 ($j=1$) &  &  & 0.78 & 473.45 & 59.23 & &  &  & 0.70 & 412.06 & 49.79 && &  &  & 0.29 & 167.21 & 22.48 & &  &  & 0.29 & 166.56 & 22.76 \\ 
			&&&& Disease 2 ($j=2$)   &  &  & 0.75 & 369.79 & 57.71&&  &  & 0.69 & 332.36 & 53.16&& &  &  & 0.80 & 403.09 & 63.15 &&  &  & 0.82 & 413.06 & 63.61\\ 
			&& &&Total  & 0.0001 & 0.75 & 1.53 & 843.23 & & & 0.0000 & 0.68 & 1.39 & 744.43 & &&  & 0.0091 & 0.52 & 1.09 & 570.30 &&& 0.0047 & 0.53 & 1.11 & 579.62 &   \\ 
			& & & &  &  &  & &&&&&&&& & & & &  &  &  & &&&&\\
			&&& $\lambda_2=0.8$ &  Disease 1 ($j=1$) &  &  & 0.74 & 442.22 & 52.66 &&  &  & 0.70 & 412.22 & 49.52 &&&  &  & 0.29 & 167.22 & 22.38 &&  &  & 0.29 & 166.27 & 22.60 \\ 
			&&&& Disease 2 ($j=2$)  &  &  & 0.75 & 370.32 & 57.75 & &  &  & 0.69 & 331.93 & 53.17 &&  &  &  & 0.75 & 370.13 & 57.78 &&  &  & 0.78 & 387.53 & 59.94\\ 
			&&& &Total & 0.0000 & 0.73 & 1.49 & 812.55 &&& 0.0000 & 0.68 & 1.39 & 744.15 & && & 0.0043 & 0.49 & 1.04 & 537.35 &  & & 0.0022 & 0.51 & 1.07 & 553.80 & \\  
			& & & &  &  &  & &&&&&&&& & & & &  &  &  & &&&&\\
			&$\Sigma_{22}=0.25$  &$\lambda_1=0.2$&$\lambda_2=0.2$& Disease 1 ($j=1$) &  &  & 0.78 & 473.34 & 59.09 &&  &  & 0.82 & 500.03 & 60.91 &&&  &  & 0.27 & 157.69 & 17.13 & & &  & 0.29 & 171.95 & 22.13 \\ 
			&&&& Disease 2 ($j=2$) &  &  & 0.35 & 162.15 & 18.33 &&  &  & 0.35 & 160.61 & 18.71  &&&  &  & 0.35 & 161.14 & 17.74 & & &  & 0.37 & 171.74 & 22.16 \\  
			&&&&Total & 0.0200 & 0.57 & 1.13 & 635.49 &&& 0.0102 & 0.59 & 1.17 & 660.64 &   && & 2.0027 & 0.30 & 0.62 & 318.83 &&& 1.0214 & 0.32 & 0.66 & 343.68 &  \\ 
			& & & &  &  &  & &&&&&&&& & & & &  &  &  & &&&&\\
			&&& $\lambda_2=0.8$ &  Disease 1 ($j=1$)  &  &  & 0.74 & 442.37 & 52.80 &&  &  & 0.78 & 474.45 & 56.61 &&&  &  & 0.29 & 168.22 & 22.67 &&  &  & 0.31 & 181.17 & 24.55 \\ 
			&&&& Disease 2 ($j=2$) &  &  & 0.35 & 162.90 & 18.28  &&  &  & 0.35 & 160.42 & 18.18 && &  &  & 0.35 & 160.71 & 17.93 &&  &  & 0.38 & 180.26 & 25.97 \\ 
			&&&&Total & 0.0091 & 0.55 & 1.09 & 605.27  &&& 0.0047 & 0.57 & 1.13 & 634.86 &   &&& 0.9131 & 0.31 & 0.64 & 328.94 &&& 0.4657 & 0.34 & 0.69 & 361.44 &   \\  
			& & & &  &  &  & &&&&&&&& & & & &  &  &  & &&&&\\
			&&$\lambda_1=0.8$&$\lambda_2=0.2$& Disease 1 ($j=1$) &  &  & 0.78 & 473.78 & 59.29 & &  &  & 0.82 & 499.35 & 60.86 &&&  &  & 0.27 & 157.93 & 16.99 &&  &  & 0.29 & 166.96 & 22.80  \\ 
			&&&& Disease 2 ($j=2$)  &  &  & 0.36 & 164.80 & 23.07 &&  &  & 0.36 & 163.39 & 23.15 &&&  &  & 0.36 & 163.81 & 23.16 &&  &  & 0.36 & 162.59 & 22.45 \\  
			&&& &Total& 0.0091 & 0.57 & 1.14 & 638.58 &&& 0.0047 & 0.59 & 1.18 & 662.74 &&&& 0.9131 & 0.31 & 0.63 & 321.73 & && 0.4657 & 0.31 & 0.65 & 329.55 & \\  
			& & & &  &  &  & &&&&&&&& & & & &  &  &  & &&&&\\
			
			&&& $\lambda_2=0.8$ &  Disease 1 ($j=1$) &  &  & 0.74 & 441.99 & 52.84 & &  &  & 0.79 & 474.78 & 56.40 &&&  &  & 0.29 & 167.49 & 22.37 & &  &  & 0.31 & 177.23 & 25.77\\  
			&&&& Disease 2 ($j=2$)  &  &  & 0.36 & 164.84 & 23.35 &&  &  & 0.36 & 163.81 & 23.24 &&&  &  & 0.36 & 163.40 & 22.90 &&  &  & 0.37 & 171.33 & 25.75\\ 
			&&&& Total & 0.0043 & 0.55 & 1.11 & 606.82 &&& 0.0022 & 0.58 & 1.15 & 638.59 &  &&& 0.4267 & 0.32 & 0.65 & 330.90 & && 0.2176 & 0.33 & 0.68 & 348.56 &   \\
			& & & &  &  &  & &&&&&&&& & & & &  &  &  & &&&&\\
			\hline
		\end{tabular}
	}
\end{sidewaystable}

\autoref{tabA5} presents the smoothing amount obtained by the L$_j$CAR spatial prior. Similar to the LCAR spatial prior, the smoothing depends on $\Sigma_{11}$, $\Sigma_{22}$ and $\rho$, but now the spatial dependence parameters are disease-specific, $\lambda_1$ and $\lambda_2$. Results are presented for $G=100$ under varying values of $\Sigma_{11}$, $\Sigma_{22}$ and $\rho$. For the L$_j$CAR spatial prior, we analyze all combinations of $\lambda_j \in {0.2, 0.8}$ for $j=1,2$. We summarize the main findings here and omit the cases with $\Sigma_{11}=0.04$ and $\Sigma_{22}=0.04$. In general, conclusions are similar to those obtained for the LCAR spatial prior.
For fixed $\rho$ and $\lambda_j$ values, the theoretical and empirical measures are consistent, both showing stronger smoothing as $\Sigma_{11}$ and $\Sigma_{22}$ decrease, and the effect of variations in these values are consistent with the previous results.
As before, the effect of $\rho$ is less stable, and the theoretical and empirical metrics do not always align. When $\Sigma_{11}$ or $\Sigma_{22}$ is large, both measures show consistent behavior: increasing $\rho$ leads to greater smoothing, although this effect differs across scenarios.
The discrepancy in the effect of the $\lambda_j$ parameters on the empirical and theoretical smoothing values is even more pronounced. This divergence is no longer observed when both $\Sigma_{11}$ and $\Sigma_{22}$ are equal to 0.25, and this pattern is consistent across all scenarios considered.

\begin{table}[ht]
	\centering
	\caption{95\% intervals of the distribution of the empirical smoothing criteria across simulation replicates for $G=100$ areas under the LCAR spatial prior, for different values of $\Sigma_{11}$, $\Sigma_{22}$, $\rho$ and $\lambda$ in Scenario 1. Results are reported for the overall model rather than separately for each disease.}\label{tabA5.1}
	\resizebox{\textwidth}{!}{
		\begin{tabular}{lllll|cccrccc}
			\hline
			&&&&&\multicolumn{3}{c}{$\rho=0$}&&\multicolumn{3}{c}{$\rho=0.7$}\\
			\cline{6-8}\cline{10-12}
			&&&&& SRSS & RSP & SP  &&  SRSS & RSP & SP \\ 
			\hline
			\multicolumn{12}{l}{$\Sigma_{11}=0.0025$}\\[1.2ex]
			&$\Sigma_{22}=0.0025$ & $\lambda_1=0.2$ & $\lambda_2=0.2$ & Total & (699.01,  1113.06) & (1.52, 1.64) & (0.76,  0.82) &  & (626.81,  1023.26) & (1.40,  1.53) & (0.70,  0.77) \\ 
			&  &  &  &  &  &  &  \\ 
			& && $\lambda_2=0.8$ & Total  & (674.43,  1080.79) & (1.48,  1.60) & (0.74,  0.80) &  & (628.80,  1033.46) & (1.39,  1.54) & (0.69,  0.77) \\ 
			&  &  &  &  &  &  &  \\ 
			&& $\lambda_1=0.8$ & $\lambda_2=0.2$ & Total & (669.07,  1061.48) & (1.46,  1.59) & (0.74,  0.80) &  & (577.12,  946.28) & (1.30, 1.47) & (0.65,  0.74) \\ 
			&  &  &  &  &  &  &  \\ 
			& && $\lambda_2=0.8$ & Total & (640.16,  1034.36) & (1.42,  1.56) & (0.71,  0.78) &  & (581.81,  953.36) & (1.30,  1.48) & (0.65,  0.74) \\ 
			&  &  &  &  &  &  &  \\ 
			&$\Sigma_{22}=0.25$ & $\lambda_1=0.2$ & $\lambda_2=0.2$ & Total    & (522.11,  788.57) & (1.05,  1.23) & (0.51,  0.66) &  & (549.02,  819.16) & (1.08,  1.25) & (0.52,  0.68) \\ 
			&  &  &  &  &  &  &  \\ 
			& && $\lambda_2=0.8$ & Total  & (501.81,  731.48) & (1.01,  1.19) & (0.49,  0.64) &  & (520.49,  787.18) & (1.05,  1.22) & (0.51,  0.66) \\ 
			&  &  &  &  &  &  &  \\ 
			&& $\lambda_1=0.8$ & $\lambda_2=0.2$ & Total & (520.11,  795.38) & (1.05,  1.24) & (0.52,  0.67) &  & (548.71,  823.39) & (1.09,  1.28) & (0.54,  0.69) \\ 
			&  &  &  &  &  &  &  \\ 
			& && $\lambda_2=0.8$ & Total  & (502.91,  740.81) & (1.01,  1.21) & (0.50,  0.64) &  & (524.84,  792.38) & (1.05,  1.26) & (0.52,  0.67) \\ 
			&  &  &  &  &  &  &  \\ 
			\multicolumn{12}{l}{$\Sigma_{11}=0.25$}\\[1.2ex]
			&$\Sigma_{22}=0.0025$ & $\lambda_1=0.2$ & $\lambda_2=0.2$ & Total & (456.97,  773.52) & (1.02,  1.14) & (0.45,  0.60) &  & (459.14,  780.44) & (1.04,  1.15) & (0.46,  0.61) \\ 
			&  &  &  &  &  &  &  \\ 
			& && $\lambda_2=0.8$ & Total & (427.97,  726.43) & (0.97,  1.11) & (0.43,  0.57) &  & (445.88,  748.87) & (0.99,  1.12) & (0.44,  0.59) \\ 
			&  &  &  &  &  &  &  \\ 
			&& $\lambda_1=0.8$ & $\lambda_2=0.2$ & Total  & (452.25,  764.07) & (1.02,  1.16) & (0.45,  0.60) &  & (458.12,  776.12) & (1.04,  1.19) & (0.46,  0.61) \\ 
			&  &  &  &  &  &  &  \\ 
			& && $\lambda_2=0.8$ & Total  & (425.29,  730.64) & (0.98,  1.12) & (0.44,  0.57) &  & (440.30,  752.17) & (1.00, 1.15) & (0.45,  0.60) \\ 
			&  &  &  &  &  &  &  \\ 
			&$\Sigma_{22}=0.25$ & $\lambda_1=0.2$ & $\lambda_2=0.2$ & Total & (282.53,  375.57) & (0.53,  0.70) & (0.26,  0.34) &  & (288.31,  433.51) & (0.56,  0.77) & (0.28,  0.38) \\ 
			&  &  &  &  &  &  &  \\ 
			& && $\lambda_2=0.8$ & Total& (274.60,  387.21) & (0.55,  0.76) & (0.27,  0.37) &  & (290.74,  476.31) & (0.57,  0.82) & (0.29,  0.41) \\ 
			&  &  &  &  &  &  &  \\ 
			&& $\lambda_1=0.8$ & $\lambda_2=0.2$ & Total  & (270.09,  387.75) & (0.55,  0.73) & (0.27,  0.36) &  & (272.03,  420.83) & (0.55,  0.75) & (0.27,  0.37) \\ 
			&  &  &  &  &  &  &  \\ 
			& && $\lambda_2=0.8$ & Total & (274.15,  403.30) & (0.55,  0.75) & (0.27,  0.37) &  & (277.49,  452.37) & (0.56,  0.80) & (0.28,  0.40) \\ 
			\hline
		\end{tabular}
	}
\end{table}

Similar to the iCAR and LCAR spatial priors, \autoref{tabA5.1} presents the 95\% intervals of the distribution of the empirical smoothing criteria across simulation replicates for $G=100$ areas under the L$j$CAR spatial prior, for different values of $\Sigma{11}$, $\Sigma_{22}$, $\rho$, and the disease-specific spatial dependence parameters $\lambda_1$ and $\lambda_2$. Overall, the interval-based results show the same patterns observed for the iCAR and LCAR priors. For fixed $\rho$ and $\lambda_j$, increasing $\Sigma_{11}$ or $\Sigma_{22}$ generally leads to narrower SRSS intervals, while for the RSP and SP criteria, increasing either parameter from 0.0025 to 0.25 generally results in no overlap between the corresponding intervals. When changing from $\rho=0$ to $\rho\neq0$, the intervals generally overlap. However, the endpoints of the SRSS intervals decrease with smaller values of $\Sigma_{11}$ or $\Sigma_{22}$, consistent with the behaviour of the mean values. The effect of $\lambda_j$ is also consistent with that observed under the LCAR prior, for smaller values of $\Sigma_{11}$ or $\Sigma_{22}$, increasing $\lambda_j$ generally results in narrower intervals, whereas larger values of these parameters tend to produce wider intervals, particularly for the SRSS.

\clearpage	
\section{Across Simulation Study}\label{SectionSM:Across}

This section presents the results of the across-prior simulation study described in Section 3.3 of the main text for $G=47$, together with the 95\% intervals for the differences between priors for the criteria reported for $G=47$, $100$, and $300$. We also provide the mean bias across the $B$ simulation replicates for the iCAR, LCAR, and L$_j$CAR priors, including scenarios and diseases not presented in the main text.

The results of the across-prior simulation study for $G=47$ are presented in \autoref{tabA6}. The results show that the LCAR and L$_j$CAR priors produce less smoothing than iCAR across all scenarios, as reflected most clearly in the SRSS values, although differences are smaller than for $G=100$ or $G=300$. Scenario~2, followed by Scenario~4, exhibits the highest smoothing proportions; these scenarios feature greater rate disparities between diseases, with disease 2 showing lower values, suggesting that larger differences in disease rates lead to stronger smoothing. Disease 1, in contrast, shows consistent smoothing across all scenarios. Therefore, conclusions are similar to those reached as the number of areas increases.

\begin{table}[b!]
	\caption{Empirical smoothing criteria values obtained by the separables iCAR and LCAR spatial priors and the non-separable L$_j$CAR spatial prior for $G=47$.\label{tabA6}}
	\centering
	\resizebox{\textwidth}{!}{
		\begin{tabular}{c|rrrrrrrrrrrrrr}
			\hline
			&\multicolumn{4}{c}{\textbf{iCAR}}&&\multicolumn{4}{c}{\textbf{LCAR}}&&\multicolumn{4}{c}{\textbf{L$_j$CAR}}\\
			\cline{2-5}\cline{7-10}\cline{12-15}
			& SRSS & maxRSS & RSP & SP && SRSS & maxRSS & RSP & SP&& SRSS & maxRSS & RSP & SP\\ 
			\hline
			\multicolumn{10}{l}{\textbf{Scenario 1}}\\
			Disease 1 ($j=1$) & 16.88 &  2.22 & 0.11 & && 16.26 &  1.93 & 0.10 &&& 15.99 & 1.82 & 0.10 &  \\ 
			Disease 2 ($j=2$)  & 14.84 &  1.99 & 0.14 &&& 14.54 &  1.79 & 0.13 &&& 14.40 &  1.74 & 0.13 &   \\ 
			Total & 31.72 &  & 0.25 &   0.12 && 30.80 &  & 0.23 & 0.11 & & 30.38 &  & 0.28 & 0.11\\ 
			\multicolumn{10}{l}{ }\\
			\multicolumn{10}{l}{\textbf{Scenario 2}}\\
			Disease 1 ($j=1$)  & 16.21 & 1.83 & 0.10 & && 15.76 &  1.65 & 0.10 && & 15.56 & 1.58 & 0.10 &   \\ 
			Disease 2 ($j=2$)& 14.90 &  2.44 & 0.23 &  && 14.51 & 2.18 & 0.22 &&& 14.53 &  2.31 & 0.22 &    \\ 
			Total  & 31.11 &  &0.33 &   0.12 & & 30.27 &   & 0.32 & 0.12 && 30.09 &   & 0.32 & 0.12\\ 
			\multicolumn{10}{l}{ }\\
			\multicolumn{10}{l}{\textbf{Scenario 3}}\\
			Disease 1 ($j=1$)   & 15.72 &  1.41 & 0.10 && & 15.13 &  1.33 & 0.09 & && 15.09 & 1.31 & 0.09 &    \\ 
			Disease 2 ($j=2$) & 15.91 &  2.00 & 0.12 &&& 15.33 &  1.89 & 0.12 &&& 15.29 & 1.89 & 0.12 &   \\ 
			Total & 31.63 &  & 0.22 &   0.10 && 30.46 &  &  0.21 & 0.10 & & 30.38 &    & 0.21 & 0.10 \\ 
			\multicolumn{10}{l}{ }\\
			\multicolumn{10}{l}{\textbf{Scenario 4}}\\
			Disease 1 ($j=1$)  & 15.52 &  1.33 & 0.10 & && 14.97 &  1.26 & 0.09 && & 15.00 & 1.27 & 0.09 &    \\ 
			Disease 2 ($j=2$)& 16.07 &  2.24 & 0.19 & && 15.39  & 2.12 & 0.18 &&& 15.36 &  2.11 & 0.18 &    \\ 
			Total  & 31.60 &  & 0.29 &   0.11 & & 30.36 &   & 0.27 & 0.11 && 30.36 &   & 0.27 & 0.11 \\ 
			
			\hline
		\end{tabular}
	}
\end{table}

\begin{sidewaystable}
		\centering
		\caption{95\% interval of the distribution of the empirical metrics across the simulation replicates for $G=47$ across all scenarios.}
		\label{tabA6.1}
		\resizebox{\textwidth}{!}{
			\begin{tabular}{llcccccccccccccc}
				\hline
				&&\multicolumn{4}{c}{iCAR vs LCAR} &&\multicolumn{4}{c}{iCAR vs L$_j$CAR} &&\multicolumn{4}{c}{LCAR vs L$_j$CAR} \\
				\cline{3-6}\cline{8-11}\cline{13-16}
				&& SRSS & maxRSS & RSP & SP&  & SRSS & maxRSS & RSP & SP & & SRSS & maxRSS & RSP & SP \\ 
				\hline
				\multicolumn{16}{l}{\bf $\bf G=47$}\\
				&\multicolumn{15}{l}{\bf Scenario 1}\\
				&Disease 1& (-1.51, 2.30) & (-0.38, 1.09) & (-0.01, 0.01) &  &  & (-0.88, 2.67) & (-0.27, 1.23) & (0, 0.02) &  &  & (-1.48, 2.52) & (-0.58, 0.86) & (-0.01, 0.02) &  \\ 
				& Disease 2 & (-1.36, 1.90) & (-0.47, 1.36) & (-0.01, 0.02) &  &  & (-1.09, 2.25) & (-0.45, 1.39) & (-0.01,  0.02) &  &  & (-1.49, 2.12) & (-0.87, 1.04) & (-0.01, 0.02) &  \\ 
				&Total& (-2.16, 3.19) &  & (-0.02, 0.03) & (-0.01, 0.01) &  & (-1.62, 3.88) &  & (-0.01, 0.03) & (-0.01, 0.02) &  & (-2.43, 3.78) &  & (-0.02, 0.03) & (-0.01, 0.01) \\ 
				&&&  &  &  &  &  &  &  &  & &&& &  \\ 
				&\multicolumn{15}{l}{\bf Scenario 2}\\
				&Disease 1 & (-0.94, 1.61) & (-0.28, 0.73) & (0,  0.01) &  &  & (-0.69, 1.94) & (-0.23, 0.87) & (0, 0.01) &  &  & (-1.13, 1.71) & (-0.44, 0.63) & (-0.01, 0.01) &  \\ 
				& Disease 2 & (-1.24, 2.35) & (-0.43, 1.50) & (-0.02, 0.03) &  &  & (-1.20, 2.33) & (-0.44, 1.43) & (-0.02, 0.03) &  &  & (-1.82, 1.90) & (-1.01, 1.04) & (-0.02, 0.03) &  \\ 
				&Total& (-2.09, 3.25) &  & (-0.02, 0.04) & (-0.01, 0.01) &  & (-1.47, 3.73) &  & (-0.01, 0.04) & (0, 0.02) &  & (-2.66, 3.48) &  & (-0.03, 0.03) & (-0.01, 0.01) \\ 
				&&&  &  &  &  &  &  &  &  & &&& &  \\ 
				&\multicolumn{15}{l}{\bf Scenario 3}\\
				&Disease 1 & (-0.27, 1.51) & (-0.31, 0.54) & (0, 0.01) &  &  & (-0.33, 1.57) & (-0.30, 0.59) & (0, 0.01) &  &  & (-1.00, 1.06) & (-0.42, 0.49) & (-0.01, 0.01) &  \\ 
				& Disease 2 & (-0.38, 1.77) & (-0.47, 0.97) & (0, 0.01) &  &  & (-0.43, 1.75) & (-0.47, 0.88) & (0, 0.01) &  &  & (-1.14, 1.26) & (-0.75, 0.80) & (-0.01, 0.01) &  \\ 
				&Total & (-0.08, 2.60) &  & (0, 0.02) & (0, 0.01) &  & (-0.52, 2.63) &  & (0, 0.02) & (0, 0.01) &  & (-1.82, 1.65) &  & (-0.01, 0.01) & (-0.01, 0.01) \\ 
				&&&  &  &  &  &  &  &  &  & &&& &  \\ 
				&\multicolumn{15}{l}{\bf Scenario 4}\\
				&Disease 1 & (-0.11, 1.28) & (-0.28, 0.46) & (0, 0.01) &  &  & (-0.23, 1.30) & (-0.32, 0.46) & (0, 0.01) &  &  & (-0.93, 0.75) & (-0.4, 0.36) & (-0.01, 0) &  \\ 
				&Disease 2 & (-0.47,  2.26) & (-0.55,  1.06) & (-0.01,  0.03) &  &  & (-0.65,  2.20) & (-0.52,  1.10) & (-0.01,  0.03) &  &  & (-1.61,  1.45) & (-0.84,  0.87) & (-0.02,  0.02) &  \\ 
				&Total & (-0.07, 2.82) &  & (0,  0.03) & (0,  0.01) &  & (-0.56,  2.85) &  & (0,  0.03) & (0,  0.01) &  & (-2.02,  1.72) &  & (-0.02,  0.02) & (-0.01,  0.01) \\ 
				\hline
			\end{tabular}
		}
\end{sidewaystable}

\begin{sidewaystable}
		\centering
		\caption{95\% interval of the distribution of the empirical metrics across the simulation replicates for $G=100$ and $G=300$ across all scenarios.}\label{tabA6.2}
		\resizebox{\textwidth}{!}{
			\begin{tabular}{llcccccccccccccc}
				\hline
				&&\multicolumn{4}{c}{iCAR vs LCAR} &&\multicolumn{4}{c}{iCAR vs L$_j$CAR} &&\multicolumn{4}{c}{LCAR vs L$_j$CAR} \\
				\cline{3-6}\cline{8-11}\cline{13-16}
				&& SRSS & maxRSS & RSP & SP&  & SRSS & maxRSS & RSP & SP & & SRSS & maxRSS & RSP & SP \\ 
				\hline
				\multicolumn{16}{l}{\bf $\bf G=100$}\\
				&\multicolumn{15}{l}{\bf Scenario 1}\\
				&Disease 1 & (-9.89,  22.16) & (-4.97, 7.52) & (-0.02, 0.04)  & &  & (-9.66,  25.11) & (-5.07, 7.24) & (-0.01, 0.04) &  &  & (-16.72, 21.84) & (-6.51, 7.40) & (-0.03, 0.04) &\\ 
				& Disease 2 & (-6.02, 31.09) & (-3.53, 10.03) & (-0.01,  0.06) & &  & (-6.77, 35.55) & (-3.52, 11.79) & (-0.01,  0.06) &  &  & (-20.22,  22.11) & (-7.39, 8.97) & (-0.04,  0.04) & \\ 
				&Total & (-8.90, 43.28) &  & (-0.02, 0.08) & (-0.01, 0.04) & & (-11.94, 50.87) &  &  (-0.02, 0.09) & (-0.01, 0.05)&  & (-29.62, 36.73) &  & (-0.06, 0.07) & (-0.03,  0.03) \\ 
				&&&  &  &  &  &  &  &  &  & &&& &  \\ 
				&\multicolumn{15}{l}{\bf Scenario 2}\\
				&Disease 1 & (-9.56, 21.98) & (-5.15, 6.66) & (-0.01, 0.04) & &  & (-7.83, 23.88) & (-4.78, 7.56) & (-0.01, 0.04) &   &  & (-18.46, 21.18) & (-5.81, 8.19) & (-0.03, 0.03) &   \\ 
				&Disease 2 & (-9.66, 26.34) & (-4.92, 12.14) & (-0.02, 0.06) & &  & (-10.19, 28.91) & (-5.06, 12.22) & (-0.02, 0.07) &   &  & (-21.07, 21.13) & (-9.73, 8.57) & (-0.05, 0.05) & \\ 
				&Total & (-15.46, 39.93) & & (-0.03, 0.09) & (-0.01, 0.04) &  & (-13.73, 44.43) &&  (-0.03, 0.1) & (-0.01, 0.04)   &  & (-30.70, 37.20) &&  (-0.07, 0.07) & (-0.03, 0.03)   \\ 
				&&  &  &  &  &  &  &  &  &  & &&& &  \\ 
				
				&\multicolumn{15}{l}{\bf Scenario 3}\\
				&Disease 1 & (-5.98, 24.36) & (-4.36, 7.63) & (-0.01,  0.04) &&  & (-3.22, 24.28) & (-3.12, 8.11) & (-0.01,  0.04) &  &  & (-16.68, 21.65) & (-6.90, 7.76) & (-0.03,  0.03) &  \\ 
				&Disease 2 & (-1.97, 33.62) & (-4.55, 13.66) & (0,  0.05) & &  & (-1.32, 33.50) & (-5.66, 14.05) & (0,  0.06) & &  & (-21.68, 22.75) & (-11.00, 10.79) & (-0.03, 0.04) \\ 
				&Total & (-7.05, 45.92) &  & (0, 0.08) & (0, 0.04) &  & (0.99, 46.28) &  & (0.01, 0.08) & (0, 0.04)  &  & (-27.02, 37.62) &  &  (-0.04, 0.06) & (-0.02, 0.03)\\ 
				&&  &  &  &  &  &  &  &  &  &  & &&& \\ 
				
				&\multicolumn{15}{l}{\bf Scenario 4}\\
				&Disease 1 & (-0.7, 23.12) & (-3.54, 7.36) & (0, 0.04) &  &  & (-0.74, 24.15) & (-3.66, 8.1) & (0, 0.04) &  &  & (-14.36, 15.99) & (-6.78, 6.51) & (-0.02, 0.03) &  \\ 
				&Disease 2 & (-4.72, 30.84) & (-5.92, 13.38) & (-0.01, 0.07) &  &  & (-9.48, 33.15) & (-6.30, 14.66) & (-0.01, 0.07) &  &  & (-25.33, 21.84) & (-10.96, 10.77) & (-0.05, 0.05) &  \\ 
				&Total & (-3.55, 43.31) &  & (0.01, 0.09) & (0, 0.04) &  & (-7.43, 43.32) &  & (0, 0.09) & (0, 0.04) &  & (-31.69, 26.89) &  & (-0.06, 0.05) & (-0.02, 0.02) \\
				&&  &  &  &  &  &  &  &  &  &  & &&& \\ 
				&&  &  &  &  &  &  &  &  &  &  & &&& \\ 
				\multicolumn{16}{l}{\bf $\bf G=300$}\\
				&\multicolumn{15}{l}{\bf Scenario 1}\\
				&Disease 1& (-428.38, 350.83) & (-165.97, 100.12) & (-0.05, 0.07) &  &  & (-483.77, 376.92) & (-194.56, 112.65) & (-0.05, 0.09) &  &  & (-601.69, 576.07) & (-203.71, 177.8) & (-0.08, 0.09) &  \\ 
				&Disease 2& (-303.19, 334.27) & (-108.27, 90.91) & (-0.03, 0.09) &  &  & (-321.71, 356.31) & (-128.16, 87.84) & (-0.03, 0.09) &  &  & (-492.54, 460.87) & (-154.01, 125.2) & (-0.08, 0.08) &  \\ 
				&Total & (-655.88, 541.69) &  & (-0.08, 0.13) & (-0.04, 0.07) &  & (-796.15, 603.00) &  & (-0.08, 0.15) & (-0.04, 0.08) &  & (-1037.29, 993.97) &  & (-0.15, 0.16) & (-0.07, 0.08) \\ 
				&  &  &  &  &  &  &  &  &  &  &  &  &  &  \\ 
				&\multicolumn{15}{l}{\bf Scenario 2}\\
				&Disease 1 & (-393.34, 345.85) & (-151.02, 91.72) & (-0.05, 0.08) &  &  & (-248.41, 387.44) & (-135.48, 88.35) & (-0.02, 0.09) &  &  & (-403.04, 646.68) & (-139.47, 151.61) & (-0.06, 0.11) &  \\ 
				&Disease 2 & (-392.16, 315.20) & (-151.72, 96.59) & (-0.04, 0.07) &  &  & (-298.28, 312.61) & (-146.62, 109.20) & (-0.02, 0.08) &  &  & (-414.54, 516.17) & (-165.40, 180.06) & (-0.06, 0.08) &  \\ 
				&Total & (-691.93, 536.8) &  & (-0.08, 0.14) & (-0.04, 0.07) &  & (-485.66, 603.45) &  & (-0.03, 0.15) & (-0.02, 0.08) &  & (-746.07, 1017.79) &  & (-0.11, 0.18) & (-0.06, 0.1) \\ 
				&  &  &  &  &  &  &  &  &  &  &  &  &  &  \\ 
				&\multicolumn{15}{l}{\bf Scenario 3}\\
				&Disease 1& (-11.01, 394.54) & (-94.75, 106.82) & (0.01, 0.08) &  &  & (-147.97, 449.20) & (-97.92, 126.26) & (-0.01, 0.09) &  &  & (-324.32, 330.2) & (-124.92, 121.33) & (-0.06, 0.06) &  \\ 
				&Disease 2 & (-44.05, 423.43) & (-108.89, 137.97) & (0.01, 0.09) &  &  & (-114.40, 445.71) & (-79.59, 140.48) & (-0.01, 0.09) &  &  & (-371.48, 334.03) & (-137.58, 163.17) & (-0.07, 0.06) &  \\ 
				&Total & (31.35, 690.32) &  & (0.03, 0.15) & (0.01, 0.08) &  & (-230.27, 800.51) &  & (-0.02, 0.17) & (-0.01, 0.09) &  & (-678.59, 547.38) &  & (-0.13, 0.10) & (-0.06, 0.05) \\ 
				&  &  &  &  &  &  &  &  &  &  &  &  &  &  \\ 
				&\multicolumn{15}{l}{\bf Scenario 14}\\
				&Disease 1 & (40.71, 428.66) & (-74.69, 126.84) & (0.02, 0.10) &  &  & (-87.14, 469.11) & (-98.25, 132.58) & (-0.01, 0.1) &  &  & (-382.39, 291.24) & (-162.18, 133.62) & (-0.07, 0.06) &  \\ 
				&Disease 2 & (-143.65, 381.24) & (-128.22, 134.88) & (0.01, 0.09) &  &  & (-263.59, 403.99) & (-146.02, 142.22) & (-0.02, 0.09) &  &  & (-458.50, 340.89) & (-197.50, 180.94) & (-0.08, 0.05) &  \\ 
				&Total & (-13.77, 677.61) &  & (0.05, 0.16) & (0.02, 0.08) &  & (-330.33, 726.03) &  & (-0.03, 0.17) & (-0.01, 0.09) &  & (-744.99, 530.74) &  & (-0.15, 0.10) & (-0.07, 0.05) \\
				\hline
			\end{tabular}
		}
\end{sidewaystable}

We also provide the empirical distribution of the differences in the metric values across replicates, which provides additional information on the variability of the reported results. In particular, we report the 95\% intervals for the differences between priors for the criteria presented in the main text. \autoref{tabA6.1} reports the corresponding intervals based on the results in \autoref{tabA6}, while \autoref{tabA6.2} presents the intervals for the results reported in the main text for $G=100$ and $G=300$.
These results suggest no significant differences between the results obtained under the LCAR and L$_j$CAR priors across all scenarios and numbers of areas considered. However, clear disparities are observed between the iCAR prior and the LCAR and L$_j$CAR priors in Scenarios 3 and 4 across different numbers of areas, as zero falls outside the corresponding intervals. This indicates that the LCAR and L$_j$CAR priors provide less smoothing in scenarios where the diseases exhibit low spatial correlation.

\begin{figure}[b!]
	\centering
	\includegraphics[width=0.9\linewidth]{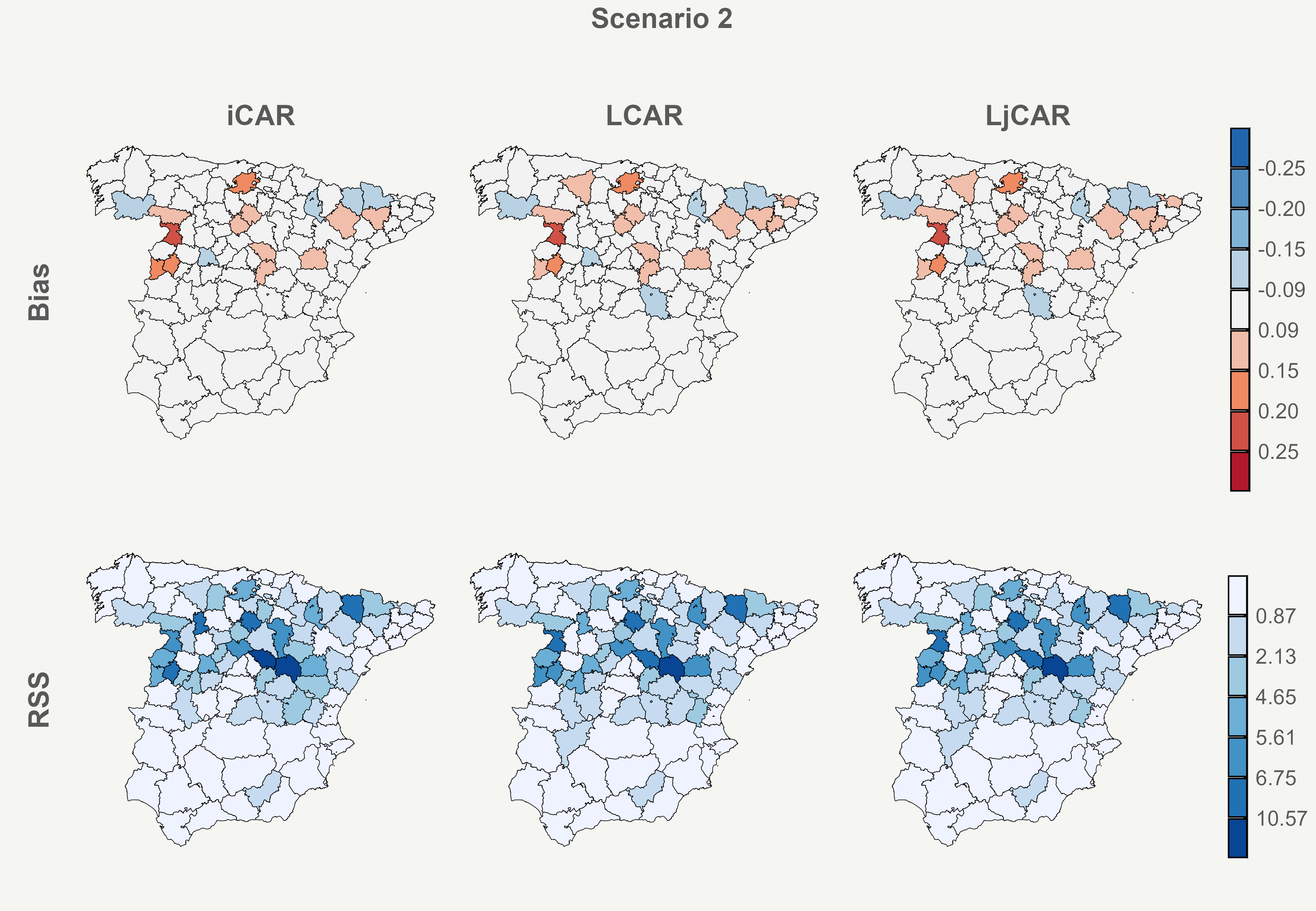}\\
	\vspace{0.4cm}
	\includegraphics[width=0.9\linewidth]{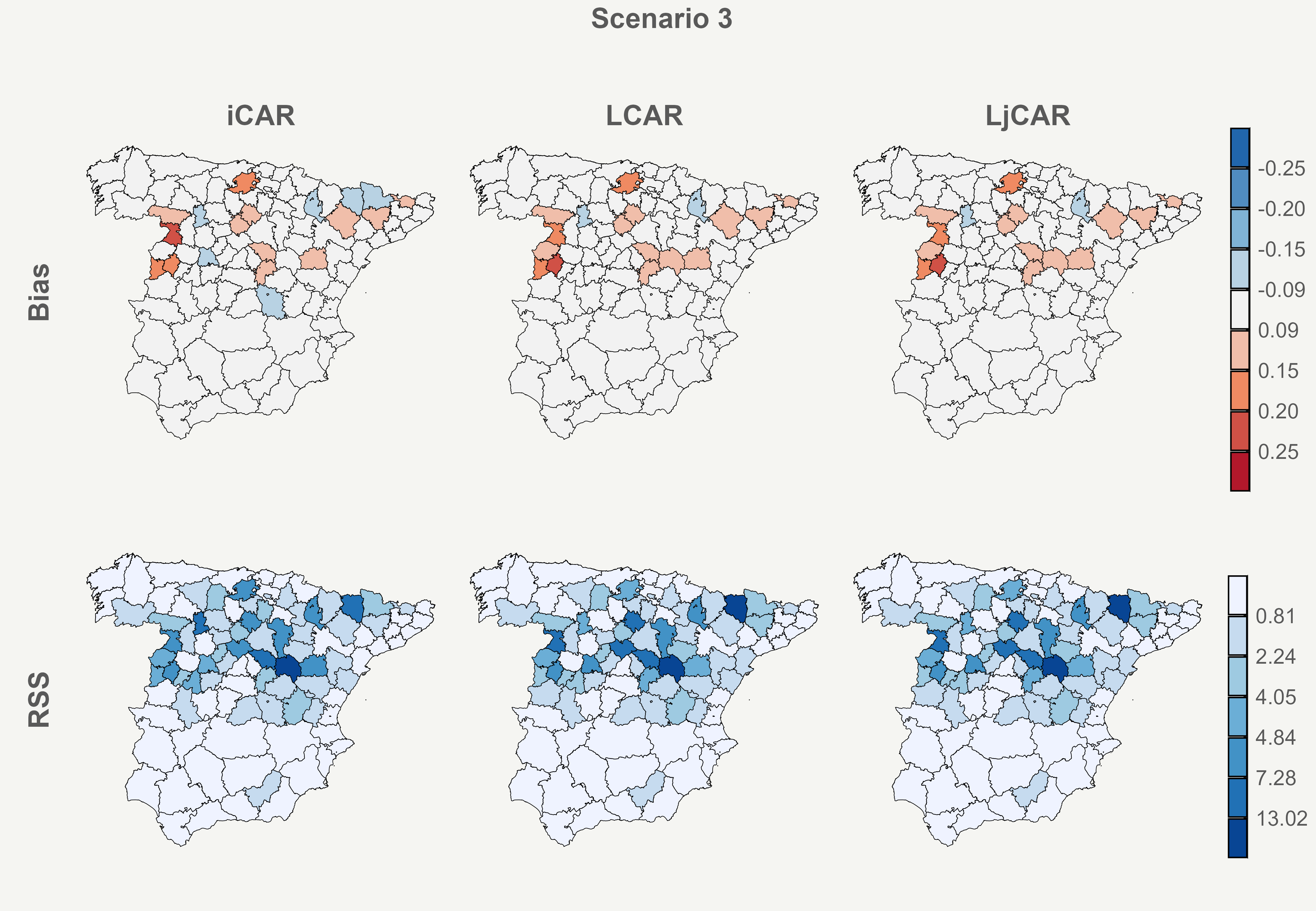}
	\caption{Mean bias across the $B$ replicates of the simulation study and the average across the $B$ replicates of the $RSS_{1i}$  values for the iCAR, LCAR and L$_j$CAR priors in Scenario 2 and 3 for disease 1, when analyzing 100 areas.}\label{figA0.1}
\end{figure}

\begin{figure}[t!]
	\centering
	\includegraphics[width=0.9\linewidth]{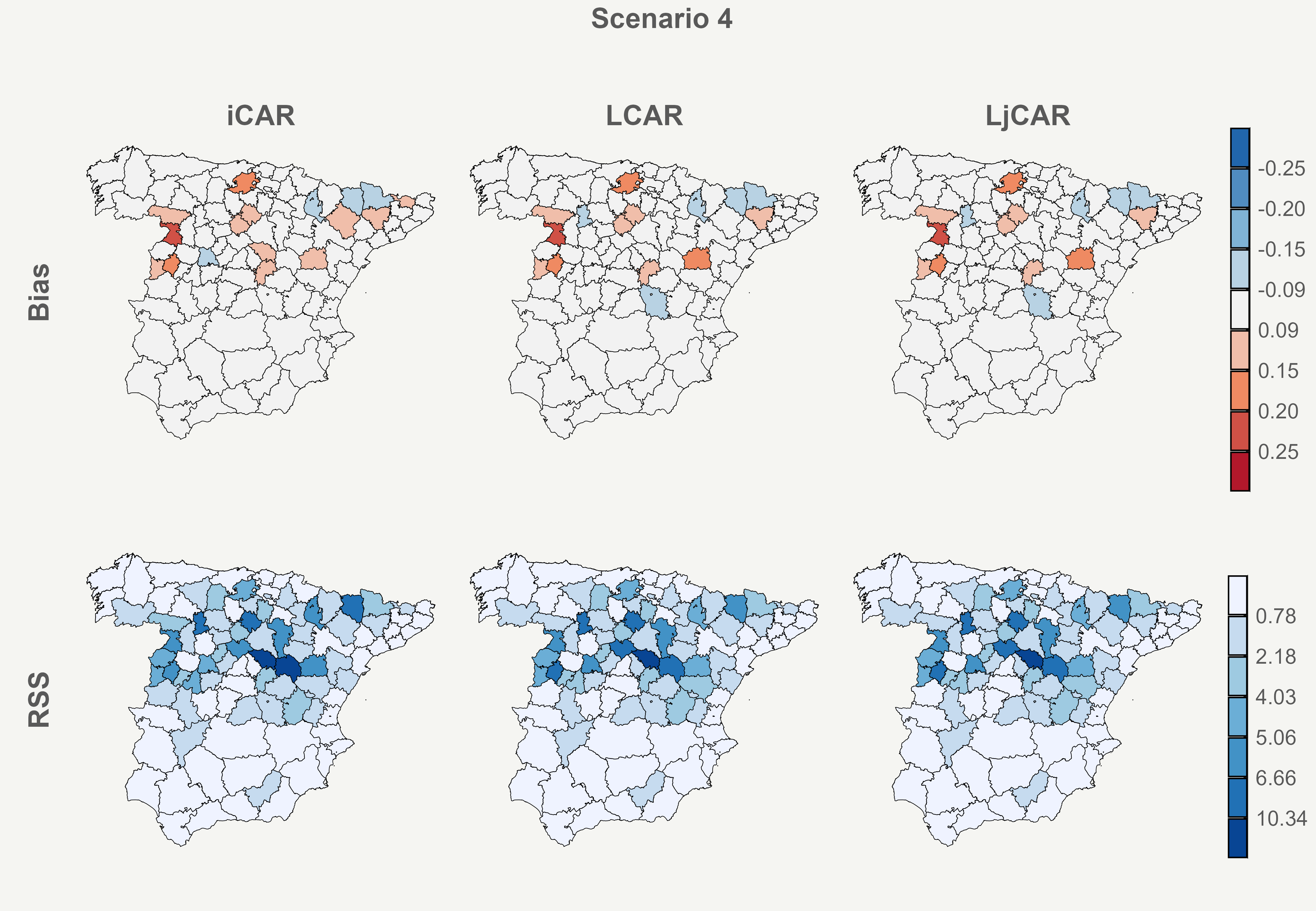}
	\caption{Mean bias across the $B$ replicates of the simulation study and the average across the $B$ replicates of the $RSS_{1i}$  values for the iCAR, LCAR and L$_j$CAR priors in Scenario 4 for disease 1, when analyzing 100 areas.}
	\label{figA0.2}
\end{figure}

As mentioned in the main paper, although the primary goal of this paper is to evaluate, both theoretically and empirically, the degree of smoothing induced by different spatial priors, the simulation study also allows us to assess whether the shift of the posterior estimates away from the crude rates is appropriate relative to the true underlying risk surface. To this end, \autoref{figA0.1} and \autoref{figA0.2} present the mean bias across the $B$ simulation replicates for the iCAR, LCAR, and L$_j$CAR priors in Scenarios 2, 3, and 4 for Disease 1, when analyzing 100 areas, together with the corresponding average across the $B$ replicates of the $RSS_{1i}$  values. As in the corresponding figure for Scenario 1 presented in the main text, the $RSS_{1i}$ maps report the 50th, 75th, 85th, 90th, 95th, and 97th percentiles in the legend, while the bias maps highlight areas with an absolute bias of at least 9\%.
Overall, the RSS and bias maps show similar patterns across the four scenarios. The areas with the highest RSS values and those with the largest absolute bias are largely the same across scenarios. Therefore, the conclusions are consistent with those reported in the main paper.

\begin{figure}[b!]
	\centering
	\includegraphics[width=0.9\linewidth]{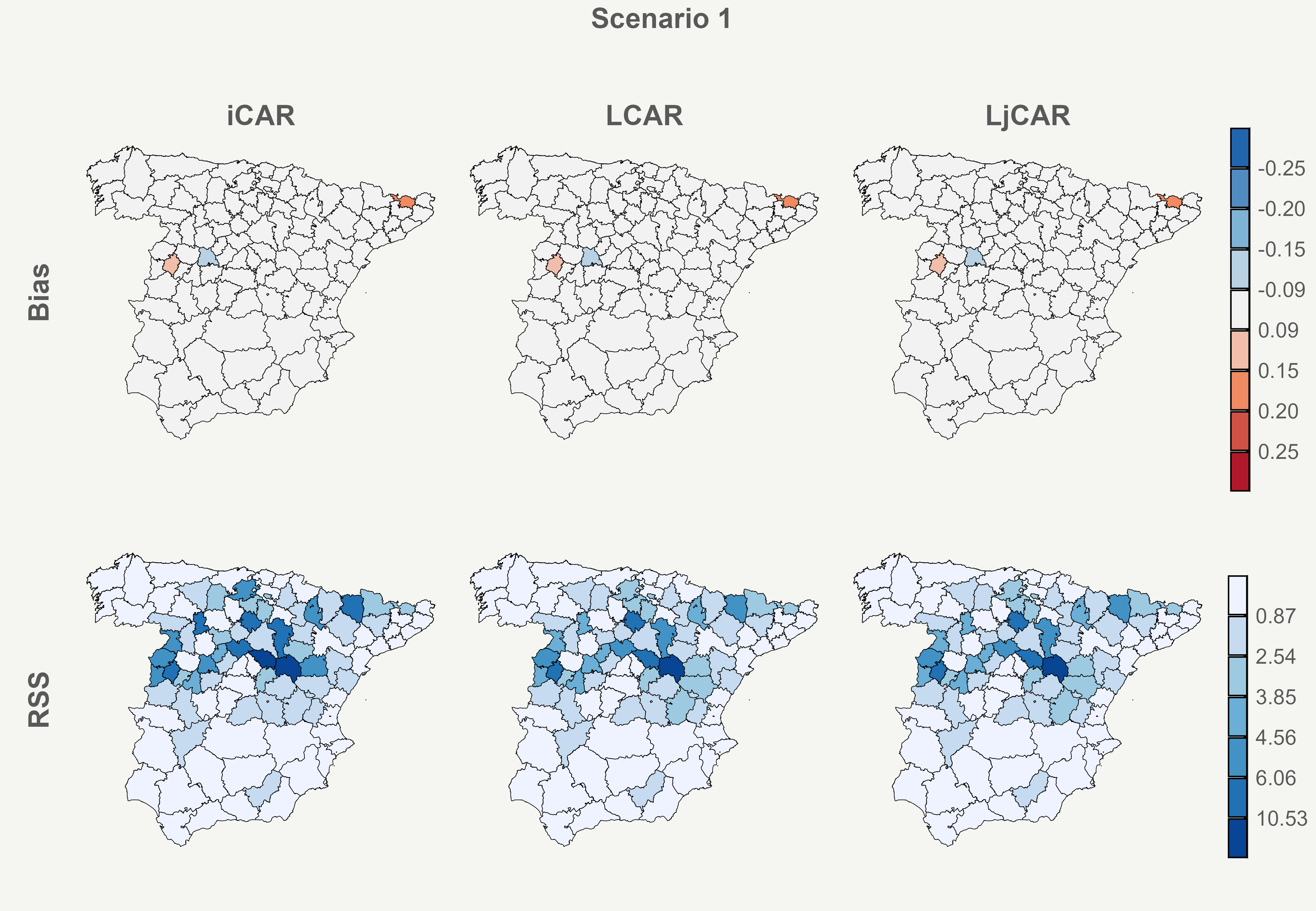}\\
	\vspace{0.4cm}
	\includegraphics[width=0.9\linewidth]{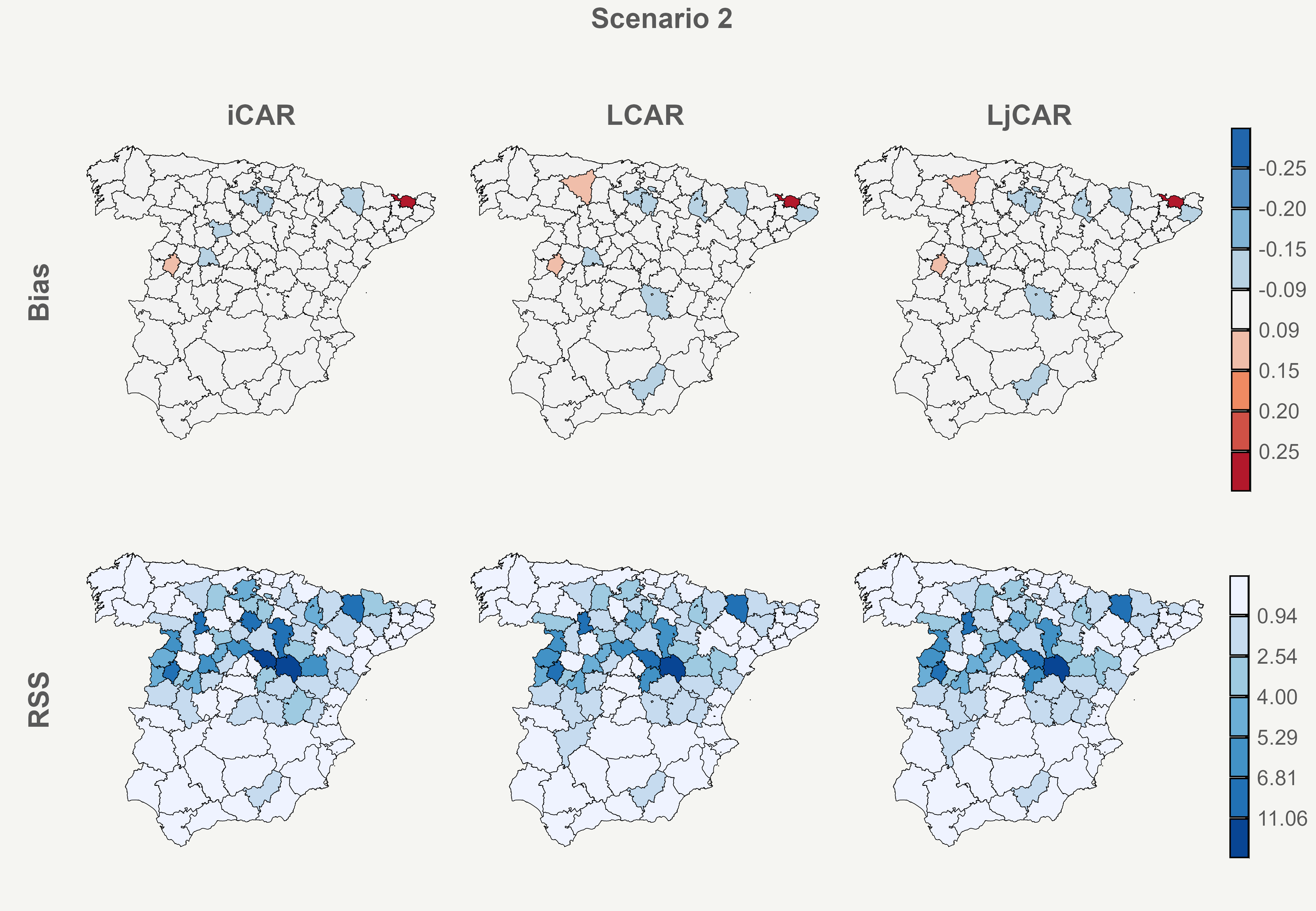}
	\caption{Mean bias across the $B$ replicates of the simulation study and the average across the $B$ replicates of the $RSS_{2i}$  values for the iCAR, LCAR and L$_j$CAR priors in Scenario 1 and 2 for disease 2, when analyzing 100 areas.}
	\label{figA0.3}
\end{figure}

\begin{figure}[b!]
	\centering
	\includegraphics[width=0.9\linewidth]{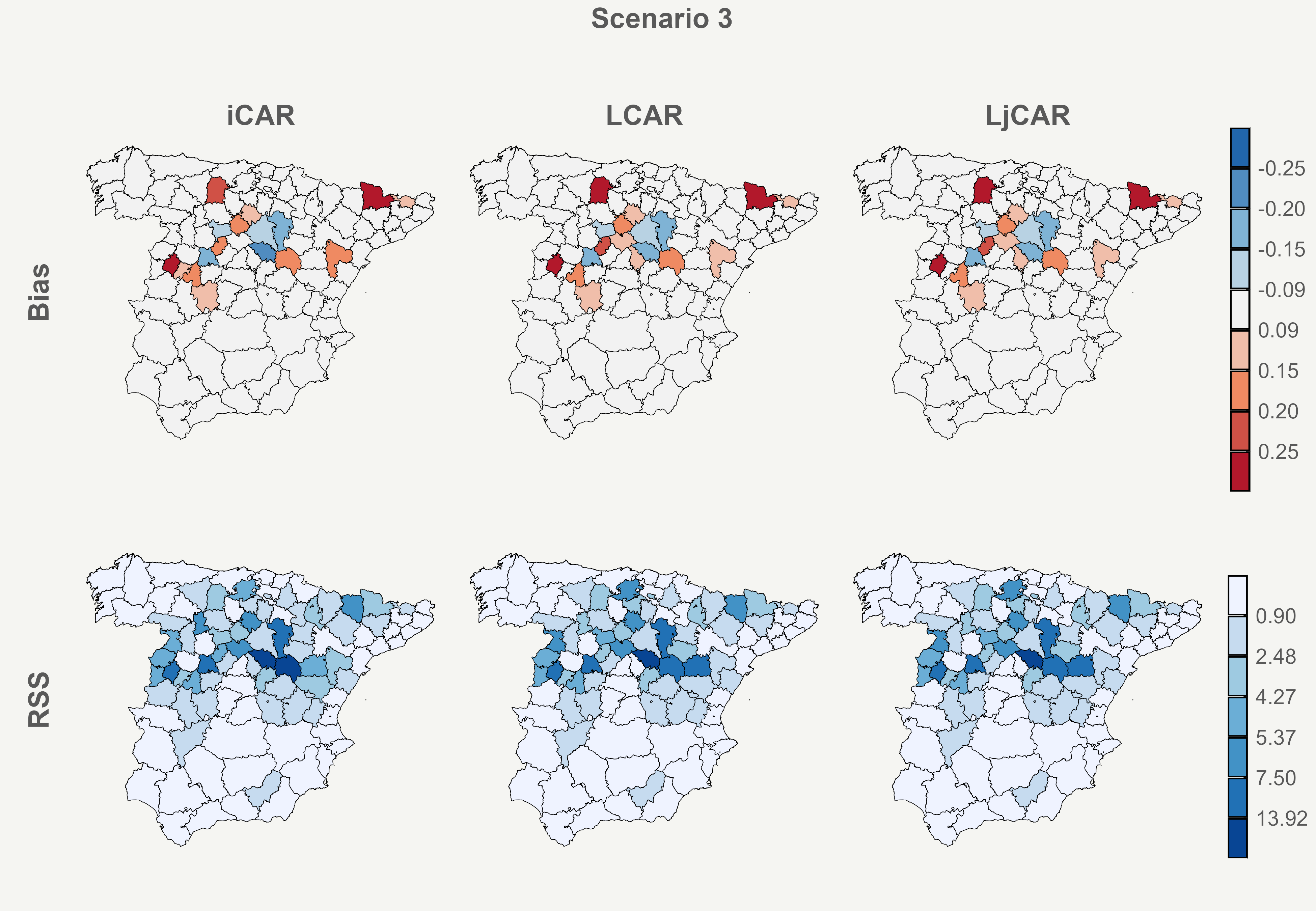}\\
	\vspace{0.4cm}
	\includegraphics[width=0.9\linewidth]{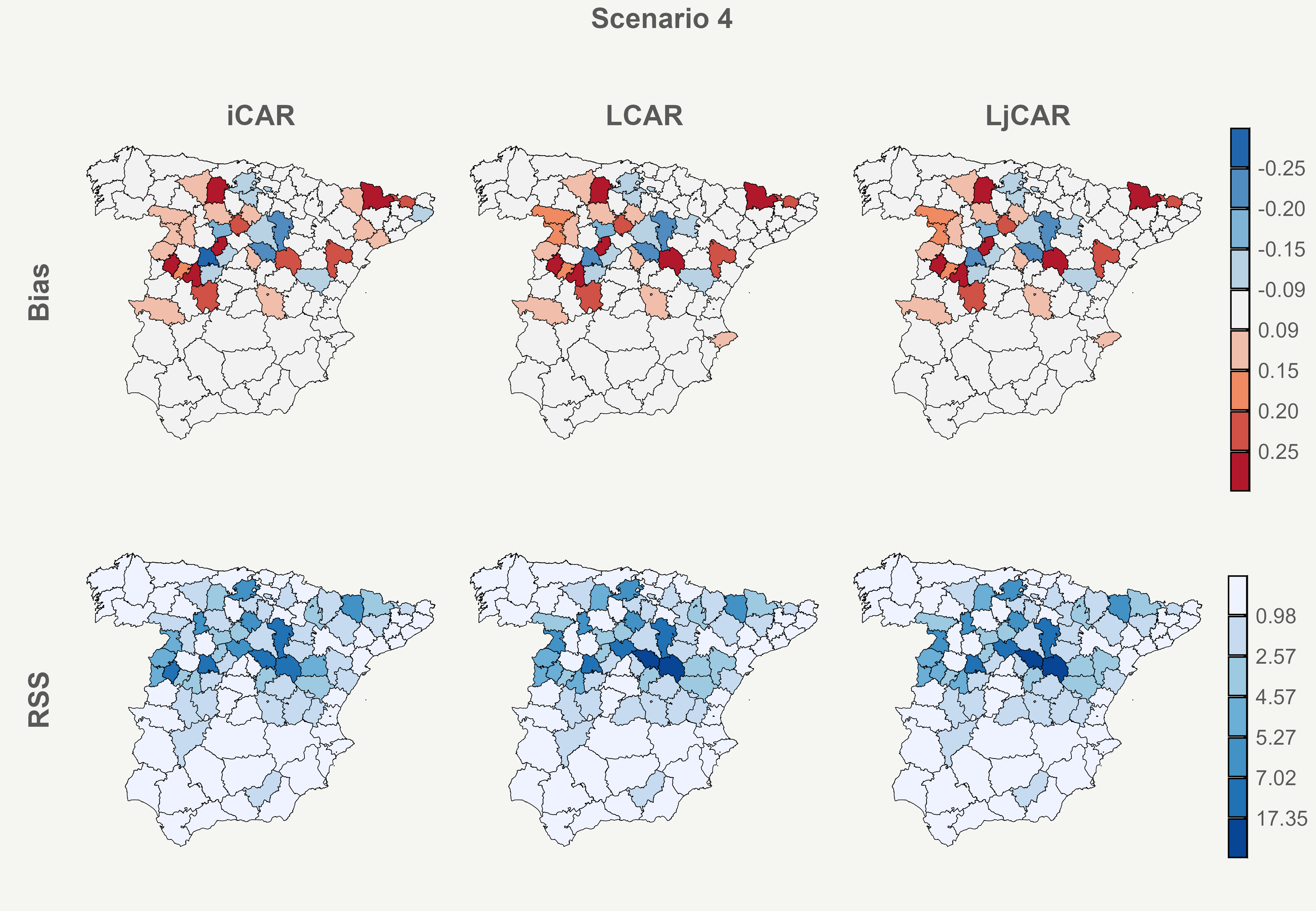}
	\caption{Mean bias across the $B$ replicates of the simulation study and the average across the $B$ replicates of the $RSS_{2i}$  values for the iCAR, LCAR and L$_j$CAR priors in Scenario 3 and 4 for disease 2, when analyzing 100 areas.}
	\label{figA0.4}
\end{figure}

For Disease 2, whose spatial pattern varies across scenarios, results are presented in \autoref{figA0.3} for Scenarios 1 and 2 and in \autoref{figA0.4} for Scenarios 3 and 4. Scenarios 1 and 2 have the same spatial pattern of crude rates, but the variability in the rates is higher in Scenario 2. The RSS maps are similar between the two scenarios, although Scenario 2 exhibits higher RSS values. However, differences are observed in the bias maps, with more areas exhibiting an absolute bias greater than 9\% in Scenario 2. The areas highlighted in Scenario 1 are also highlighted in Scenario 2. Thus, higher variability in the rates tends to result in larger bias induced by smoothing. More areas exhibit high bias in Scenarios 3 and 4 than in Scenarios 1 and 2. This difference is consistent with the correlation between diseases: Scenarios 1 and 2 have a high correlation between diseases, whereas Scenarios 3 and 4 have a low correlation. Thus, lower correlation between diseases tends to be associated with higher bias, as a result of the greater degree of smoothing induced, as observed in the main paper. Moreover, Scenario 4 has higher variability in the rates than Scenario 3, and, again, more areas exhibit high bias compared with Scenario 3.
Across all scenarios and for both diseases, areas with high positive bias, where the estimated rate exceeds the true rate, tend to be surrounded by areas with higher rates than the area itself. Conversely, areas with high negative bias tend to be surrounded by areas with lower rates than the area itself.

\section{Real Data}\label{SectionSM:RealData}

\begin{figure}[h!]
	\centering
	\includegraphics[width=0.7\linewidth]{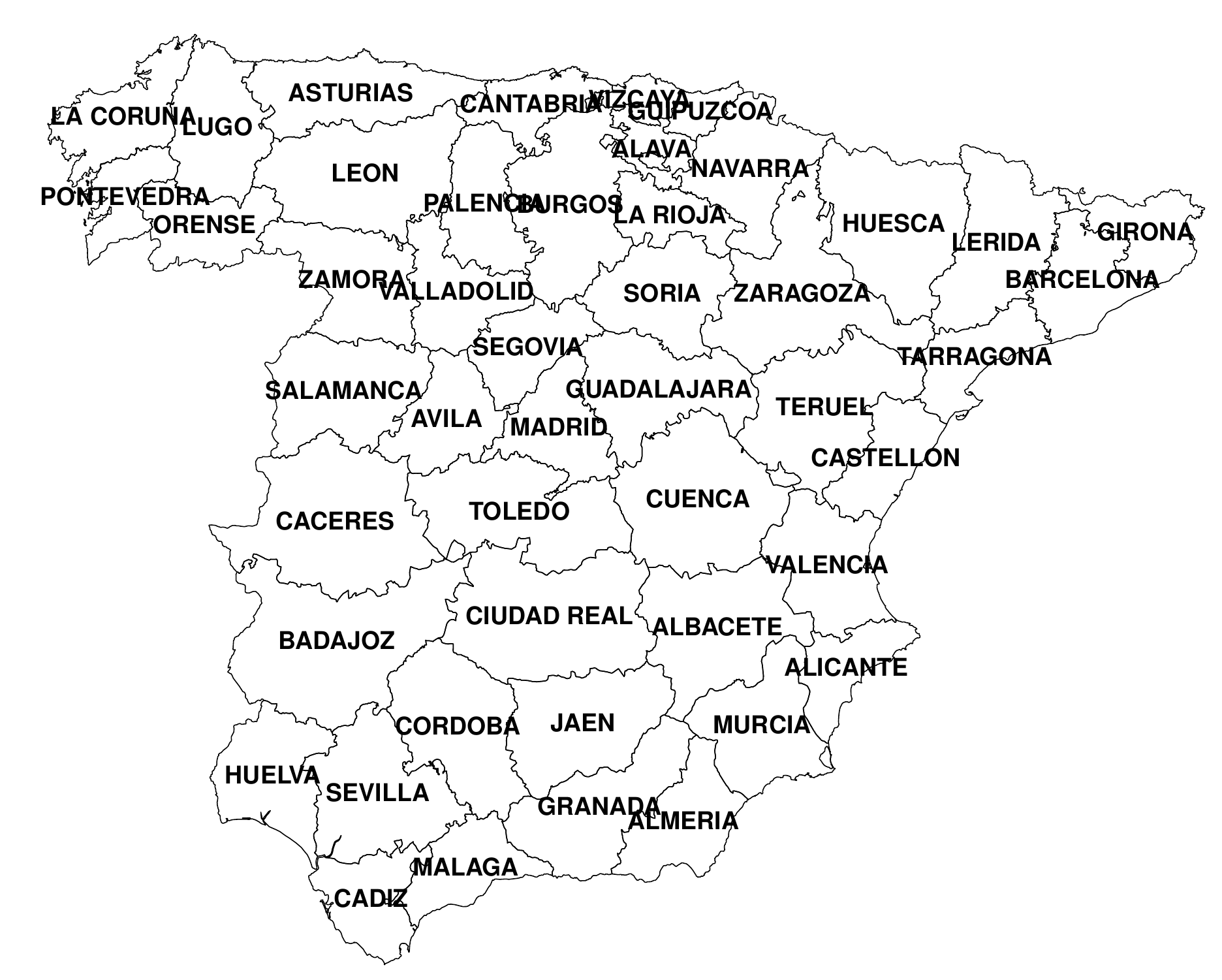}
	\caption{Names of the 47 provinces of continental Spain included in the study.}
	\label{figS6}
\end{figure}

This section presents tables and figures discussed in the main text but omitted due to space constraints. The datasets include mortality rates for three cancers: colon, pancreatic, and stomach, and population-at-risk figures for Spanish females aged 50+, covering 2020–2022.  \autoref{figS6} shows the provincial names and their geographic locations, which are referenced in the results.

To assess whether the results from real data are consistent with the simulation findings, we first fitted the multivariate models pairwise, analyzing two diseases at a time. The results are summarized in \autoref{tabA7}, \autoref{tabA7.2} and \autoref{tabA7.3} for $G=47$, $G=100$ and $G=300$, respectively. \autoref{tabA7}, \autoref{tabA7.2} and \autoref{tabA7.3} show both the empirical metrics and the theoretical multivariate TCV, along with the posterior means of the hyperparameters.
The LCAR and L$_j$CAR yield a lower amount of smoothing than the iCAR, with generally similar smoothing between them. Note that the theoretical metric is not comparable among different data sets or spatial priors.  Regarding the number of areas, smoothing increases as $G$ increases. However, the smoothing does not increase equally for both diseases, since the proportion between them is not preserved as the number of areas increases; instead, the smoothing amount becomes more similar for the two diseases. 
Examining the RSP$_j$ values further shows that each disease exhibits similar smoothing proportions across different pairwise modelings, a pattern consistent with the findings from the across simulation study.

\begin{table}[h!]
	\centering
	\caption{\label{tabA7} Empirical smoothing metrics for the pairwise analysis of real data ($G=47$) using the separable iCAR and LCAR spatial priors, and the non-separable L$_j$CAR prior.}
	\resizebox{\textwidth}{!}{
		\begin{tabular}{cc|rr|r|crr|r|crr|r|}
			\hline
			&&\multicolumn{3}{c}{iCAR} &&\multicolumn{3}{c}{LCAR} &&\multicolumn{3}{c}{L$_j$CAR}\\
			\cline{2-4}\cline{6-8}\cline{10-12}
			&&\multirow{2}{1.6cm}{\centering Disease 1 ($j=1$)} & \multirow{2}{1.6cm}{\centering Disease 2 ($j=2$)} &\multirow{2}{1cm}{\centering Total} &  & \multirow{2}{1.6cm}{\centering Disease 1 ($j=1$)} & \multirow{2}{1.6cm}{\centering Disease 2 ($j=2$)} &\multirow{2}{1cm}{\centering Total} & & \multirow{2}{1.6cm}{\centering Disease 1 ($j=1$)}& \multirow{2}{1.6cm}{\centering Disease 2 ($j=2$)} & \multirow{2}{1cm}{\centering Total} \\ 
			&&&&&&&&&&&&\\
			\hline
			\multicolumn{13}{l}{$\mathbf{G=47}$} \\
			&\multicolumn{12}{l}{\bf Colon-Pancreas}\\
			&SRSS  & 17.88 & 17.67 & 35.55 &  & 16.88 & 16.89 & 33.77 &  & 17.43 & 17.33 & 34.76   \\ 
			&maxRSS & 2.83 & 4.03 &  &  & 2.51 & 3.63 &  &  & 2.75 & 4.07 &    \\ 
			&RSP & 0.31 & 0.35 &  &  & 0.29 & 0.34 &  &  & 0.30 & 0.34 &    \\ 
			&SP &  &  & 0.33 &  &  &  & 0.31 &  &  &  & 0.32   \\ 
			&MultiTCV &&&0.006&&&&0.009&&&&0.008\\
			&$\Sigma_{jj}$& 0.041 & 0.052 &  &  & 0.040 & 0.052 &  &  & 0.033 & 0.049 &\\
			&$\rho$&&&0.240&&&&0.278&&&&0.235\\
			&\multicolumn{12}{l}{ }\\
			&\multicolumn{12}{l}{\bf Colon-Stomach}\\
			&SRSS & 15.46 & 16.52 & 31.97 &  & 15.67 & 15.05 & 30.72 &  & 15.54 & 14.91 & 30.45   \\ 
			&maxRSS & 1.92 & 2.60 &  &  & 1.94 & 2.14 &  &  & 1.95 & 2.01 &    \\ 
			&RSP & 0.27 & 0.19 &  &  & 0.27 & 0.18 &  &  & 0.27 & 0.17 &    \\ 
			&SP &  &  & 0.24 &  &  &  & 0.24 &  &  &  & 0.23   \\ 
			&MultiTCV &  &  & 0.016 &  &  &  & 0.020 &  &  &  & 0.022   \\ 
			&$\Sigma_{jj}$ & 0.068 & 0.121 &  &  & 0.047 & 0.135 &  &  & 0.044 & 0.140 &    \\ 
			&$\rho$ &  &  & 0.531 &  &  &  & 0.529 &  &  &  & 0.552   \\ 
			&\multicolumn{12}{l}{ }\\
			&\multicolumn{12}{l}{\bf Pancreas-Stomach}\\
			&SRSS & 18.55 & 16.86 & 35.41 &  & 17.89 & 15.58 & 33.46 &  & 18.69 & 15.16 & 33.85   \\ 
			&maxRSS  & 4.15 & 2.88 &  &  & 3.56 & 2.16 &  &  & 4.19 & 2.08 &    \\ 
			&RSP & 0.37 & 0.20 &  &  & 0.36 & 0.18 &  &  & 0.37 & 0.18 &    \\ 
			&SP &  &  & 0.29 &  &  &  & 0.27 &  &  &  & 0.28   \\ 
			&MultiTCV  &  &  & 0.014 &  &  &  & 0.026 &  &  &  & 0.018   \\ 
			&$\Sigma_{jj}$ & 0.049 & 0.119 &  &  & 0.074 & 0.163 &  &  & 0.034 & 0.137   &  \\ 
			&$\rho$  &  &  & 0.228 &  &  &  & 0.441 &  &  &  & 0.365   \\ 
			\hline
		\end{tabular}
	}
\end{table}

\begin{table}[h!]
	\centering
	\caption{\label{tabA7.2} Empirical smoothing metrics for the pairwise analysis of real data ($G=100$) using the separable iCAR and LCAR spatial priors, and the non-separable L$_j$CAR prior.}
	\resizebox{\textwidth}{!}{
		\begin{tabular}{cc|rr|r|crr|r|crr|r|}
			\hline
			&&\multicolumn{3}{c}{iCAR} &&\multicolumn{3}{c}{LCAR} &&\multicolumn{3}{c}{L$_j$CAR}\\
			\cline{2-4}\cline{6-8}\cline{10-12}
			&&\multirow{2}{1.6cm}{\centering Disease 1 ($j=1$)} & \multirow{2}{1.6cm}{\centering Disease 2 ($j=2$)} &\multirow{2}{1cm}{\centering Total} &  & \multirow{2}{1.6cm}{\centering Disease 1 ($j=1$)} & \multirow{2}{1.6cm}{\centering Disease 2 ($j=2$)} &\multirow{2}{1cm}{\centering Total} & & \multirow{2}{1.6cm}{\centering Disease 1 ($j=1$)}& \multirow{2}{1.6cm}{\centering Disease 2 ($j=2$)} & \multirow{2}{1cm}{\centering Total} \\ 
			&&&&&&&&&&&&\\
			\hline
			\multicolumn{13}{l}{$\mathbf{G=100}$} \\
			&\multicolumn{12}{l}{\bf Colon-Pancreas}\\
			&SRSS & 289.28 & 219.27 & 508.55 &  & 284.23 & 215.05 & 499.27 &  & 264.09 & 216.33 & 480.42   \\ 
			&maxRSS  & 74.99 & 39.37 &  &  & 72.70 & 40.23 &  &  & 69.24 & 35.95 &    \\ 
			&RSP & 0.68 & 0.76 &  &  & 0.67 & 0.74 &  &  & 0.63 & 0.76 &    \\ 
			&SP &  &  & 0.71 &  &  &  & 0.69 &  &  &  & 0.67   \\ 
			&MultiTCV &  &  & 0.008 &  &  &  & 0.014 &  &  &  & 0.012\\
			&$\Sigma_{jj}$ & 0.046 & 0.041 &  &  & 0.043 & 0.055 &  &  & 0.053 & 0.042 & \\
			&$\rho$  &  &  & 0.343 &  &  &  & 0.228 &  &  &  & 0.375\\
			&\multicolumn{12}{l}{ }\\
			&\multicolumn{12}{l}{\bf Colon-Stomach}\\
			&SRSS & 259.51 & 205.02 & 464.53 &  & 256.24 & 179.66 & 435.90 &  & 248.56 & 196.31 & 444.86  \\ 
			&maxRSS  & 65.89 & 24.11 &  &  & 65.20 & 23.31 &  &  & 63.29 & 24.08 &    \\ 
			&RSP  & 0.63 & 0.46 &  &  & 0.62 & 0.41 &  &  & 0.60 & 0.44 &    \\ 
			&SP  &  &  & 0.57 &  &  &  & 0.55 &  &  &  & 0.55   \\ 
			&MultiTCV&  &  & 0.027 &  &  &  & 0.045 &  &  &  & 0.050 \\
			&$\Sigma_{jj}$ & 0.068 & 0.115 &  &  & 0.060 & 0.157 &  &  & 0.072 & 0.144 & \\
			&$\rho$&  &  & 0.511 &  &  &  & 0.456 &  &  &  & 0.364 \\
			&\multicolumn{12}{l}{ }\\
			&\multicolumn{12}{l}{\bf Pancreas-Stomach}\\
			&SRSS & 222.45 & 210.69 & 433.14 &  & 202.74 & 212.90 & 415.64 &  & 205.79 & 201.85 & 407.65   \\ 
			&maxRSS & 39.91 & 24.90 &  &  & 40.04 & 25.94 &  &  & 34.48 & 22.70 &    \\ 
			&RSP & 0.77 & 0.47 &  &  & 0.71 & 0.48 &  &  & 0.72 & 0.46 &    \\ 
			&SP &  &  & 0.63 &  &  &  & 0.60 &  &  &  & 0.60   \\ 
			&MultiTCV&  &  & 0.021 &  &  &  & 0.036 &  &  &  & 0.038\\
			&$\Sigma_{jj}$& 0.045 & 0.113 &  &  & 0.072 & 0.099 &  &  & 0.069 & 0.104 &  \\
			&$\rho$&  &  & 0.187 &  &  &  & 0.102 &  &  &  & 0.230\\
			\hline
		\end{tabular}
	}
\end{table}

\begin{table}[h!]
	\centering
	\caption{\label{tabA7.3} Empirical smoothing metrics for the pairwise analysis of real data ($G=300$) using the separable iCAR and LCAR spatial priors, and the non-separable L$_j$CAR prior.}
	\resizebox{\textwidth}{!}{
		\begin{tabular}{cc|rr|r|crr|r|crr|r|}
			\hline
			&&\multicolumn{3}{c}{iCAR} &&\multicolumn{3}{c}{LCAR} &&\multicolumn{3}{c}{L$_j$CAR}\\
			\cline{2-4}\cline{6-8}\cline{10-12}
			&&\multirow{2}{1.6cm}{\centering Disease 1 ($j=1$)} & \multirow{2}{1.6cm}{\centering Disease 2 ($j=2$)} &\multirow{2}{1cm}{\centering Total} &  & \multirow{2}{1.6cm}{\centering Disease 1 ($j=1$)} & \multirow{2}{1.6cm}{\centering Disease 2 ($j=2$)} &\multirow{2}{1cm}{\centering Total} & & \multirow{2}{1.6cm}{\centering Disease 1 ($j=1$)}& \multirow{2}{1.6cm}{\centering Disease 2 ($j=2$)} & \multirow{2}{1cm}{\centering Total} \\ 
			&&&&&&&&&&&&\\
			\hline
			\multicolumn{13}{l}{$\mathbf{G=300}$} \\
			&\multicolumn{12}{l}{\bf Colon-Pancreas}\\
			&SRSS & 3876.86 & 2557.47 & 6434.33 &  & 3545.68 & 2539.82 & 6085.50 &  & 3653.38 & 2456.56 & 6109.93  \\ 
			&maxRSS & 606.95 & 305.47 &  &  & 547.51 & 309.41 &  &  & 518.89 & 262.82 &   \\ 
			&RSP  & 0.89 & 0.93 &  &  & 0.84 & 0.91 &  &  & 0.86 & 0.90 &   \\ 
			&SP  &  &  & 0.90 &  &  &  & 0.86 &  &  &  & 0.87   \\ 
			&MultiTCV &  &  & 0.016 &  &  &  & 0.068 &  &  &  & 0.075   \\ 
			&$\Sigma_{jj}$ & 0.037 & 0.044 &  &  & 0.087 & 0.060 &  &  & 0.074 & 0.085 &    \\ 
			&$\rho$ &  &  & 0.494 &  &  &  & 0.205 &  &  &  & 0.352   \\ 
			&\multicolumn{12}{l}{ }\\
			&\multicolumn{12}{l}{\bf Colon-Stomach}\\
			&SRSS & 3584.42 & 3371.04 & 6955.46 &  & 3268.91 & 2898.08 & 6166.99 &  & 3449.44 & 3109.31 & 6558.75   \\ 
			&maxRSS & 566.68 & 442.21 &  &  & 481.87 & 370.62 &  &  & 564.80 & 366.65 &    \\ 
			&RSP & 0.85 & 0.79 &  &  & 0.81 & 0.71 &  &  & 0.83 & 0.74 &    \\ 
			&SP &  &  & 0.83 &  &  &  & 0.78 &  &  &  & 0.80   \\
			&MultiTCV &  &  & 0.052 &  &  &  & 0.828 &  &  &  & 0.161   \\ 
			&$\Sigma_{jj}$ & 0.060 & 0.104 &  &  & 0.160 & 0.291 &  &  & 0.074 & 0.222 &    \\ 
			&$\rho$ &  &  & 0.571 &  &  &  & 0.413 &  &  &  & 0.591   \\ 
			&\multicolumn{12}{l}{ }\\
			&\multicolumn{12}{l}{\bf Pancreas-Stomach}\\
			&SRSS & 2633.81 & 3481.92 & 6115.73 &  & 2549.47 & 2930.96 & 5480.43 &  & 2564.72 & 2931.19 & 5495.91   \\ 
			&maxRSS & 318.68 & 466.22 &  &  & 295.85 & 378.31 &  &  & 293.84 & 348.57 &    \\ 
			&RSP & 0.94 & 0.80 &  &  & 0.92 & 0.72 &  &  & 0.91 & 0.72 &    \\ 
			&SP &  &  & 0.87 &  &  &  & 0.82 &  &  &  & 0.82  \\ 
			&MultiTCV&  &  & 0.046 &  &  &  & 0.202 &  &  &  & 0.231   \\ 
			&$\Sigma_{jj}$ & 0.040 & 0.102 &  &  & 0.062 & 0.246 &  &  & 0.074 & 0.233  & \\ 
			&$\rho$  &  &  & 0.032 &  &  &  & 0.225 &  &  &  & 0.279   \\ 
			\hline
		\end{tabular}
	}
\end{table}

\begin{table}[b!]
	\centering
	\caption{\label{tabA8} Empirical smoothing criteria values obtained by the separables iCAR and LCAR spatial priors and the non-separable L$_j$CAR spatial prior for real data fitting multivariate models with three diseases.}
	\resizebox{\textwidth}{!}{
		\begin{tabular}{c|rrr|r|crrr|r|crrr|r|}
			\hline
			&\multicolumn{4}{c}{iCAR} &&\multicolumn{4}{c}{LCAR} &&\multicolumn{4}{c}{L$_j$CAR}\\
			\cline{2-5}\cline{7-10}\cline{12-15}
			&\multirow{2}{1.2cm}{\centering Colon ($j=1$)} & \multirow{2}{1.2cm}{\centering Pancreas ($j=2$)}&\multirow{2}{1.2cm}{\centering Stomach ($j=3$)}  &\multirow{2}{1cm}{\centering Total} &  & \multirow{2}{1.2cm}{\centering Colon ($j=1$)} & \multirow{2}{1.2cm}{\centering Pancreas ($j=2$)}&\multirow{2}{1.2cm}{\centering Stomach ($j=3$)}  &\multirow{2}{1cm}{\centering Total} & & \multirow{2}{1.2cm}{\centering Colon ($j=1$)} & \multirow{2}{1.2cm}{\centering Pancreas ($j=2$)}&\multirow{2}{1.2cm}{\centering Stomach ($j=3$)}   & \multirow{2}{1cm}{\centering Total} \\ 
			&&&&&&&&&&&&&&\\ 
			&&&&&&&&&&&&&&\\
			\hline
			\multicolumn{15}{l}{\bf G=47}\\
			SRSS& 16.30 & 18.78 & 15.52 & 50.60 &  & 16.73 & 16.82 & 15.38 & 48.93 &  & 15.45 & 18.46 & 14.93 & 48.84   \\ 
			maxRSS  & 1.99 & 4.16 & 2.12 &  &  & 2.19 & 3.50 & 2.48 &  &  & 1.78 & 3.75 & 2.00 &    \\ 
			RSP  & 0.29 & 0.38 & 0.18 &  &  & 0.29 & 0.33 & 0.18 &  &  & 0.27 & 0.37 & 0.18 &    \\ 
			SP  &  &  &  & 0.28 &  &  &  &  & 0.27 &  &  &  &  & 0.27   \\ 
			MultiTCV &  &   & & 0.193 &  &  &  && 0.442 &  & & &  & 0.295     \\ 
			$\Sigma_{jj}$& 0.055 & 0.041 & 0.149 &&  & 0.043 & 0.064 & 0.132 &&  & 0.053 & 0.039 & 0.138 &        \\ 
			$\rho_{12}$ &  &  &  &0.172 & &  &  &  &0.187&  &  &  &  &  0.330      \\ 
			$\rho_{23}$ & & &  &0.189 & & &  &  &0.105&  &  &  &  &   0.280     \\ 
			$\rho_{13}$ && & &  0.421& &  && & 0.160 &  & &  &  & 0.436       \\ 
			\multicolumn{15}{l}{ }\\
			\multicolumn{15}{l}{\bf G=100}\\
			SRSS & 264.10 & 221.83 & 195.53 & 681.47 &  & 273.11 & 212.77 & 180.62 & 666.5 &  & 245.88 & 210.22 & 175.77 & 631.87   \\ 
			maxRSS & 67.37 & 38.73 & 23.05 &  &  & 72.25 & 39.12 & 22.07 &  &  & 58.66 & 39.07 & 21.20 &    \\ 
			RSP & 0.64 & 0.77 & 0.44 &  &  & 0.65 & 0.74 & 0.42 &  &  & 0.60 & 0.73 & 0.41 &    \\ 
			SP &  &  &  & 0.63 &  &  &  &  & 0.62 &  &  &  &  & 0.59   \\ 
			MultiTCV&  & & & 0.000 &  &  &  && 0.001 &  &  &  && 0.000 \\
			$\Sigma_{jj}$& 0.057 & 0.034 & 0.142 &&  & 0.051 & 0.059 & 0.147 & & & 0.057 & 0.046 & 0.168 &\\
			$\rho_{12}$ &  &  &  &0.338 & &  &  &  & 0.134&  &  &  &  &  0.231   \\ 
			$\rho_{23}$ & & &  & 0.114 & & &  &  &0.112&  &  &  &  &   0.247    \\ 
			$\rho_{13}$ && & & 0.418& &  && & 0.258 &  & &  &  & 0.615       \\ 
			\multicolumn{15}{l}{ }\\
			\multicolumn{15}{l}{\bf G=300}\\
			SRSS & 3721.07 & 2620.43 & 3287.15 & 9628.65 &  & 3479.57 & 2428.92 & 3048.38 & 8956.87 &  & 3321.67 & 2400.36 & 3068.40 & 8790.43  \\ 
			maxRSS & 581.27 & 322.17 & 420.47 &  &  & 526.32 & 275.77 & 415.18 &  &  & 518.34 & 259.59 & 365.84 &    \\ 
			RSP & 0.87 & 0.94 & 0.77 &  &  & 0.83 & 0.88 & 0.73 &  &  & 0.81 & 0.89 & 0.75 &    \\ 
			SP &  &  &  & 0.87 &  &  &  &  & 0.82 &  &  &  &  & 0.81   \\ 
			MultiTCV&  &  &  & 0.000 &  &&  &  & 0.005 &  &  &&  & 0.006     \\ 
			$\Sigma_{jj}$& 0.048 & 0.033 & 0.133 &  && 0.081 & 0.098 & 0.236 & & & 0.107 & 0.099 & 0.231 & \\
			$\rho_{12}$ &  &  &  & 0.305  & &  &  &  & 0.213&  &  &  &  &   0.272   \\ 
			$\rho_{23}$ & & &  & 0.005  & & &  &  &0.307&  &  &  &  &   0.436   \\ 
			$\rho_{13}$ && & & 0.217 & &  && & 0.142 &  & &  &  & 0.351     \\ 
			\hline
		\end{tabular}
	}
\end{table}

We also fitted the models including three diseases to further assess the amount of smoothing achieved. The results are presented in \autoref{tabA8}. Consistent with the pairwise analyses, the iCAR prior generally produces the highest level of smoothing. However, in this setting the L$_j$CAR produces the lowest amount of smoothing as we increase $G$. As in the pairwise analysis, smoothing increases with $G$, and the smoothing amount becomes more similar among diseases. Interestingly, increasing the number of diseases modeled jointly does not lead to a higher proportion of smoothing. Comparing the SP results in \autoref{tabA7} and \autoref{tabA8}, both remain at similar levels (around 0.25 for $G=47$ and 0.8 for $G=300$). Examining the disease-specific RSP$_j$ values shows a similar pattern, with each disease exhibiting similar smoothing levels whether modeled in pairs or jointly. As the number of areas increases, a slight improvement in the RSP$_j$ values is observed when modeling all three diseases together.

\begin{figure}[b!]
	\centering
	\includegraphics[width=0.95\linewidth]{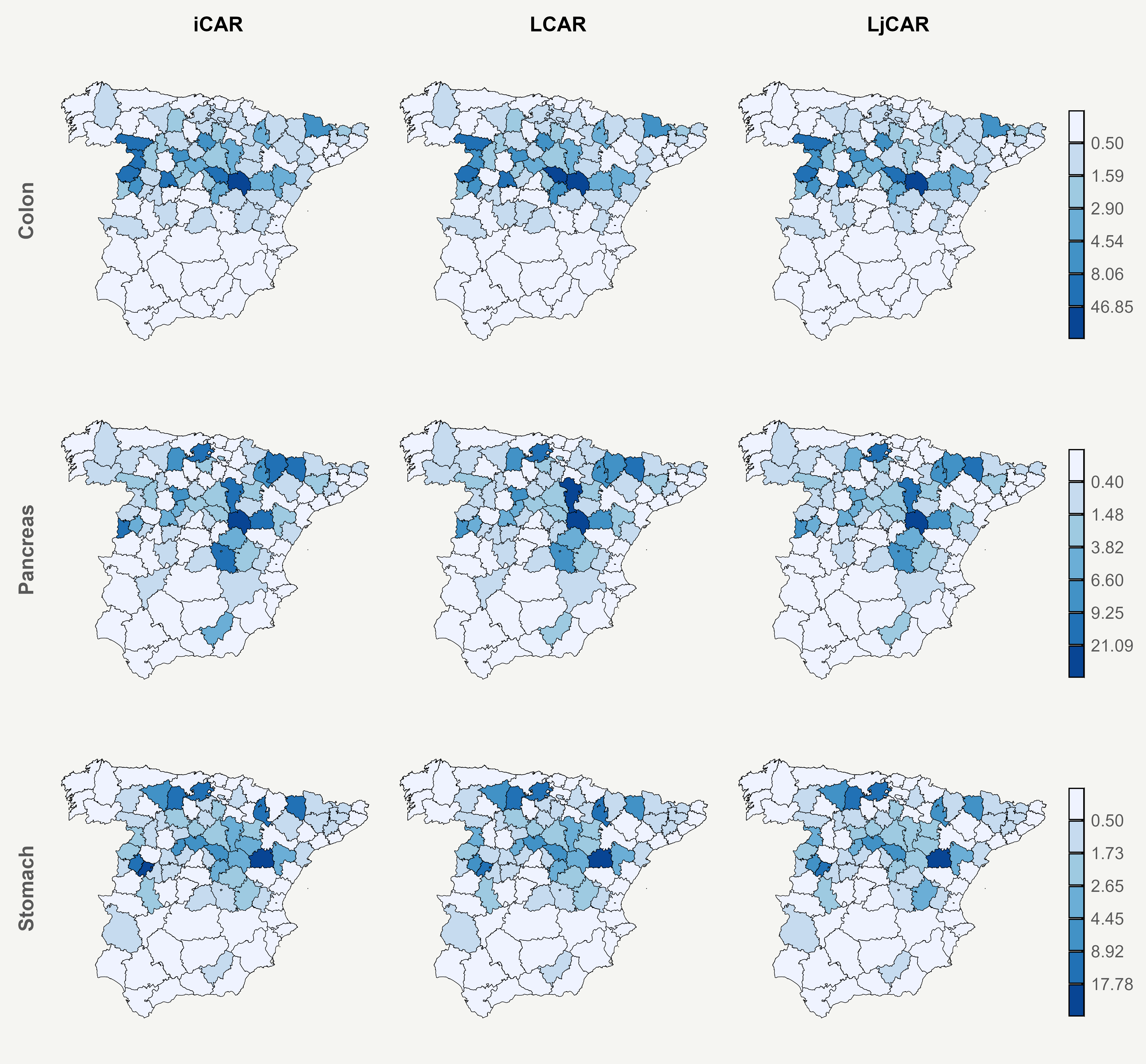}
	\caption{Area-specific components of the SRSS for each geographical unit $i$ for colon, pancreas and stomach cancer in continental Spain, shown for the 47 real provinces ($G = 100$).The legend reports the 50th, 75th, 85th, 90th, 95th, and 97th percentiles.}
	\label{fig6}
\end{figure}

\begin{figure}[t!]
	\centering
	\includegraphics[width=0.95\linewidth]{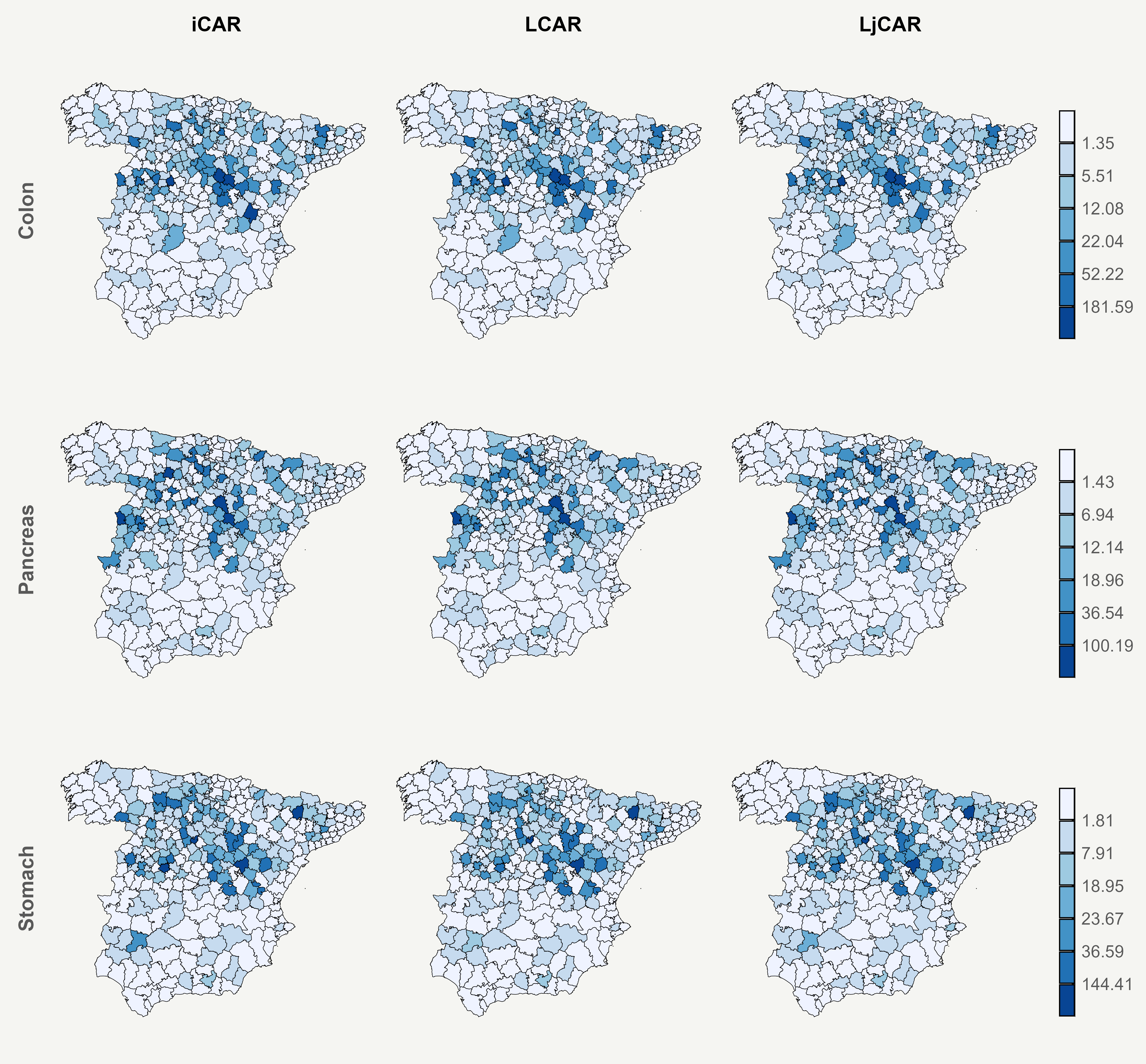}
	\caption{Area-specific components of the SRSS for each geographical unit $i$ for colon, pancreas and stomach cancer in continental Spain, shown for the 47 real provinces ($G = 300$).The legend reports the 50th, 75th, 85th, 90th, 95th, and 97th percentiles.}
	\label{fig7}
\end{figure}

As noted in the main paper, beyond comparing overall smoothing levels, \autoref{fig6} and \ref{fig7} show the area-specific components of the SRSS for each geographical unit $i$. Results are presented for models fitting the three diseases jointly at $G=100$ (see \autoref{fig6}) and $G=300$ (see \autoref{fig7}), with the legend reporting the 50th, 75th, 85th, 90th, 95th, and 97th percentiles. As the number of areas increases, it becomes more challenging to draw general conclusions about the spatial distribution of smoothing. Nevertheless, the pattern observed for $G=47$ is largely preserved, with areas that differ markedly from their neighbors tending to exhibit higher levels of smoothing.

	\end{appendices}
	
\end{document}